\documentclass[trackchanges, twocolumn]{aastex701}
\usepackage{enumitem}
\usepackage{subcaption}
\usepackage{tikz}
\usetikzlibrary{positioning}
\usetikzlibrary{angles,quotes}


\newcommand{\chandra}{\mbox{\em Chandra \/}}
\newcommand{\gaia}{\mbox{\em Gaia \/}}
\newcommand{\HST}{\mbox{\em HST \/}}
\newcommand{\Euclid}{\mbox{\em Euclid \/}}

\newcommand{\gravlens}{\mbox{\textsc{Gravlens} \/}}
\newcommand{\marx}{\mbox{\textsc{Marx} \/}}
\newcommand{\saotrace}{\mbox{\textsc{SAOTrace} \/}}
\newcommand{\ciao}{\mbox{\textsc{CIAO}\/}}
\newcommand{\sherpa}{\mbox{\textsc{Sherpa}\/}}
\newcommand{\ds}{\mbox{\textsc{SAOImage DS9}\/}}

\begin{document}

\title{Probing the X-Ray Structure of the $z = 2.38$ Strongly Lensed Quasar J1721+8842, the First Einstein Zig-Zag Lens, Down to $140\ \mathrm{pc}$}

\author[orcid=0009-0000-9392-3557,gname='Charlotte', sname='Eades']{Charlotte A. Eades}
\affiliation{Center for Astrophysics $|$ Harvard \& Smithsonian, 60 Garden Street, Cambridge, MA 02138, USA}
\affiliation{School of Physics and Astronomy, University of Southampton, Southampton, SO17 1BJ, UK}
\affiliation{Department of Physics, Astrophysics, University of Oxford, Denys Wilkinson Building, Keeble Road, Oxford, OX1 3RH, UK}
\email[show]{charlotte.eades@physics.ox.ac.uk}

\author[orcid=0000-0003-3814-6796,gname='Julia', sname='Sisk Reynes']{J\'{u}lia M.~Sisk Reyn\'{e}s}
\affiliation{Center for Astrophysics $|$ Harvard \& Smithsonian, 60 Garden Street, Cambridge, MA 02138, USA}
\affiliation{Department of Physics, University of Maryland, Baltimore County, 1000 Hilltop Circle, Baltimore, MD 21250, USA}
\email[hide]{jsiskre1@umbc.edu}  

\author[orcid=0000-0001-8252-4753,gname='Daniel', sname='Schwartz']{Daniel Schwartz}
\affiliation{Center for Astrophysics $|$ Harvard \& Smithsonian, 60 Garden Street, Cambridge, MA 02138, USA}
\email[hide]{dschwartz@cfa.harvard.edu }

\author[orcid=0000-0001-5655-4158,gname='Anna', sname='Barnacka']{Anna Barnacka}
\affiliation{Center for Astrophysics $|$ Harvard \& Smithsonian, 60 Garden Street, Cambridge, MA 02138, USA}
\email[hide]{abarnacka@cfa.harvard.edu}

\begin{abstract}
Sub-kiloparsec X-ray structures in active galactic nuclei (AGN), including outflows, jets, and dual AGN, provide key insights into their evolution and influence on host galaxies. In the local universe, \textit{Chandra X-ray Observatory's} state-of-the-art $0\farcs5$ angular resolution has resolved these structures, but at $z > 0.1$, milliarcsecond resolution is required to spatially resolve sub-kpc scales. We leverage the spatial amplification by strong gravitational lensing to investigate the X-ray structure of J1721+8842, a $z=2.38$ quasar lensed into six images. To reconstruct the optical emission position, we model the lensing mass distribution using the six quasar images, five detected by \gaia Data Release 3 (A-E) and one by \HST (F). Our model reproduces the five \gaia image positions to milliarcsecond precision. To locate the X-ray emission relative to the optical, we combine ray-trace simulations of archival \chandra observations with the lens model image position predictions using a double maximum-likelihood analysis. We constrain the X-ray centroid within a $0\farcs017\times0\farcs004\ (139\times30\ \mathrm{pc})$ ellipse centered $0\farcs005\ (44\ \mathrm{pc})$ away from the optical source position. Source-plane reconstructions of the soft-X-ray and hard-X-ray to optical sources demonstrate these emission regions are cospatial, allowing the optical source at $1\sigma$ confidence. These results indicate no evidence for kpc-scale X-ray jets, outflows, or dual AGN with projected separations exceeding $140\ \mathrm{pc}$. This method improves \textit{Chandra's} effective angular resolution by a factor of 30 at cosmological distances, enabling studies of sub-kpc AGN X-ray structure at cosmic noon.
\end{abstract}

\keywords{\uat{Active galactic nuclei}{16} --- \uat{Black hole physics}{159} --- \uat{Galaxies}{573} --- \uat{High energy astrophysics}{739} --- \uat{Strong gravitational lensing}{1643} --- \uat{X-ray astronomy}{1810}}

\section{Introduction}
\label{sec:introduction}

Active galactic nuclei (AGN) are compact regions found in $1-10\%$ of massive galaxies, powered by accretion onto supermassive black holes (SMBHs). Resolving and interpreting kpc- and sub-kpc-scale X-ray structure provides crucial insights into SMBH growth and SMBH--galaxy co-evolution. At $z\approx2 -3$, the `cosmic noon', both star formation and black hole accretion peak \citep{ForsterSchreiber2020}, producing a higher quasar number density than in the local universe, making this epoch central to our understanding of SMBH and galaxy evolution. 

X-ray jets and outflows could underlie self-regulating systems in which AGN feedback may inject sufficient energy into the host galaxy to quench both star formation and AGN fueling by unbinding galactic gas \citep{Wyithe2003}. SMBH pairs\footnote{Following \citet{Burke-Spolaor2018}, here we refer to dual AGN as SMBH pairs separated by $<10\ \mathrm{kpc}$, and to binary AGN as SMBH pairs with separations $<10\ \mathrm{pc}$.} trace hierarchical galaxy mergers and structure formation. AGN mergers with masses $>10^8\ M_\odot$ at cosmic noon are thought to be progenitors of the nHz stochastic gravitational wave background detected by \textit{NANOGrav} \citep{Agazie2023a}. Dual AGN systems where only one SMBH is active may appear as offset AGN, while other offset AGN may instead represent recoiling SMBHs displaced after mergers or gravitational interactions \citep{Blecha2011, Blecha2016}. Since these phenomena are expected to be common at cosmic noon and yet remain unresolved at high $z$ in the X-rays, spatially-resolved low-$z$ ($z\ll1$) analogues are essential for interpreting their possible high-$z$ counterparts. 

At these low redshifts, the \textit{Chandra X-ray Observatory} has revealed a wide range of AGN phenomena relevant to SMBH growth. In 3C 353, \citet{Kataoka2008} detected twin X-ray jets extending out $\approx5\ \mathrm{kpc}$ from the core with several luminous X-ray knots. Relativistic X-ray outflows launched $\sim10-100$ gravitational radii from the accretion disk have also been observed, for example in the Seyfert I galaxies NGC 3783 \citep{Gu2025}, NGC 4051 \citep{Reeves2026}, and NGC 4151 \citep{Xiang2025}. In addition, searches for offset and dual AGN candidates in galaxies at $z<0.4$ have identified systems with separations $1-100\ \mathrm{kpc}$ \citep{Koss2012, Barrows2016a, Hou2020a, Foord2020a, Foord2021a}. However, at $z>0.1$ the spatial resolution required to probe sub‑kpc structure lies beyond \textit{Chandra's} capabilities.

By exploiting the flux magnification and spatial amplification effects from strong gravitational lensing, we can probe spatial scales of $10\ \mathrm{pc}-1\ \mathrm{kpc}$ at cosmic noon -- otherwise inaccessible at such redshifts \citep{Barnacka2018a}. Caustics define positions in the source plane where the flux magnification by lensing is maximal and where nonlinear spatial amplification is most notable. That is, milliarcsecond-level changes in the positions of sources (in the source plane) close to caustics result in lensed image position changes of tens to hundreds of milliarcseconds in the image plane \citep[see figure 17 of][]{Barnacka2018a}. This constitutes the basis of the `caustic method' applied herein.  

We apply the caustic method to \textit{Gaia}, \textit{HST}, and \chandra observations of J1721+8842 (J1721 hereafter), the first Einstein zig-zag lens. This $z=2.38$ quasar is lensed into six images by two galaxies at $z=1.885$ and $z=0.184$, respectively. Previously, this method has been used to conduct astrometric analyses on five other quadruply lensed AGN, locating the X-ray emission relative to the optical and/or radio emission regions with milliarcsecond precision \citep{Schwartz2021a, Spingola2022a, Rogers2025a, Sisk-Reynes2026a}. This makes J1721 the sixth lensed AGN studied with this technique. Utilizing this method on highly magnified, quadruply lensed quasars can boost \textit{Chandra's} effective angular resolution by factors of $\sim10-100$.

The caustic method exploits the achromaticity of (macro)lensing, where the X-ray and optical lensed images will be spatially coincident \textit{if and only if} their emission regions are cospatial in the source plane. Any offset between the X-ray and optical source positions therefore produces a corresponding, amplified offset between their lensed image positions. In applications of the caustic method, a lens model is inferred from optical observations to reconstruct the spatial origin of the optical emission in the source plane. A high-fidelity simulation of the \chandra observation (using the absolute positions of the optical lensed images) is then generated. Trial X-ray sources are placed on a grid around the optical source, and each is propagated through the fixed lens model to predict its lensed image positions. The simulated X-ray lensed images are then placed at these predicted positions and compared against the real \chandra observation using a double maximum-likelihood (ML) analysis. The first ML analysis aligns the \chandra X-ray data to the ICRS system. The second ML procedure selects the trial source position across the full grid that best matches the \chandra\ observation. This method subsequently resolves the spatial origin of the X-ray emission in the quasar relative to the optical with milliarcsecond precision.

The above analysis enables AGN X-ray structure on the scale of $10\ \mathrm{pc}-1\ \mathrm{kpc}$ to be resolved at high redshifts. In addition, we extend the caustic method to separately determine the astrometric origin of soft and hard X-ray emission in lensed quasars \citep[following][]{Sisk-Reynes2026a}. This work demonstrates the first use of extending the caustic method to more complex multi-plane lens systems and is the first published use of \gaia Data Release 3 astrometry to constrain a lens model for J1721. 

This paper is organized as follows. In Section \ref{sec:J1712+8842} we present an overview of the discovery and previous works on J1721. Section \ref{sec:observations} outlines the archival \textit{Gaia}, \textit{HST}, and \chandra observations used in this work alongside the spectral analysis of these \chandra data. In Section \ref{sec:lens_model} we present the lens model constructed using the optical data. In Section \ref{sec:xray_method}, we outline our \chandra ray-trace simulations, which we then use to determine the spatial origin of soft, hard, and broadband X-ray emission in J1721 relative to the optical as outlined in Section \ref{sec:results}. Sections \ref{sec:discussion} and \ref{sec:conclusions} present our discussions (including future prospects) and conclusions, respectively.

Throughout this work we adopt the following cosmological parameter values: $H_0=67.5\ \mathrm{kms}^{-1}\mathrm{Mpc}^{-1}$, $\Omega_m=0.308$, $\Omega_\Lambda=0.692$ \citep{PlanckCollaboration2016}.

\section{J1721+8842 Discovery and Previous Works}\label{sec:J1712+8842}

\begin{figure}
    \centering
    \includegraphics[width=1.0\linewidth]{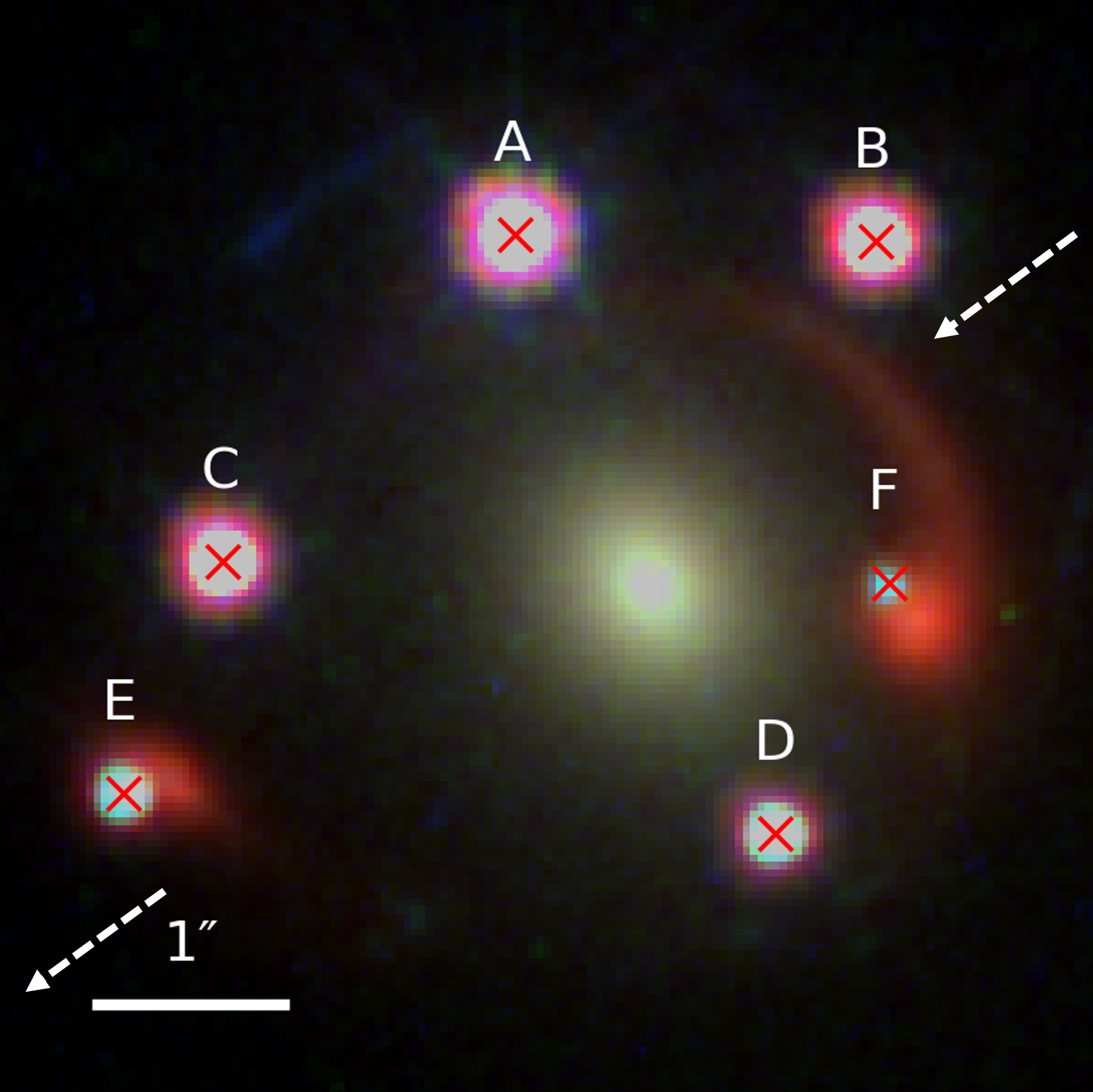}
    \caption{Composite \HST image of J1721 showing the F160W (red), F475X (blue), and F814W (green) filters. The red crosses correspond to the lensed quasar images with the astrometric positions reported by \gaia DR3 (images A -- E), and \HST for image F. The two red arcs are lensed images of the $z = 1.885$ deflector by the $z = 0.184$ deflector, as described in the text (Section \ref{sec:J1712+8842}). The white dashed arrows show the direction Figure \ref{fig:lens_model_diagram} looks across the system.}
    \label{fig:J1721_RGB}
\end{figure}

J1721 is a radio-quiet quasar at $z=2.38$ that is lensed by two galaxies at $z=1.885$ and $z=0.184$, respectively. The distant ($z=1.885$) galaxy is lensed by the nearby ($z=0.184$) galaxy, producing the two red arcs shown in Figure \ref{fig:J1721_RGB}. The nearby deflector lies in a galaxy group.

J1721 was first classified as a quadruply lensed quasar following the discovery of lensed images A--D in \gaia DR1 as part of a search for lensed quasars using \textit{WISE} colors \citep{Stern2012a} and \textit{PanSTARRS} imaging \citep{Chambers2018} by \citet{Lemon2018a}. Follow-up observations by \citet{Lemon2022a} with the \textit{Nordic Optical Telescope} (\textit{NOT}) revealed two additional lensed images (E-F), establishing J1721 as a unique six-image system. In parallel, the six lensed quasar images were also resolved at radio wavelengths with the \textit{Jansky Very Large Array} by \citet{Mangat2021a}, who suggested that the radio emission stems either from weak jets or star formation-related processes. 

Lens modeling by \citet{Lemon2022a} adopted a 2-source, 1-deflector scenario under the interpretation that the red arcs in Figure \ref{fig:J1721_RGB} were associated with the lensed host galaxy of a potential dual AGN. Under this scenario, the authors inferred a projected separation of $6\ \mathrm{kpc}\ (0\farcs73)$ for the putative AGN pair, where one of the AGN would be quadruply lensed (yielding images A -- D), while the other would be doubly lensed into images E and F. The radio study of \citet{Mangat2021a} also proposed a dual AGN scenario, suggesting that each SMBH had been ejected out of the host galaxy in a post-merger scenario. 

However, an imaging analysis of two-year continuous \textit{NOT} observations by \citet{Dux2025a} revealed that the light curves of all six lensed images correspond to the same underlying astrophysical source, disfavoring the dual AGN interpretation. Subsequent \textit{JWST}/NIRSpec observations revealed that the red arcs are located at $z=1.885$, confirming that they correspond to a distant, lensed deflector and not to the quasar host galaxy. \citet{Dux2025a} thus concluded that J1721 is a quasar lensed by two deflectors at different redshifts, yielding six quasar lensed images.

The zig-zag configuration -- where two images cross the optical axis between multi-plane deflectors -- was first predicted by \citet{Collett2016a} in a study into exotic lensing configurations that may be discovered in large-scale surveys with \Euclid and the \textit{Vera Rubin Legacy Survey of Space and Time} (\textit{LSST}). The authors predicted that the probability of a six-image zig-zag system occurring at $z>2$ was $10^{-8}$, making J1721 the first system found exhibiting this rare configuration. 

\section{Optical and X-ray observations}\label{sec:observations}

In this section we detail both the optical and X-ray observations used to perform the caustic method on J1721. Section \ref{sec:optical_observations} introduces the optical observations we used to conduct lens modeling to determine the spatial origin of optical emission in J1721. Section \ref{sec:xray_observations} introduces the X-ray observations used throughout our analysis. 

\subsection{\textit{Gaia} DR3 and \textit{HST} Observations}\label{sec:optical_observations}

\begin{table}
	\centering
    \setlength{\tabcolsep}{3pt}
	\caption{Image positions of J1721 observed in the optical band. These positions and their uncertainties are used as constraints for the lens model inferred in this work. The shifts in RA and Decl. are calculated from image A as in \citet{Sisk-Reynes2026a}, where the shifts in RA are multiplied by cos(Decl.). The ICRS coordinates of image A are $260.4541941^\circ,\ 88.70621466^\circ$.}
	\label{tab:optical_obs}
    \hspace*{-1.0cm}
	\begin{tabular}{lccr}
		\hline
		Image & R.A. Offset (mas) & Decl. Offset (mas) & Observation\\
		\hline
		A & $0.00\pm0.100$ & $0.00\pm0.102$ & \gaia DR3 \\
		B & $1331.40\pm0.187$ & $1401.45\pm0.177$ & \gaia DR3 \\
		C & $237.88\pm0.116$ & $-2348.28\pm0.119$ & \gaia DR3 \\
        D & $3320.64\pm0.432$ & $-1142.81\pm0.367$ & \gaia DR3 \\
		E & $795.33\pm1.555$ & $-3580.32\pm0.890$ & \gaia DR3 \\
		F & $2738.79\pm30$ & $217.20\pm30$ & \HST F475X \\
		\hline
	\end{tabular}
\end{table}

To reconstruct the optical source position with milliarcsecond precision, we require a lens model with highly precise astrometry for the observed lensed quasar images. The state-of-the-art astrometric precision is provided by \textit{Gaia}, since it has absolute astrometric errors $<0\farcs001$ for quasar lensed images.

Five of the lensed images (A-E) in J1721 are detected in the 3rd \gaia Data Release (DR3) with astrometric uncertainties of $0\farcs0001-0\farcs002$ as shown in Table \ref{tab:optical_obs}. Image F and the nearby deflector are not detected by \textit{Gaia} since their $g$-band magnitudes fall below \textit{Gaia's} detection threshold \citep[$g=20.7\ \mathrm{mag}$;][]{GaiaCollaboration2016, GaiaCollaboration2023}. 

J1721 was observed by \HST in 2019 (PI: T. Treu, Proposal ID: 15652), shown in Figure \ref{fig:J1721_RGB}. All six images, the nearby deflector, and the lensed arcs of the distant deflector are detected in \HST imaging using the Wide Field Camera 3 instrument in the F160W, F475X, and F814W filters. Currently, \HST achieves a baseline astrometric precision of $\sim0\farcs1$, based on the Guide Star Catalog 2.4.3 calibrated against \gaia DR3 positions \citep{Lasker2008, Sahu2021}. However, the relative astrometric precision achievable within the \HST image is significantly better than this absolute precision, as we demonstrate below.

To conduct lens modeling, we consider the \gaia DR3 image positions for A-E and take the \HST positions for image F and the nearby deflector. The \HST positions were measured in the F475X filter. The centroid of image F and its associated uncertainty were calculated with a Gaussian PSF fit using the \texttt{curve\_fit} command from the \textsc{scipy.optimize} package \citep{Virtanen2020}. We calculated the centroids of images A--E in the \HST image using the same method. The ICRS positions of images A--E from \gaia were then converted into pixel coordinates in the \HST image using the \HST WCS in \textsc{SAOImage\ DS9} 8.7 \citep{Joye2003a}. The positional offsets between the \gaia-derived and \HST measured centroids were calculated for images A-E, and the root-mean-square (RMS) of these residuals was used to quantify the astrometric alignment between the two telescopes. A total astrometric uncertainty of $0\farcs03$ for image F was found by adding the RMS error and the centroid error in quadrature, consistent with the uncertainty previously reported by \citet{Dux2025a}. This is the best available public astrometric measurement at the time of writing for the position of image F and is used throughout this analysis.

\subsection{Chandra X-Ray Observations}\label{sec:xray_observations}
\begin{table}
	\centering
    \setlength{\tabcolsep}{3pt}
	\caption{Archival \chandra observations of J1721. Only ObsIDs 23827 and 26415 are used for the astrometric analysis described in Section \ref{sec:xray_method}.}
	\label{tab:cxc_obs}
	\begin{tabular}{lccc}
		\hline
		ObsID & Start Date & Exposure Time (ks) & PI \\
		\hline
		ID 22020 & 2019-05-30 01:57:14 & 1.59 & Pooley \\
		ID 23827 & 2022-05-09 04:13:38 & 21.67 & Pooley \\
		ID 26415 & 2022-05-09 17:03:51 & 22.66 & Pooley \\
		\hline
	\end{tabular}
\end{table}

\chandra has observed J1721 in three separate segments using the ACIS-S instrument in VFAINT mode as summarized in Table \ref{tab:cxc_obs}. We reprocessed these data using the \texttt{chandra\_repro} command in \ciao\ 4.17 \citep{Fruscione2006a,Fruscione2026}.

In ObsIDs 23827 and 26415\footnote{DOI for the observations used: ~\href{https://doi.org/10.25574/cdc.568}{doi:10.25574/cdc.568}.}, a total of 771 counts were detected in the $0.5-7\ \mathrm{keV}$ band, with 456 counts in the soft band ($0.5-2\ \mathrm{keV}$), and 315 counts in the hard band ($2-7\ \mathrm{keV}$) in the regions containing images A-F (shown for ObsID 26415 in Figure \ref{fig:chandra_obs_and_sim}). Images A-E are detected at high significance above the background level of $3.59\times10^{-3}$ counts per ACIS pixel. Image F is detected at $3.3\sigma$ significance. Neither the nearby deflector nor the lensed arcs of the distant deflector are detected in the \chandra observations. Due to its poor S/N and negligible photon yield ($\sim 20$ X-ray counts), ObsID 22020 is unusable for our astrometric analysis and is therefore excluded.

We used \textsc{SAOImage\ DS9} 8.7 to place regions containing 90\% of the counts around each lensed image. The $0.5-7\ \mathrm{keV}$ observation from ObsID 26415 with these regions is shown in the left panel of Figure \ref{fig:chandra_obs_and_sim}. We then used the \texttt{specextract} tool in \ciao\ 4.17 to create the appropriate response files for each lensed image in each ObsID and extracted their spectra. 

The individual spectra of each lensed image (A-F) were found to be well-fitted by a primary power-law continuum modified by line-of-sight neutral Galactic absorption. The spectra were modeled with the \texttt{powlaw1d} and \texttt{xsphabs} models in \sherpa\ 4.18.0 \citep{Burke2025}, using a line-of-sight Galactic absorption of $5.85\times10^{20}\ \mathrm{cm}^{-2}$ in the direction of J1721 \citep{HI4PICollaboration:2016}. The best-fit models were found by finding suitable spectral models to minimize the Cash statistic of the fit in \sherpa\ \citep{Cash1979a}. The Cash statistic is the maximum likelihood value, assuming an underlying Poisson distribution of events. However, as shown by \citet{wilks1938} the maximum likelihood ratio test is almost independent of the underlying distribution. The inclusion of astrophysical complexity (such as the Fe K$\alpha$ emission line) is not statistically required by the data for either ObsID given their photon statistics. 

The summed fit for all images combined, averaged over ObsIDs 26415 and 23827, returns a photon index of $\Gamma_X=2.12\pm0.11$, and an unabsorbed flux of $2.95\pm0.05\times10^{-13}\ \mathrm{erg}\ \mathrm{s}^{-1}\ \mathrm{cm}^{-2}$. We note that these numbers are quoted for reference and are not used in the astrometric analysis. 

The ML analysis is performed on each individual ObsID (see Section \ref{sec:xray_method}). Step (iv) of that section describes how during the first ML test, simulated \chandra data is best aligned to the observed \chandra data to nullify any offset from the inaccuracy of \textit{Chandra's} pointing shown in Figure \ref{fig:chandra_obs_and_sim}. The second ML test is then used to define the best X-ray source position.

\section{Lens Model}\label{sec:lens_model}

\begin{figure*}[!t]
    \centering

    \hspace*{-1.3cm}
    \begin{tikzpicture}[scale=1.1,>=latex]
    
    \node at (-7.4,-4) {\large \textbf{Observer}};
    \node at (-2.0,-4) {\large \textbf{$z=0.184$}};
    \node at (6,-4) {\large \textbf{$z=1.885$}};
    \node at (9,-4) {\large \textbf{$z=2.38$}};

    \filldraw[fill=gray!30,draw=black] (-2,0.0) circle (0.15);
    \filldraw[fill=gray!30,draw=black] (6.0,0.25) circle (0.15);
    \filldraw[fill=black!100,draw=black] (9, 0.2) circle (0.05);
    \draw[thick,black] (-7.4, 0) -- (-7.0, 0.3);
    \draw[thick,black] (-7.4, 0) -- (-7.0, -0.3);
    \draw[thick,black] (-7.4,0) ++(37:0.4) arc[start angle=37, end angle=-37, radius=0.4];

    \draw[thick,black] (6.0, 0.57) -- (-2, 1.25);
    \draw[thick,black] (6.0, 0.57) -- (-2, 0.937);
    \draw[thick,black] (6.0, 0.57) -- (-2, 0.725);
    \draw[thick,black] (6.0, 0.57) -- (-2, -0.93);
    \draw[thick,black] (6.0, -0.29) -- (-2, -0.315);
    \draw[thick,black] (6.0,-0.29) -- (-2, 0.315);

    \draw[thick,black] (-2, 1.25) -- (-7,0);
    \draw[thick,black] (-2, 0.937) -- (-7,0);
    \draw[thick,black] (-2, 0.725) -- (-7,0);
    \draw[thick,black] (-2, 0.315) -- (-7,0);
    \draw[thick,black] (-2, -0.315) -- (-7,0);
    \draw[thick,black] (-2, -0.93) -- (-7,0);
    \draw[thick,blue, dash dot] (6, 0.57) -- (-7,0);
    \draw[thick,blue, dash dot] (6, -0.29) -- (-7,0);

    \draw[dashed,black] (9, 4.0) -- (-7,0);
    \draw[dashed,black] (9, 3.0) -- (-7,0);
    \draw[dashed,black] (9, 2.3) -- (-7,0);
    \draw[dashed,black] (9, 1.0) -- (-7,0);
    \draw[dashed,black] (9, -1.0) -- (-7,0);
    \draw[dashed,black] (9, -3.0) -- (-7,0);

    \draw[thick,black] (9,0.2) -- (6, 0.57);
    \draw[thick,black] (9,0.2) -- (6, -0.29);

    \filldraw[fill=red!60,draw=black] (9.0, 4.0) circle (0.08);
    \filldraw[fill=red!60,draw=black] ((9.0, 3.0) circle (0.08);
    \filldraw[fill=red!60,draw=black] ((9.0, 2.3) circle (0.08);
    \filldraw[fill=red!60,draw=black] ((9.0, -3.0) circle (0.08);
    \filldraw[fill=red!60,draw=black] ((9.0, 1.0) circle (0.08);
    \filldraw[fill=red!60,draw=black] ((9.0, -1.0) circle (0.08);
    \filldraw[fill=blue!60,draw=black] ((6.0, 0.56) circle (0.08);
    \filldraw[fill=blue!60,draw=black] ((6.0, -0.29) circle (0.08);

    \node[left] at (9.5, 4.0) {\small A};
    \node[left] at (9.5, 3.0) {\small B};
    \node[left] at (9.5, 2.3) {\small C};
    \node[left] at (9.5, -3.0) {\small D};
    \node[left] at (9.5, 1.0) {\small E};
    \node[left] at (9.5, -1.0) {\small F};

    \end{tikzpicture}

    \caption{Edge-on view (in the direction of the white dashed arrows in Figure \ref{fig:J1721_RGB}) of the lens model configuration assumed in this work. The gray circles represent the two deflectors, the black dot represents the source quasar, the blue circles represent the virtual images, and the red circles represent the six lensed images. The blue dot-dash lines correspond to the virtual images created at the distant deflector, while the solid black lines demonstrate the path light is taking through the system. The dashed black lines trace to where the six lensed images are observed. The distances and angles are not to scale.}
    \label{fig:lens_model_diagram}

\end{figure*}
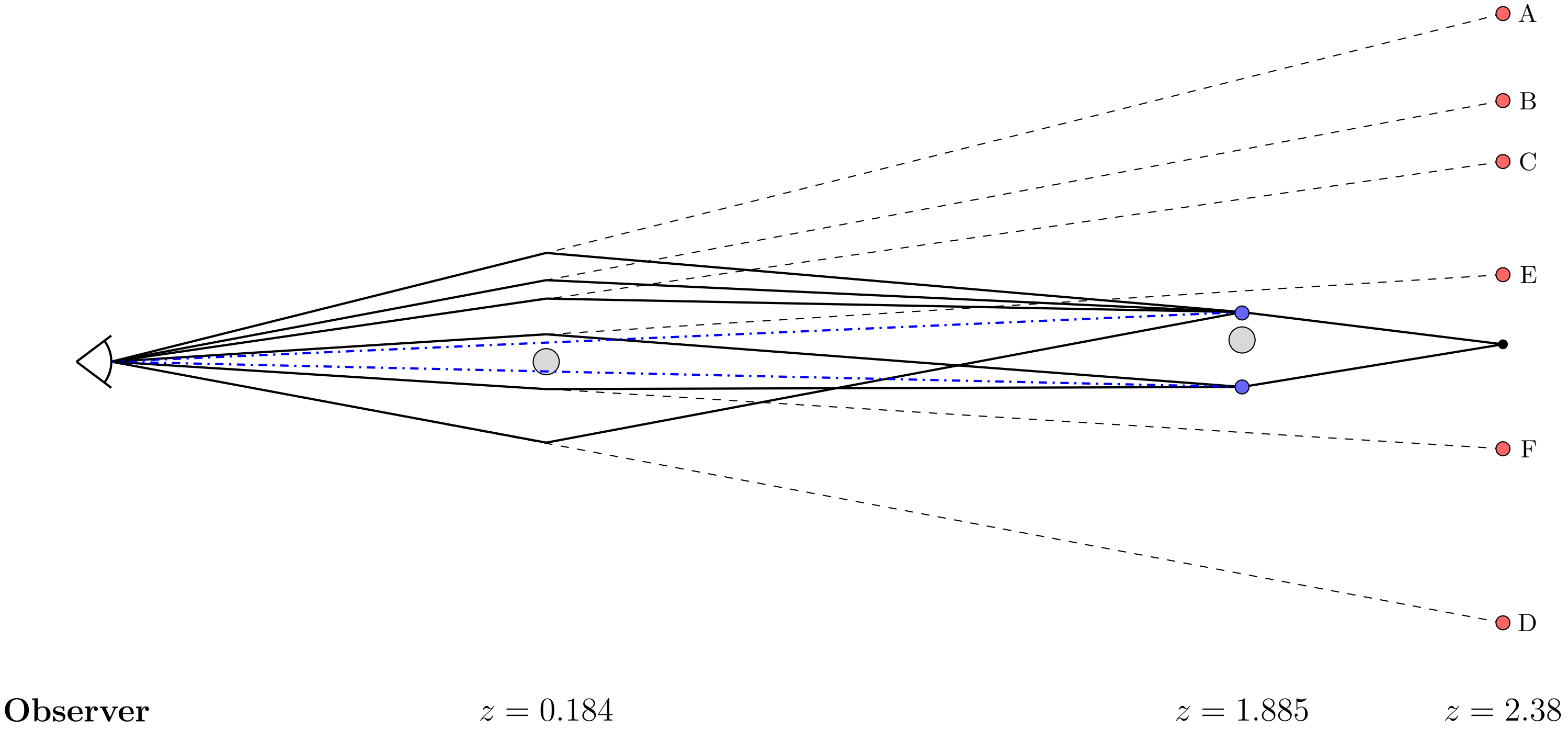

\begin{figure*}[!t]
    \centering

    \includegraphics[width=\linewidth]{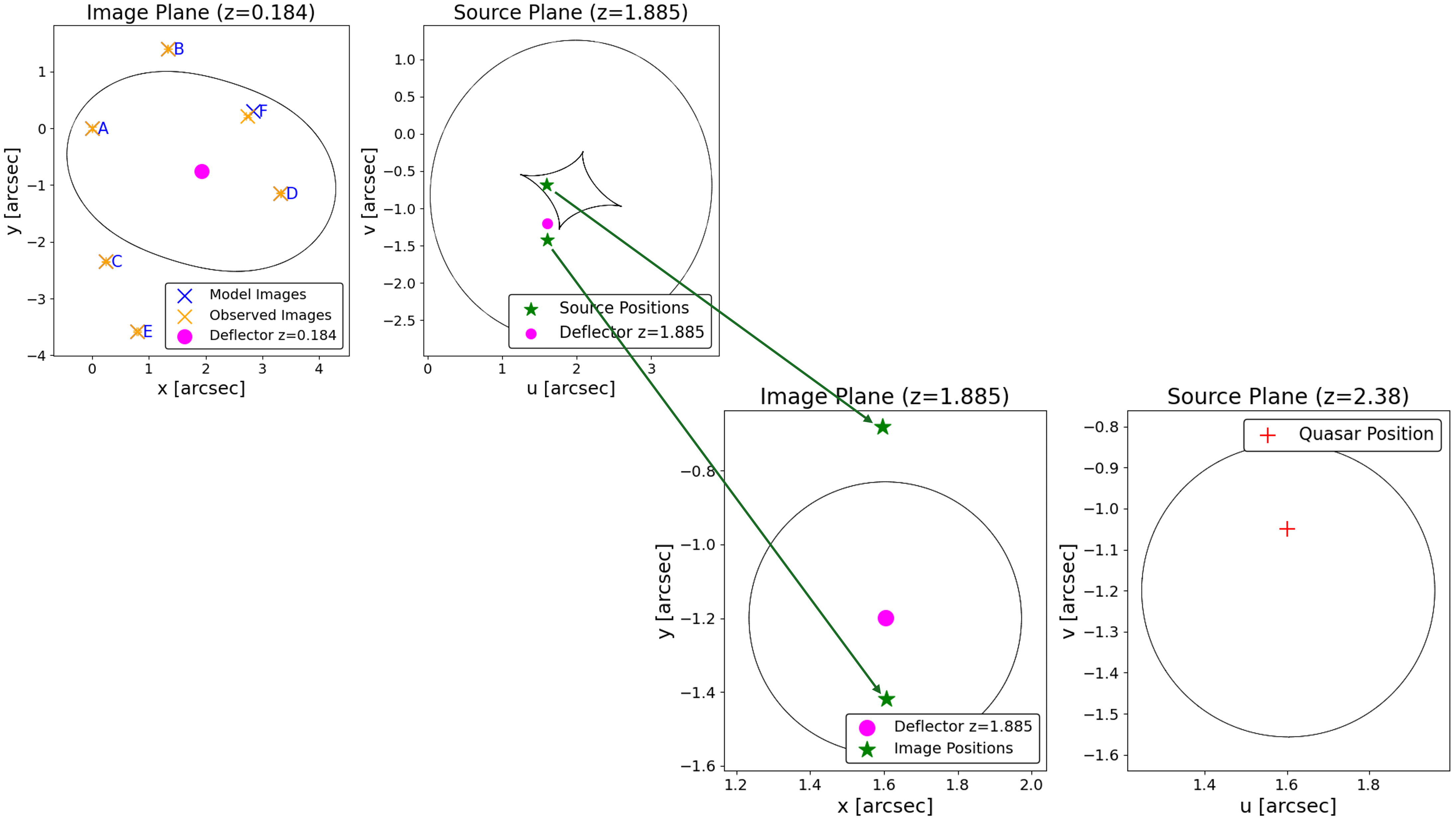}

    \caption{Source-plane and image-plane reconstructions for the lens model inferred in this work. Upper left: reconstructions for the $z = 0.184$ deflector, where the image plane shows the observed (orange) and model-predicted (blue) lensed image positions. The blue and orange crosses overlap for the five lensed images detected by \textit{Gaia}, while the astrometric uncertainties are too small to appreciate. The model images for A-E are reproduced within these uncertainties. Images A-D map to the source position (green star) inside the inner caustic, while images E and F map to the other source position. Lower right: reconstructions for the $z = 1.885$ deflector, showing how the two virtual images (green stars) from the previous model map into the quasar plane to provide the optical source position.}
    \label{fig:lens_model_combined}

\end{figure*}

\begin{table*}
	\centering
    \small
    \setlength{\tabcolsep}{1pt}
	\caption{Best-fit parameters for the nearby deflector (D1), distant deflector (D2), quadruply lensed virtual image (V1), doubly lensed virtual image (V2), and the source quasar returned from the lens model. $\theta_E$ is calculated from the mass parameter $b'$ in \gravlens with the relationship $\theta_E=b'\sqrt{1-e}$. The error bars come from the MCMC analysis as discussed in Section \ref{sec:lens_model_method_opt}. $\Delta\mathrm{R.A.}$ and $\Delta\mathrm{Decl.}$ are measured from the position of image A in the $z=0.184$ image plane, which is set to (0,0). PA$_\epsilon$ and PA$_\gamma$ are measured east of north.}
	\label{tab:params_table}
    \hspace*{-2.2cm}
	\begin{tabular}{lcccccccc} 
		\hline
		Component & z & $\theta_E$ ($\arcsec$) & $\Delta\mathrm{R.A.}$ ($\arcsec$) & $\Delta$Decl. ($\arcsec$) & $\epsilon$ & PA$_\epsilon$ ($^\circ$) & $\gamma$ & PA$_\gamma$ ($^\circ$) \\
		\hline
		D1 & 0.184 & $1.7662\pm0.0002$ & $1.9247\pm0.0004$ & $-0.7595\pm0.0001$ & $0.191\pm0.001$ & $76.828\pm0.009$ & $0.0864\pm0.0001$ & $69.326\pm0.010$\\
		D2 & 1.885 & $0.3692\pm0.0001$ & $1.6035\pm0.0001$ & $-1.995\pm0.001$ & 0.0 & 0.0 & 0.0 & 0.0\\
		Source & 2.38 & - & $1.6010\pm0.0001$ & $-1.0502\pm0.0001$ & - & - & - & -\\
        V1 & 1.885 & - & $1.5949\pm0.0001$ & $-0.6811\pm0.0001$ & - & - & - & -\\
        V2 & 1.885 & - & $1.6077\pm0.0001$ & $-1.4192\pm0.0001$ & - & - & - & -\\
        
		\hline
	\end{tabular}
\end{table*}

In this section we introduce the lens model used throughout our analysis. This is the first lens model of J1721 that incorporates the milliarcsecond and sub-milliarcsecond astrometric constraints provided by \gaia DR3 for images A -- E, which subsequently locates the spatial origin of optical emission in the source plane with milliarcsecond astrometric precision. While the position of image F comes from \textit{HST}, the lens model is set by the precise astrometry of images A -- E. The optical source provides the astrometric reference against which the position of the X-ray emission region is subsequently measured (as discussed in Section \ref{sec:xray_method}).

Several lens models for J1721 have been used in the literature under a quadruply lensed assumption \citep{Luhtaru2021, Schmidt2022, Schechter2024}, a dual AGN hypothesis \citep{Mangat2021a, Lemon2022a, Ertl2023}, and the zig-zag model \citep{Dux2025a, Schmidt2025a}, some of which are discussed further in Section \ref{sec:model_discussion}.

We infer a model for the effective gravitational potential lensing J1721 (i.e. a `lens model') using the \gaia and \HST image positions for images A--E and F, respectively, to reconstruct the position of optical emission in the source quasar at $z=2.38$ with the best possible astrometric precision. Specifically, we model the system plane by plane, firstly by tracing the six quasar images (A--F) back to two virtual images in the deflecting $z=1.885$ plane, and secondly by tracing these two virtual images back to the $z=2.38$ source quasar plane. This is illustrated in Figure \ref{fig:lens_model_diagram}. Since the quartet of lensed images A--D and the E--F doublet are both initially deflected by the $z=1.885$ deflector, yet comprise two distinct and compact groups of light rays at this redshift, as pointed out by \citet{Dux2025a} and further demonstrated by the lens model of \citet{Salama2026}, we interpret them as follows. Specifically we trace images A--D to one of the virtual images, lying inside the caustic of the nearby deflector, and images E and F to the other virtual source lying outside the caustic.

Our lens model is constrained so as to reproduce the six lensed image positions within astrometric uncertainties. The \gaia astrometry for images A--E drives the fit and location of the source. This subsequently allows us to determine the spatial origin of optical emission to milliarcsecond precision. We note that our lens model does not attempt to reproduce the image fluxes, since these could be affected by microlensing. We stress that microlensing has a negligible impact on image positions, inducing shifts only on microarcsecond scales \citep{Vernardos2024}, well below \textit{Gaia's} astrometric precision.

We model the nearby deflector with a singular isothermal ellipse plus external shear model (SIE $+\ \gamma$). From the extended \HST image, and as discussed in \citet{Schmidt2025a}, this deflector has several nearby perturbers, so the external shear is added to account for these. We fix the power-law index ($\alpha$) to 1.0, setting an isothermal model. The distant deflector is modeled as a singular isothermal sphere (SIS), where we fix both the ellipticity and shear to 0 for simplicity, since the physical nature of this deflector is not well known.

To determine the position of the distant deflector, we initially adopted the same $\Delta\mathrm{R.A.}$ and $\Delta\mathrm{Decl.}$ as the nearby deflector, motivated by the description of the two deflectors as being `nearly perfectly aligned with the lensed quasar source along the line of sight' by \citet{Schmidt2025a}. We assigned a large positional uncertainty of $1\farcs0$ to the distant deflector, allowing its position to vary freely to best fit the two virtual images. The positions of the red arcs were not used to constrain the position of the distant deflector.

\subsection{Lens model optimization}\label{sec:lens_model_method_opt}

We optimize the lens model plane by plane using the \textit{lensmodel} optimization routines within the \gravlens software of \citet{Keeton2001a}. The two deflectors are not optimized simultaneously. In the first optimization, only the mass model of the nearby deflector is varied, mapping the positions of the six images through the lensing potential of the nearby deflector to produce two virtual images in the $z=1.885$ plane. The second optimization is then performed, using the virtual image positions as constraints and varying only the mass model of the distant deflector to map the virtual images to the optical source position at $z=2.38$ (Figures \ref{fig:lens_model_diagram} and \ref{fig:lens_model_combined}). The lens model is optimized by allowing the parameters in Table \ref{tab:params_table} to vary (excluding redshift and parameters previously specified to be fixed) to minimize $\chi^2$, calculated from the difference between the model-predicted image positions and the observed image positions.

We subsequently used the Markov chain Monte Carlo (MCMC) sampler in \gravlens to find the $1\sigma$ uncertainties on the model parameters. These are shown in Table \ref{tab:params_table}. Their associated posterior distributions are shown in Appendix \ref{app:parameter_distributions}. Although the uncertainties are extremely tight, this behavior is consistent with the small astrometric constraints from \textit{Gaia}.

The left-hand panel of Figure \ref{fig:lens_model_combined} shows that images A-E are well reproduced by the lens model, with all model positions lying within the observational uncertainties derived from \textit{Gaia}. However, the position of image F is not reproduced within \textit{HST's} uncertainty. This is not a concern for the remainder of the analysis as image F contains so few X-ray counts, having a negligible effect on the final ML calculation and hence the X-ray emission position. We explore this further in Section \ref{sec:results}.

The model-predicted total enclosed masses within their respective Einstein radii are $2.9550\pm0.0003\times10^{11}\\ M_\odot$ and $2.5097\pm0.0007\times10^{11}\ M_\odot$ for the nearby and distant deflectors respectively. Our lens model predicts a total flux magnification (for images A--F) of $\mu = 70.9$. However, this magnification value can vary by approximately an order of magnitude, and should be taken with caution due to the presence of possible systematic uncertainties between lens models as discussed further in Section \ref{sec:model_discussion}.

\section{X-Ray Astrometric Analysis}\label{sec:xray_method}

\begin{figure*}
    \centering
    \figsettitle{Observed and Simulated 0.5-7 keV View of J1721+8842}
    \par\vspace{6pt}
    \includegraphics[width=0.9\linewidth]{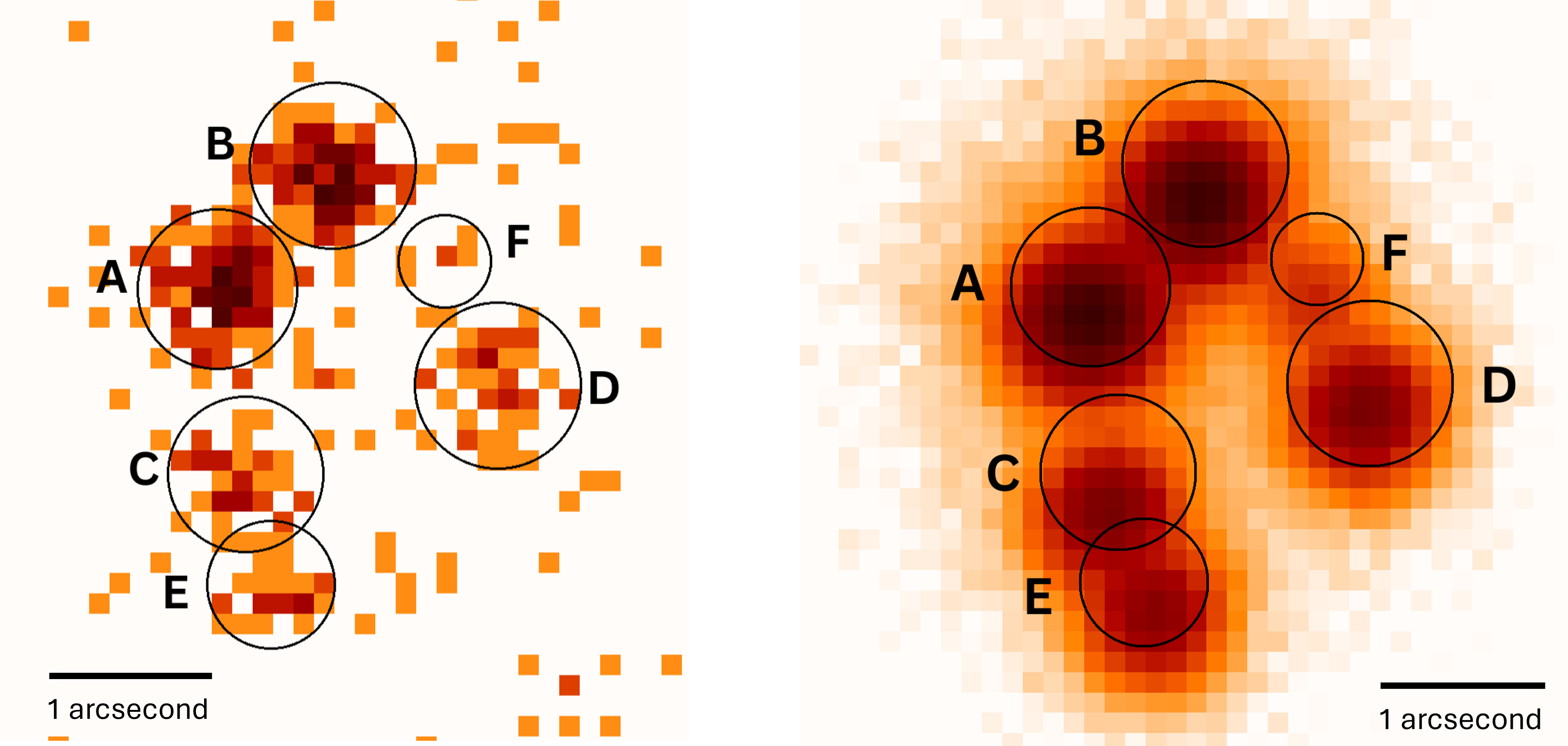}
    \caption{Left: $0.5-7\ \mathrm{keV}$ \chandra observation (ObsID 26415) of J1721 binned in $0\farcs25$ pixels. The regions used to extract the spectra described in Section \ref{sec:xray_observations} are shown by the black circles. Right: The result of 1\,000 merged simulations of ObsID 26415 placed at the positions given by \gaia and \textit{HST}. This configuration of simulated X-ray images would be the same as the observed X-ray image (left) if and only if the optical and X-ray sources are coincident. The black circles correspond to the same regions as the left plot. The systematic offset between the regions and the simulated image positions is due to the error in \textit{Chandra's} absolute pointing ($0\farcs4$ in this case). To resolve this offset we register the observed data relative to the predicted data in ICRS coordinates to a precision of about 15 mas by the first ML analysis for each putative source position as detailed in step (iv) in Section \ref{sec:xray_method}.}
    \label{fig:chandra_obs_and_sim}
\end{figure*}

\begin{table*}
	\centering
    \setlength{\tabcolsep}{7pt}
	\caption{Fluxes and photon indices of the X-ray images in ObsIDs 26415 and 23827. The broadband ($0.5-7\ \mathrm{keV}$), hard band ($2-7\ \mathrm{keV}$), and soft band ($0.5-2\ \mathrm{keV}$) fluxes are shown. The uncertainties in $\Gamma_X$ are the $1\sigma$ confidence level, and the flux uncertainties are taken from $\sqrt{N}$ (where $N$ is the number of X-ray counts). ObsID 26415 and the extraction regions are shown in Figure \ref{fig:chandra_obs_and_sim}.}
    \hspace*{-2.0cm}
	\label{tab:photon_index}
	\begin{tabular}{lcccccc} 
		\hline
		Parameter & ImA & ImB & ImC & ImD & ImE & ImF \\
		\hline
        26415 Region Radius ($\farcs$) & 0.962 & 1.002 & 0.938 & 1.000 & 0.772 & 0.558 \\
        23827 Region Radius ($\farcs$) & 1.191 & 1.182 & 0.944 & 1.210 & 0.915 & 0.769 \\
		26415 $F_{0.5-7} (10^{-14}\ \mathrm{erg}\ \mathrm{s}^{-1}\ \mathrm{cm}^{-2})$ & $9.61\pm0.81$ & $9.57\pm0.80$ & $3.03\pm0.49$ & $2.95\pm0.43$ & $2.12\pm0.41$ & $0.51\pm0.17$ \\
		23827 $F_{0.5-7}\ (10^{-14}\ \mathrm{erg}\ \mathrm{s}^{-1}\ \mathrm{cm}^{-2})$ & $10.89\pm0.88$ & $9.67\pm0.82$ & $2.27\pm0.38$ & $4.72\pm0.59$ & $1.22\pm0.35$ & $0.39\pm0.16$ \\
		26415 $F_{2-7}\ (10^{-14}\ \mathrm{erg}\ \mathrm{s}^{-1}\ \mathrm{cm}^{-2})$ & $4.54\pm0.69$ & $5.71\pm0.71$ & $1.14\pm0.29$ & $1.39\pm0.37$ & $0.78\pm0.23$ & $0.24\pm0.12$ \\
        23827 $F_{2-7}\ (10^{-14}\ \mathrm{erg}\ \mathrm{s}^{-1}\ \mathrm{cm}^{-2})$ & $5.80\pm0.8$ & $5.44\pm0.71$ & $1.22\pm0.38$ & $2.61\pm0.44$ & $0.48\pm0.16$ & $0.23\pm0.12$ \\
        26415 $F_{0.5-2}\ (10^{-14}\ \mathrm{erg}\ \mathrm{s}^{-1}\ \mathrm{cm}^{-2})$ & $5.07\pm0.54$ & $3.86\pm0.46$ & $1.89\pm0.35$ & $1.56\pm0.30$ & $1.34\pm0.34$ & $0.27\pm0.14$ \\
        23827 $F_{0.5-2}\ (10^{-14}\ \mathrm{erg}\ \mathrm{s}^{-1}\ \mathrm{cm}^{-2})$ & $5.09\pm0.51$ & $4.25\pm0.49$ & $1.05\pm0.25$ & $2.11\pm0.37$ & $1.39\pm2.77$ & $1.63\pm1.09$ \\
        26415 $\Gamma_X$ & $2.14\pm0.20$  & $1.74\pm0.20$  & $2.45\pm0.41$  & $2.15\pm0.38$ & $2.49\pm0.51$ & $2.15\pm1.00$ \\
		23827 $\Gamma_X$ & $1.95\pm0.19$  & $1.86\pm0.20$  & $1.94\pm0.44$  & $1.89\pm0.29$ & $2.39\pm0.67$ & $1.80\pm1.18$ \\
		\hline
	\end{tabular}
\end{table*}

The caustic method was first demonstrated by \citet{Barnacka2018a} and first applied to a quadruply lensed quasar by \citet{Spingola2020a}. Here, we describe the steps of the method we follow to reconstruct the source-plane X-ray emission in J1721 with milliarcsecond precision relative to the optical source.

\begin{figure*}
    \centering
    \includegraphics[width=1.0\linewidth]{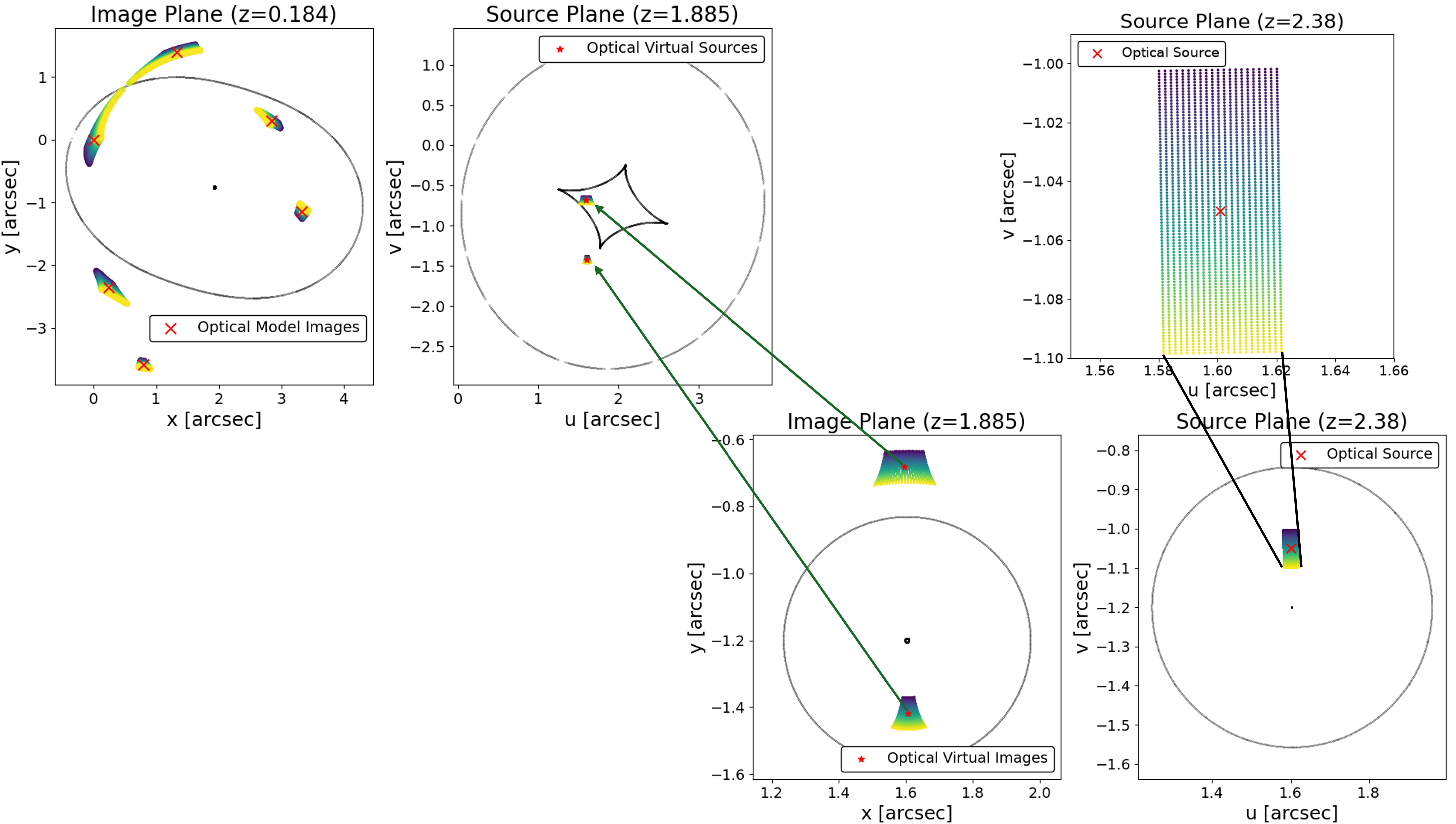}
    \caption{Top Right: The grid of trial sources spaced $0\farcs001$ perpendicular and $0\farcs002$ parallel to the caustic. Bottom Right: The trial sources as they propagate past the distant deflector to produce two virtual images. The red stars correspond to the virtual images produced by the optical source centroid, where the red cross in the $z=2.38$ source plane is the optical source centroid. Top Left: The trial sources as they propagate past the nearby deflector to produce six images. The color map indicates which trial source position maps to each set of image positions as they propagate through the mass model.}
    \label{fig:putative_model}
\end{figure*}

\begin{enumerate}[label=(\roman*)]
    \item \saotrace 2.0.6.1 \citep{Jerius2004a} and \marx 6.0.1 \citep{Davis2012} are used to create high-fidelity simulations of the six X-ray lensed images. Each simulation propagates X-ray photons through \textit{Chandra's} optics and onto the ACIS detector.
    
    \saotrace first takes the spectrum from each lensed X-ray image of a given observation from \sherpa\ (Section \ref{sec:xray_observations}) and the actual aspect solution, incorporating the dither pattern of the telescope, start time, and duration for that observation. This produces a simulation of the observation through \textit{Chandra's} optics. \marx\ subsequently projects the rays generated in \saotrace to the detector plane to create an event file that can then be analyzed. 
    
    To reduce their respective centroid error, 1\,000 simulations are performed for each X-ray image in each ObsID at the measured \gaia positions (A-E) and \HST position (F) and combined, yielding high-fidelity X-ray images, reducing the error to just 3\% of that in the observed image \citep{Auchettl2015}.
    \item To locate the position of X-ray emission relative to the optical in the source plane, a grid of 2\,017 trial positions is constructed around the optical source predicted by the lens model (Section \ref{sec:lens_model}). 
    
    The grid is constructed such that the trial sources have spacings of $0\farcs001$ perpendicular to the caustic and $0\farcs002$ parallel to it (top right plot in Figure \ref{fig:putative_model}). This choice of spacing reflects that nonlinear changes in spatial amplification and flux magnification are more pronounced in directions perpendicular to the caustic \citep[see Figure 17 of][]{Barnacka2018a}.
    \item Each trial source in the grid is forward-modeled through the fixed lens model found in Section \ref{sec:lens_model_method_opt} and its six image positions and magnifications are recorded.
    
    The simulated X-ray images are placed at these image positions and binned into a $20\times20$ array of $0\farcs246$ half-ACIS pixel bins. The counts are normalized to the expected total number of counts found in the \chandra observations. The number of counts per pixel from the trial source image positions is compared to the observed data by calculating the Cash statistic, $C$ \citep{Cash1979a}, suited for Poisson-distributed data:

    \begin{equation}
        C=-2P=-2\sum^{N=400}_{i=1}ln(\lambda^{n_i}_iexp(-\lambda_i)/n_i!)
        \label{eq:Cash_stat}
    \end{equation}

    where $N=400$ is the number of pixels, $n_i$ is the observed number of counts per pixel, and $\lambda_i$ is the number of model-predicted counts per pixel.
    \item We perform a first ML test by rastering the six images in steps of $0\farcs015$, and then calculating $C$ for each trial source using Equation \ref{eq:Cash_stat} to find the best match to the observed \chandra data. This step mitigates any offset in \textit{Chandra's} absolute astrometry. 

    To refine the X-ray source localization, we interpolate within the raster grid and recalculate the Cash statistic from Equation \ref{eq:Cash_stat}. We then calculate $P=\frac{1}{2}C$, to find the probability the trial source produces the observed X-ray images.

    These calculations are made for each trial position in the grid to find the source with the maximum likelihood of producing the observed X-ray images, defining the most likely X-ray source. Finally, confidence regions of 68\%, 95.5\%, and 99.7\% are derived from the likelihood differences, distributed as $\chi^2$ for two degrees of freedom (R.A. and Decl. of each trial source) \citep{Cash1979a}.

    \item To find the total likelihood, steps (i), (iii)-(iv) are repeated for both ObsIDs and their Cash statistics are summed.
    
    \item We also repeat this analysis for the hard X-ray band and soft X-ray band to investigate the scales of their respective emitting regions. Due to the lower number of counts, we use a coarser grid of trial sources in step (ii). This grid consists of 745 trial sources spaced $0\farcs003$ both perpendicular and parallel to the caustic.
\end{enumerate}

\section{Results}\label{sec:results}

\begin{figure*}
    \centering
    \begin{subfigure}{0.9\linewidth}
        \centering
        \caption{Broadband X-ray source, $0.5-7\ \mathrm{keV}$}
        \includegraphics[width=\linewidth]{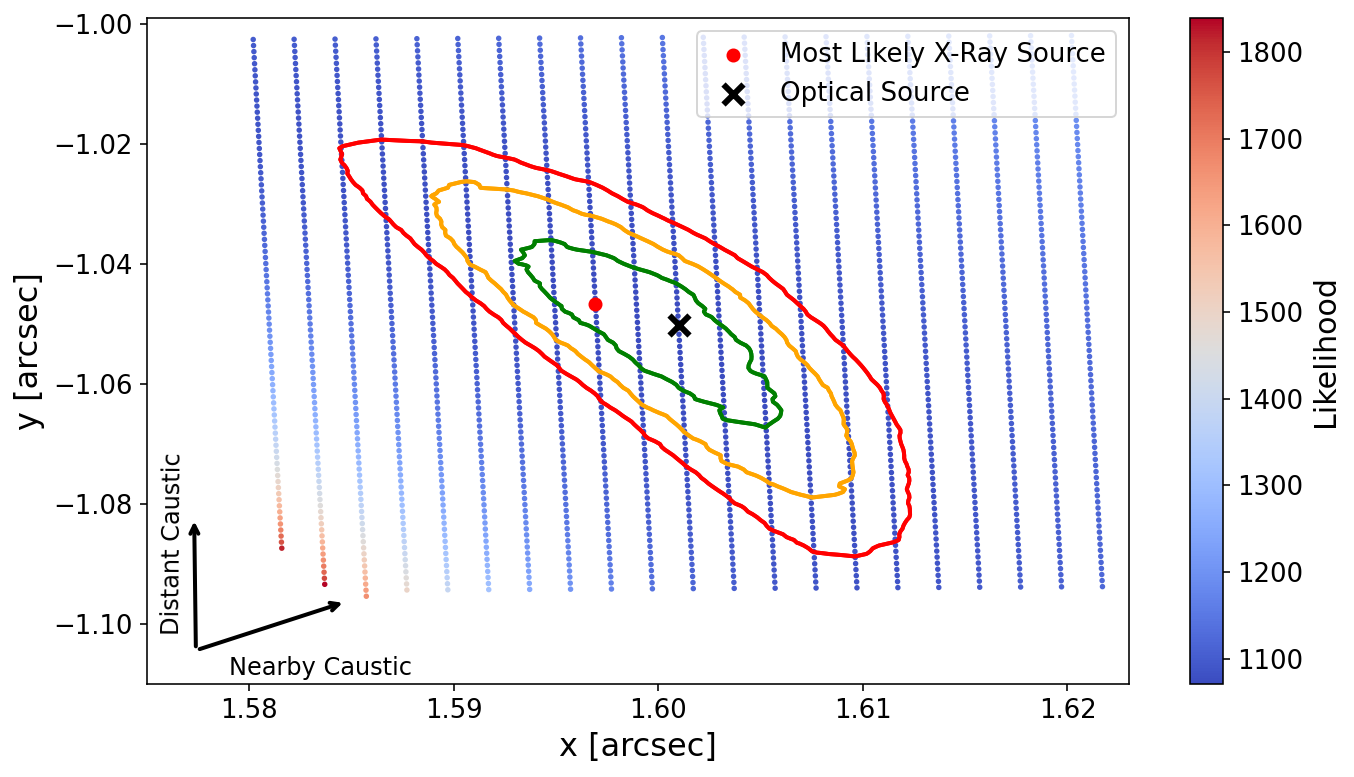}
        \label{fig:contour_plot}
    \end{subfigure}

    \vspace{-0.8cm}
    
    \begin{subfigure}{0.49\linewidth}
        \centering
        \caption{Hard X-ray source, $2-7\ \mathrm{keV}$}
        \includegraphics[width=\linewidth]{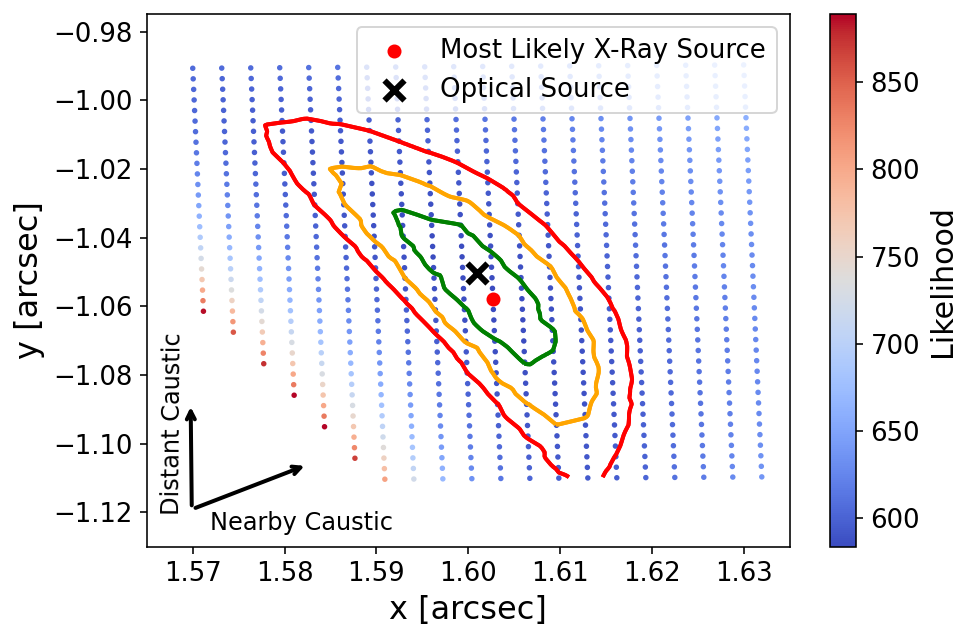}
        \label{fig:hard_contour_plot}
    \end{subfigure}
    \hfill
    \begin{subfigure}{0.49\linewidth}
        \centering
        \caption{Soft X-ray source, $0.5-2\ \mathrm{keV}$}
        \includegraphics[width=\linewidth]{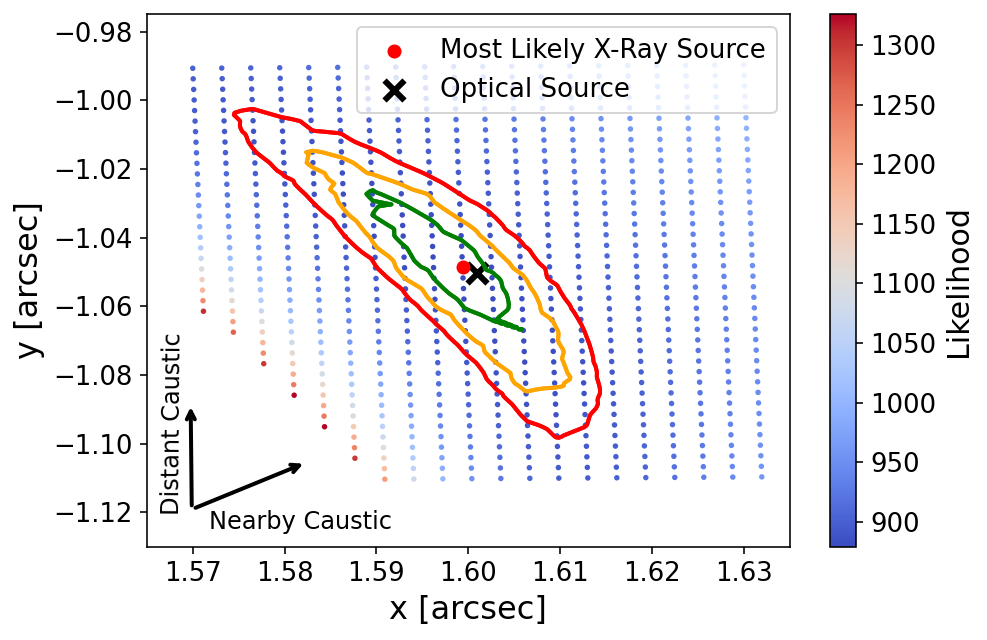}
        \label{fig:soft_contour_plot}
    \end{subfigure}
    \caption{Source-plane reconstructions for the X-ray emission relative to the optical source. The contours are at 68\% (green), 95.5\% (orange), and 99.7\% (red) confidence levels. Each trial source in the grid is color-coded by the ML resulting from the analysis of both ObsIDs (Equation \ref{eq:Cash_stat}), with the most likely position for the X-ray source corresponding to the point with the lowest likelihood value. The grid of trial positions used for the soft X-ray and hard X-ray reconstructions is coarser than the one used for the broadband X-ray reconstruction (see step (v) of Section \ref{sec:xray_method}). The arrows indicate the directions towards the caustic from the distant deflector and the inner caustic from the nearby deflector, respectively.}
    \label{fig:contour}
\end{figure*}

The results from the merged X-ray simulations are shown in the right panel of Figure \ref{fig:chandra_obs_and_sim}. A diagram showing the grid of trial X-ray positions, and their respective lensed image positions given the fixed geometry of the lens model, is shown in Figure \ref{fig:putative_model}.

Following the steps outlined in Section \ref{sec:xray_method}, we determined the spatial origin of X-ray emission to milliarcsecond precision relative to the optical. Figure \ref{fig:contour_plot} shows the 68\%, 95.5\%, and 99.7\% confidence regions for the position of the broadband X-ray emission region in J1721. The 68\% positional contour is well-described by a $0\farcs017\times0\farcs004$ ellipse (where the axes' dimensions reflect that nonlinear variations in spatial amplification are more pronounced perpendicular to the caustic). The optical source lies within this contour, $0\farcs005$ away from the center, indicating the emission regions are cospatial at the 1$\sigma$ level. At the source redshift of $z=2.38$, the angular scale is $0\farcs122\ \mathrm{kpc}^{-1}$, indicating that the most likely X-ray and optical emission regions are offset by $43.54\ \mathrm{pc}$ (projected separation), cospatial within the $1\sigma$ ellipse of $139\times30\ \mathrm{pc}$.

Figures \ref{fig:hard_contour_plot} and \ref{fig:soft_contour_plot} show that the soft and hard X-ray emission regions, located from the spectrally resolved astrometric reconstruction, are also found to be cospatial with both the optical emission and with each other at the $1\sigma$ level. The offset between the most likely hard-X-ray source and optical source is $0\farcs008\ (64.02\ \mathrm{pc})$ within a $1\sigma$ ellipse of $0\farcs024\times0\farcs005\ (196\times39\ \mathrm{pc})$. The offset between the most likely soft X-ray source and optical source is $0\farcs002\ (17.97\ \mathrm{pc})$ within a $1\sigma$ ellipse of $0\farcs022\times0\farcs004\ (177\times33\ \mathrm{pc})$. The offset between the hard and soft X-ray positions is $0\farcs010\ (79.58\ \mathrm{pc})$ at the quasar redshift. 

Equation \ref{eq:Cash_stat} indicates that the ML is dominated by the lensed images with the highest number of X-ray counts. Since image F contributes only 11 of the 771 total counts, its less accurate predicted position in our lens model is not expected to significantly affect the astrometric X-ray reconstructions presented here. However, to test this, we carried out the individual broadband-X-ray-to-optical, soft-X-ray-to-optical, and hard-X-ray-to-optical source reconstructions excluding the counts in image F. We subsequently verified that the exclusion of its corresponding counts from the double ML calculation does not affect the inference of the X-ray source position. Specifically, the individual projected offsets between the three X-ray emission regions and the optical source do not move outside of their previously inferred $1\sigma$ ellipses, shown in Appendix \ref{app:F_removed}. 

To further verify that our X-ray--to--optical source reconstruction remains insensitive to whether the X-ray counts in image F are considered, we also performed a caustic-method-reconstruction by changing the position of image F used in the ML calculation from the \HST position to that predicted by our lens model (Section \ref{sec:lens_model}). Unsurprisingly, this had no significant effect on the final result. This is shown in Appendix \ref{app:F_moved}.

\section{Discussion}\label{sec:discussion}

Here we discuss the results of our lens model, X-ray spectral analysis, and X-ray astrometric analysis of J1721. In Section \ref{sec:model_discussion} we discuss our choice of lens model and compare it to previous models of the system. Section \ref{sec:xray_discussion} interprets the X-ray spectral properties of J1721, including its intrinsic luminosity and the flux variability observed between images. In Section \ref{sec:astrometry_discussion}, we discuss the implications of our constraint on the position of the X-ray source relative to the optical, specifically for the presence of a dual AGN, an extended X-ray jet, and other compact-scale X-ray structures, placing J1721 in the context of the broader sample of lensed AGN studied with the caustic method. Finally, Section \ref{sec:future} presents the future prospects of the caustic method with upcoming missions.

\subsection{Lens Model}\label{sec:model_discussion}

Our lens model is the only published model of J1721 that takes advantage of the astrometric precision of \gaia DR3 to reconstruct the quasar source-plane emission position with milliarcsecond precision, shown in Table \ref{tab:params_table}.

An SIE was found to be a good parametrization for the nearby deflector as it represents the massive elliptical galaxy revealed by \HST imaging (Figure \ref{fig:J1721_RGB}). The external shear was added to account for any additional perturbations from nearby deflectors as the nearby deflector lies in a galaxy group. Time-delay studies have also modeled individual galaxies alongside an external shear component. \citet{Dux2025a} and \citet{Schmidt2025a} individually model four nearby galaxies as SIS to take into account any additional effects that are not well described by external shear for J1721. We tested the inclusion of these four perturbers in the $z=0.184$ plane of our model, taking $\theta_E$ for each perturber from \citet{Schmidt2025a}, allowing their $\Delta\mathrm{R.A.}$, $\Delta\mathrm{Decl.}$ and $\theta_E$ to vary alongside the nearby deflector parameters previously specified in Section \ref{sec:lens_model_method_opt}. Ultimately the inclusion of these perturbers did not improve the model fit, producing an image plane rms of $126.8\  \mathrm{mas}$ compared to $55.4\ \mathrm{mas}$ for the model without the perturbers.

However, the specific choice of lens model is not critical for the caustic method. Since gravitational lensing is achromatic, coincident optical and X-ray emission regions will produce identical absolute image positions, while any source-plane offset leads to corresponding shifts in the image positions independent of the lens model. Hence, the same offset should consistently be found within astrometric precision by the double ML calculation.~This was explored in \citet{Rogers2025a}, where the authors tested both an SIE and a Navarro–Frenk–White \citep{Navarro1997a} profile for the quadruply lensed quasar HE 0435-1223.~Both models agreed that the X-ray position was offset from the optical position in the source quasar at above the $3\sigma$ level.

Images A-E were successfully reproduced to milliarcsecond precision by the lens model, with only image F failing to be accurately reproduced. However, image F has positional uncertainty taken from \textit{HST}, whereas images A-E dominate the uncertainty due to their astrometric constraints from \textit{Gaia}. These astrometric uncertainties are seen in Figure \ref{fig:corner_6image}. We found that both deflectors cannot simultaneously reproduce both sets of images well. Fitting A-D prevented an accurate fit to E and F, and vice versa. The failure to reproduce image F could be due to the need for added complexity in the lens model \citep[such as additional deflectors;][]{Schmidt2025a}, or ``a combination of extinction, millilensing, and microlensing'' as suggested by \citet{Lemon2022a}. As a result, and since image F has so few X-ray counts, we prioritized fitting images A-E by placing a larger positional uncertainty of $0\farcs1$ on image F to reduce its constraining power under the assumption that image F would not affect the final astrometric analysis. This was justified by the various tests described in Section \ref{sec:results} and shown in Appendices \ref{app:F_removed} and \ref{app:F_moved}, which demonstrated that both the lens model position and inclusion of image F had no significant effect on the final result.

\citet{Lemon2022a} constructed a lens model for J1721 using the six image positions from the \textit{NOT} under a dual AGN scenario. Interestingly, their predicted values for $\theta_E$, $\epsilon$, $\mathrm{PA}_\epsilon$, $\gamma$, and $\mathrm{PA}_\gamma$ for the nearby deflector are consistent with the values found in this work. However, this was anticipated as although we do not use the dual AGN assumption for our model, the reconstruction of the nearby deflector to produce two sources (AGN in \citet{Lemon2022a}; virtual images in this work) from images A-D and E-F is the same in both models. \citet{Lemon2022a} find a projected separation of $0\farcs7$ ($6\ \mathrm{kpc}$) between the two putative AGN. Both \citet{Mangat2021a} and \citet{Dux2025a} recovered this separation in their respective models. This separation is also reproduced by our model. The two virtual images have a separation of $0\farcs68$.

Once J1721 had been determined to be a single AGN by \citet{Dux2025a}, both \citet{Dux2025a} and \citet{Schmidt2025a} constructed models using \textit{HST} imaging, modeling the nearby deflector with a power-law elliptical mass distribution and the distant deflector as a SIS. The authors obtained a total enclosed mass of $2.74\pm0.04\times10^{11}\ M_\odot$ for the distant deflector, similar to the value of $2.5097\pm0.0007\times10^{11}\ M_\odot$ found in this work. However, the difference between these two values is larger than the quoted statistical uncertainties, highlighting the presence of possible systematic uncertainties associated with the lens models. 

In particular, although the $\Delta\mathrm{Decl.}$ of the distant deflector is tightly constrained within our lens model, the virtual image position constraints do not uniquely determine its $\Delta\mathrm{Decl.}$ across the full range of plausible lens models. The $\Delta\mathrm{Decl.}$ of the deflector can vary by $\sim1\farcs0$ while still reproducing the virtual image positions, shifting the caustic closer or further from the virtual images, resulting in different magnifications. These differences in magnification translate into systematic uncertainties in the source-plane spatial scales inferred from the lens model. As we find the X-ray and optical sources in J1721 to be cospatial in the source plane, this systematic uncertainty does not affect our conclusion that there is no significant offset between them, but it does affect the physical scale and hence the upper limit placed on their separation. We have therefore attempted to model J1721 using a mass model that is supported by the available data and consistent with the Einstein zig-zag theory of \citet{Collett2016a}, thereby minimizing this effect. 

As a self-consistency check of our mass model, we use the observed (image-plane) positions of the red arcs in Figure \ref{fig:J1721_RGB} to infer the position of the distant deflector in the $z = 1.885$ plane. Specifically,~by keeping our lens model fixed,~we solve the lens equation for several image-plane positions along the arcs and compare the resulting source positions with the $z = 1.885$ deflector position predicted by our lens model. We find that the source positions found from the arcs are slightly offset from the source position obtained from the mass model, with offsets ranging from $\sim0\farcs05-0\farcs2$. This astrometric offset is not a cause for concern as the arcs are extended and we do not model their full extent, meaning the reconstructed source points may not trace the center of light of the distant deflector. We explore this further in Appendix \ref{app:distant_deflector}.

\subsection{X-Ray Spectrum}\label{sec:xray_discussion}

\begin{figure}
    \centering
    \hspace*{0.6cm}\footnotesize\figsettitle{Merged Spectrum of Images A-F from ObsID 26415}
    \makebox[\linewidth][c]{%
        \includegraphics[width=1\linewidth]{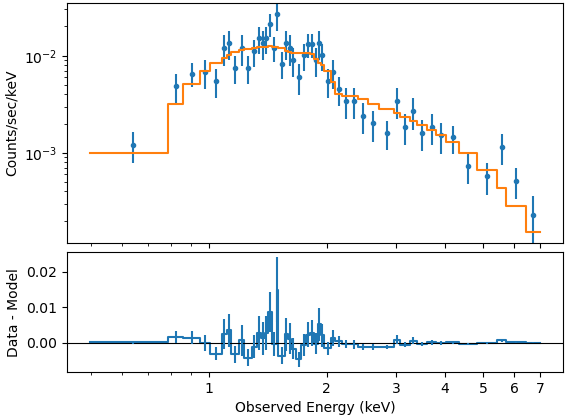}%
    }
    \caption{Top: The merged broadband ($0.5-7\ \mathrm{keV}$) X-ray spectrum extracted from images A-F in ObsID 26415 binned in 8 counts per bin using \sherpa. The blue points represent the binned observed data and their uncertainty. The orange line is the best-fit power-law with Galactic absorption model. Bottom: The residuals from the data - model.}
    \label{fig:spectra}
\end{figure}

The merged X-ray spectrum of images A-F in ObsID 26415 is shown in Figure \ref{fig:spectra}. This spectrum is well-described by a primary power-law continuum modified by the effects of neutral Galactic absorption. Whereas there is stochastic variation in the photon index across the six individual lensed images due to statistical noise (Table \ref{tab:photon_index}), these average to photon index $\Gamma_X=2.12\pm0.11$. This value of $\Gamma_X$ is consistent with radio-quiet quasars \citep{Reeves2000}. This interpretation is also consistent with our X-ray-to-optical source-plane reconstruction, determining that these emission regions are cospatial at the $1\sigma$ level within a $139\times30\ \mathrm{pc}$ ellipse, and providing evidence that there is \emph{no} significant X-ray emission beyond a projected maximum scale of $139\ \mathrm{pc}$ (e.g. due to a compact X-ray jet). The fact that J1721 is only weakly detected at radio wavelengths further supports this interpretation \citep{Mangat2021a}, since the lack of strong X-ray jet signatures would naturally be expected for a system without direct strong radio jet signatures.

The magnified, observed broadband X-ray luminosity of J1721 (adding the contribution from each individual lensed image) is $1.31\times10^{46}\ \mathrm{erg}\ \mathrm{s}^{-1}$. After correcting for absorption and the range of magnifications from both our lens model and the model of \citet[][]{Schmidt2025a} -- $\mu=70.9$~vs.~$\mu=52.1$, respectively --, the intrinsic $1.7-24\ \mathrm{keV}$ luminosity range of the lensed quasar is found to be $1.85-2.52\times10^{44}\ \mathrm{erg}\ \mathrm{s}^{-1}$. This places J1721 at the lower-luminosity end of the quasar regime \citep{Reines2016}. However, we note that this value must be taken with caution as the observed flux ratios were not used to constrain our lens model, consequently the magnification value and therefore the magnification-corrected luminosity may be inaccurate.

Due to limited photon statistics in the archival \chandra data, no additional features (including the Fe K$\alpha$ emission line) are detected at high statistical significance. Therefore, interpretation of J1721's X-ray spectrum cannot provide significant insights into the geometry of features such as the X-ray corona and accretion disk. While the caustic method alone also cannot resolve the spatial extent of these emission regions, differences in flux ratios between X-ray images may still be indicative of microlensing. As the X-ray corona is far more compact than the accretion disk, it is more susceptible to exhibiting stronger microlensing variability by stars in the lensing galaxies. 

Table \ref{tab:photon_index} shows the fluxes of each image across both ObsIDs, in the broad, hard, and soft X-ray bands. The fractional flux uncertainties are calculated as $1/\sqrt{N}$, where $N$ is the number of counts in each image. Image D exhibits a significant change in hard-band flux, decreasing by a factor of 1.88 at the $2\sigma$ level over the 13 hours between observations, while the soft flux remains consistent within statistical uncertainties (Table \ref{tab:photon_index}). However, the probability of obtaining at least one $2\sigma$ fluctuation from statistical noise is 24\% for the six hard X-ray band measurements, suggesting the fluctuation is unlikely to be caused by microlensing or intrinsic quasar variability, especially on such short timescales.

\subsection{X-Ray Astrometry of J1721}\label{sec:astrometry_discussion}

\begin{figure*}
    \centering
    \includegraphics[width=1\linewidth]{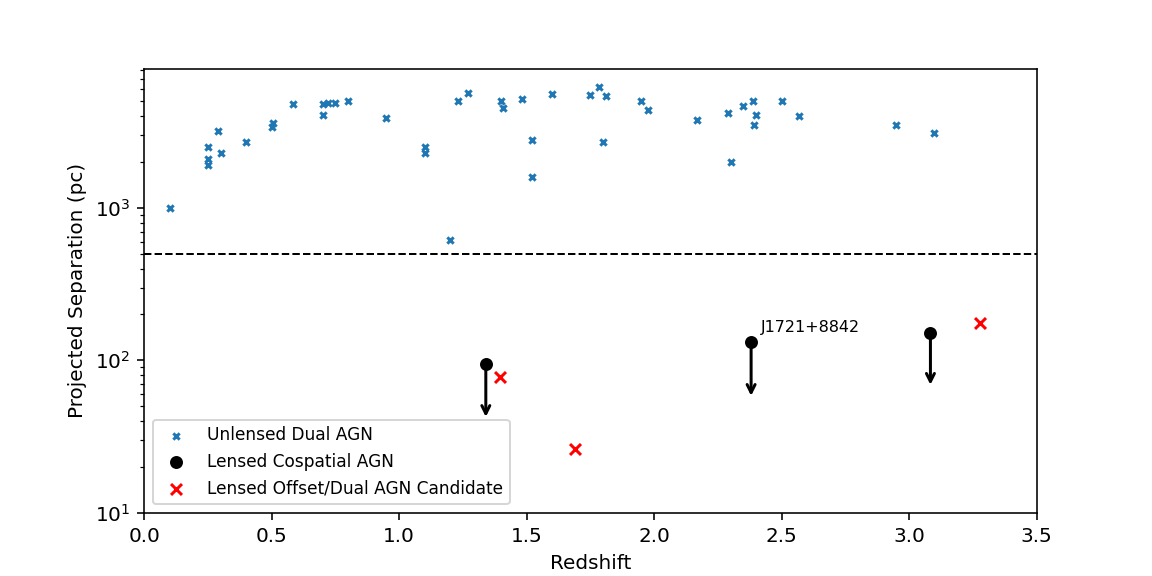}
    \caption{The blue crosses correspond to unlensed dual AGN candidates found by \citet{Chen2025c} with \HST data. The black circles and red crosses are the current selection of lensed AGN that have been investigated using the caustic method. The red crosses are AGN where the optical/radio emission has been found to be offset from the X-ray emission. These correspond to MG B2016+112, HE 0435-1223, and B1608+656 (mentioned in Sections \ref{sec:astrometry_discussion} and \ref{sec:future_dual}). The black circles are AGN where the optical/radio emission and X-ray emission have been found to be cospatial, where the projected separation is the maximum separation that could be resolved. These correspond to J1721 (this work), J065904.1+162909, and B0712+472.}
    \label{fig:current_dual_AGN}
\end{figure*}

By combining \chandra imaging, \textit{Gaia's} state-of-the-art astrometry, and caustics as nonlinear spatial amplifiers, we have constrained the X-ray emission position in J1721 to milliarcsecond precision relative to the optical. \textit{Chandra's} angular resolution has been improved by a factor of 30 with the caustic method.

The spatial origins of X-ray and optical emission from J1721 are coincident within $1\sigma$; therefore there is no evidence for the existence of X-ray phenomena at scales $\sim \mathrm{few} \times 100 \ \mathrm{pc}$ that would shift the centroid of X-ray emission beyond the $139\times30\ \mathrm{pc}$ positional ellipse for the broadband X-ray source (Figure \ref{fig:contour_plot}). This suggests that there is no significant extended X-ray jet, consistent with the lack of a strong radio detection reported by \citet{Mangat2021a}. The lack of any significant offset beyond the $1\sigma$ confidence region further disfavors a dual AGN with separation larger than $\sim 140\ \mathrm{pc}$, consistent with \citet{Dux2025a} and \citet{Schmidt2025a}. While the results make it unlikely that the AGN is significantly displaced from the host galaxy's center, either in the aftermath of a gravitational interaction or in a dual SMBH system, the latter cannot be excluded as the AGN may remain in the galactic center, with an inactive SMBH companion.

The hard and soft X-ray emission regions are cospatial at the $1\sigma$ level, suggesting a common compact spatial origin. However, if the hard-to-soft X-ray flux discrepancy in image D were due to microlensing, this would be indicative of structural differences between the two regions on smaller spatial scales that cannot be resolved using the caustic method with the currently available X-ray observations.

Overall, these results suggest that J1721 lacks significant, extended X-ray structures which would emanate from jets or large-scale outflows beyond $\sim140\ \mathrm{pc}$. The absence of these features indicates that J1721 does not present significant AGN-feedback-related structures at hundreds of parsecs. However, we stress that X-ray phenomena below the spatial scales resolvable by this work may still be present, particularly compact dual AGN systems, as well as jet knots like HST-1, offset by $60\ \mathrm{pc}$ \citep{Harris2003a, Harris2006}. The limitation of the caustic method in this work is the number of photon counts in the X-ray lensed images. As the astrometric precision scales as $1/\sqrt{N}$, where $N$ is the number of X-ray counts, with $\sim5400$ more X-ray counts, a $60\ \mathrm{pc}$ offset could be detected. However, if an M87-like jet were present and aligned perpendicular to the nearby caustic, the effective spatial resolution of $30\ \mathrm{pc}$ achieved in this work perpendicular to the nearby caustic (the minor axis of the $1\sigma$ ellipse in Figure \ref{fig:contour_plot}) with the current \chandra X-ray data would be able to resolve an offset.

In strongly lensed systems where offsets are revealed by the caustic method (e.g. \citealt{Spingola2019a, Schwartz2021a, Spingola2022a, Rogers2025a}), the origin of the offset can be further investigated with ancillary multi-wavelength observations. \citet{Spingola2022a} performed the caustic method on lensed AGN, CLASS B1608+656, finding that the X-ray and radio emission regions are cospatial, and both are offset from the optical. Host galaxy observations of this system show that it is a post-starburst galaxy. The offset is therefore indicative of either a dual AGN-inactive SMBH system, or a recoiling AGN, both stemming from a past galaxy merger which also triggered the starburst. \citet{Rogers2025a} use spectral information to determine that HE 0435-1223 is most likely not a dual AGN due to the lack of split emission lines in the X-ray spectrum, and instead belongs to a jet/outflow due to the scale of the offset ($26\ \mathrm{pc}$).

The spatial resolution achieved in this work has demonstrated the power of the caustic method to probe sub-kpc X-ray structure in AGN at cosmic noon. This adds to the growing sample of lensed AGN studied using this method. Of the AGN studied, we have revealed a diverse range of features and structures at cosmic noon which would be unresolvable without strong gravitational lensing.

\subsection{Future Prospects}\label{sec:future}

\subsubsection{Dual and Binary AGN Candidates}\label{sec:future_dual}

Current instruments, such as \textit{HST}, can identify dual AGN candidates at cosmic noon with projected distances $\gtrsim \mathrm{kpc}$ \citep{Chen2022, Chen2025c}, some of which are shown in Figure \ref{fig:current_dual_AGN}. The \textit{Laser Interferometer Gravitational-wave Observatory} (\textit{LIGO}) routinely detects merging binary stellar-mass BHs in the final seconds before coalescence \citep{Abbott2016}, and pulsar timing arrays (PTAs) have detected a nHz stochastic gravitational wave background (STGWB) consistent with coalescing SMBH binary systems at $z\approx2$ with masses $>10^{8-10}M_\odot$ and separations of $\sim0.1\ \mathrm{pc}$ \citep{Burke-Spolaor2019}. The future \textit{Laser Interferometer Space Antenna} (\textit{LISA}) is planned to search for mHz gravitational wave signals from $10^6-10^7M_\odot$ binary systems, probing the $\sim1\,000~R_G\ (0.01\ \mathrm{pc})$ scale of binary SMBHs \citep{Katz2019}.

The caustic method may offer a crucial bridge between these populations by allowing dual AGN to be identified at $10\ \mathrm{pc}-1\ \mathrm{kpc}$ separations at cosmic noon, lying between the wide kpc-scale pairs identified with optical and radio observations \citep{Gross2025} and the sub-pc, coalescing SMBH population probed by PTAs, making them a key link between galaxy merger-driven dual AGN and the eventual triggering of the gravitational wave coalescence of SMBH pairs \citep{Enoki2004}. Source-plane reconstruction enabled by strong lensing and the caustic method has so far identified one strong candidate dual AGN out of the six systems studied: the radio-loud, multiply-imaged system MG B2016+112 at $z=3.3$ discovered by \citet{Lawrence1984}. This has a projected separation of $\sim175\ \mathrm{pc}$, making it a potential progenitor of a gravitationally bound system \citep{Spingola2019a, Schwartz2021a, Schwartz2026}. 

\subsubsection{High-Redshift AGN X-Ray Structure}\label{sec:future_structure}

\gaia DR4, scheduled for December 2026 \citep{GaiaCollaboration2023}, is expected to improve on DR3 astrometry by a factor of 1.4, placing tighter constraints on lens models through reduced astrometric uncertainties, and is also predicted to discover hundreds more lensed quasars \citep{Ducourant2026}.

The current and upcoming \textit{Euclid}, \textit{LSST}, and \textit{Roman} missions will detect thousands more lensed AGN, enabling the construction of large samples to improve our knowledge of AGN demographics and structure at cosmic noon and beyond \citep{Collett2015a}. At radio wavelengths, particularly with the 2000-antenna \textit{Deep Synoptic Array} (\textit{DSA-2000}) and the \textit{Square Kilometre Array} (\textit{SKA}), \citet{McCarty2025} predict the discovery of $10^4-10^5$ strong lenses (containing $\sim10^1-10^2$ lensed AGN). 

Overall, thousands of quadruply lensed quasars with magnifications large enough to perform the caustic method will be discovered in the next few years. This will allow us to build up a population of AGN X-ray structures across cosmic time, allowing their growth, evolution, and interaction with their host galaxies to be probed.

\section{Conclusions}\label{sec:conclusions}

We have located the X-ray emission relative to the optical emission in the lensed quasar J1721 at $z=2.38$ with milliarcsecond precision, exploiting the flux magnification and spatial amplification effects of strong gravitational lensing to improve \textit{Chandra's} effective resolution by $30$ times, allowing for parsec-scale X-ray structures to be resolved at cosmic noon.

In J1721, the broadband X-ray emission is cospatial with the optical emission in the source quasar within a $1\sigma$ ellipse of $0\farcs017\times0\farcs004\ (139\times30\ \mathrm{pc})$. The soft and hard X-ray emission regions are also cospatial with the optical emission and with each other at the $1\sigma$ level. These results suggest that J1721 does not contain any significant extended X-ray structures (e.g.~jets, outflows), or a dual AGN with a projected separation above $\sim140\ \mathrm{pc}$.

This work adds to several previous analyses using the caustic method, reinforcing the importance of this method to resolve quasar structure at high redshift. Looking ahead, \textit{Gaia}, \textit{Euclid}, \textit{LSST}, \textit{SKA}, \textit{DSA-2000}, and \textit{Roman} will discover thousands of new quasar lenses, many of which will have magnifications large enough to reveal sub-kpc structures with the caustic method at X-ray, optical, and radio wavelengths. This will allow identification of dual and binary AGN, placing constraints on accretion and feedback processes, and probing the geometry of the X-ray structures at cosmic noon.
 
\begin{acknowledgments}

CE thanks Daniel Castro for his support through grant GO1-22074X from the \chandra X-Ray Center (CXC). JSR acknowledges support from a NASA ADAP Program Grant 80NSSC24K0617. This work has been supported by the National Aeronautics and Space Administration through the contract NAS8-03060 to SAO and grant GO4-25053X from the CXC.

We thank the anonymous referee for useful suggestions. CE thanks Diab Jerius for valuable discussions and help with the \saotrace software. The authors thank Martin Millon and Fr\'ed\'eric Dux for useful discussions on prior and ongoing lens modeling efforts of J1721. CE and JSR thank Tania Barone for useful discussions on \HST\ imaging. 

This work was conducted as part of CE's master's thesis, completed through the SAO/University of Southampton Astrophysics Master's Program within the partnership between the Center for Astrophysics $|$ Harvard \& Smithsonian and the University of Southampton. CE thanks Sandra Raimundo, Diego Altamirano, Doug Burke, and Daniel Castro for the organization of the program.

This research has made use of data obtained from the \chandra Data Archive, as well as software provided by the CXC in the application packages \ciao, \sherpa, \ds, and the high-fidelity simulation packages \saotrace and \marx. The lens modeling was performed using the \gravlens software. This work has also employed data from the \gaia Archive and from the \textit{Hubble Space Telescope} through the Mikulski Archive for Space Telescopes (MAST). The Space Telescope Science Institute is operated by the Association of Universities for Research in Astronomy, Inc., under NASA contract NAS 5-26555.

This work has made use of data from the European Space Agency (ESA) mission
{\it Gaia} (\url{https://www.cosmos.esa.int/gaia}), processed by the {\it Gaia}
Data Processing and Analysis Consortium (DPAC,
\url{https://www.cosmos.esa.int/web/gaia/dpac/consortium}). Funding for the DPAC
has been provided by national institutions, in particular the institutions
participating in the {\it Gaia} Multilateral Agreement.

This research has made use of the Astrophysics Data System, funded by NASA under Cooperative Agreement 80NSSC21M0056.

\textit{Facilities}: European Space Agency (ESA) \textit{Gaia} Satellite Mission, NASA \textit{Chandra X-Ray Observatory} (CXO), NASA/ESA \textit{Hubble Space Telescope}

\textit{Software}: \gravlens \citep{Keeton2001a}, \ciao - 4.17 \citep{Fruscione2006a}, \sherpa - 4.18.0 \citep{Siemiginowska2024,Burke2025}, \ds - 8.7 \citep{Joye2003a}, \saotrace - 2.0.6.1 \citep{Jerius2004a}, \marx - 6.0.1 \citep{Davis2012}

\end{acknowledgments}

\bibliography{J1721_ApJ}

@article{Agazie2023a,
  title = {The {{NANOGrav}} 15 Yr {{Data Set}}: {{Evidence}} for a {{Gravitational-wave Background}}},
  shorttitle = {The {{NANOGrav}} 15 Yr {{Data Set}}},
  author = {Agazie, Gabriella and Anumarlapudi, Akash and Archibald, Anne M. and Arzoumanian, Zaven and Baker, Paul T. and B{\'e}csy, Bence and Blecha, Laura and Brazier, Adam and Brook, Paul R. and {Burke-Spolaor}, Sarah and Burnette, Rand and Case, Robin and Charisi, Maria and Chatterjee, Shami and Chatziioannou, Katerina and Cheeseboro, Belinda D. and Chen, Siyuan and Cohen, Tyler and Cordes, James M. and Cornish, Neil J. and Crawford, Fronefield and Cromartie, H. Thankful and Crowter, Kathryn and Cutler, Curt J. and DeCesar, Megan E. and DeGan, Dallas and Demorest, Paul B. and Deng, Heling and Dolch, Timothy and Drachler, Brendan and Ellis, Justin A. and Ferrara, Elizabeth C. and Fiore, William and Fonseca, Emmanuel and Freedman, Gabriel E. and {Garver-Daniels}, Nate and Gentile, Peter A. and Gersbach, Kyle A. and Glaser, Joseph and Good, Deborah C. and G{\"u}ltekin, Kayhan and Hazboun, Jeffrey S. and Hourihane, Sophie and Islo, Kristina and Jennings, Ross J. and Johnson, Aaron D. and Jones, Megan L. and Kaiser, Andrew R. and Kaplan, David L. and Kelley, Luke Zoltan and Kerr, Matthew and Key, Joey S. and Klein, Tonia C. and Laal, Nima and Lam, Michael T. and Lamb, William G. and W. Lazio, T. Joseph and Lewandowska, Natalia and Littenberg, Tyson B. and Liu, Tingting and Lommen, Andrea and Lorimer, Duncan R. and Luo, Jing and Lynch, Ryan S. and Ma, Chung-Pei and Madison, Dustin R. and Mattson, Margaret A. and McEwen, Alexander and McKee, James W. and McLaughlin, Maura A. and McMann, Natasha and Meyers, Bradley W. and Meyers, Patrick M. and Mingarelli, Chiara M. F. and Mitridate, Andrea and Natarajan, Priyamvada and Ng, Cherry and Nice, David J. and Ocker, Stella Koch and Olum, Ken D. and Pennucci, Timothy T. and Perera, Benetge B. P. and Petrov, Polina and Pol, Nihan S. and Radovan, Henri A. and Ransom, Scott M. and Ray, Paul S. and Romano, Joseph D. and Sardesai, Shashwat C. and Schmiedekamp, Ann and Schmiedekamp, Carl and Schmitz, Kai and Schult, Levi and {Shapiro-Albert}, Brent J. and Siemens, Xavier and Simon, Joseph and Siwek, Magdalena S. and Stairs, Ingrid H. and Stinebring, Daniel R. and Stovall, Kevin and Sun, Jerry P. and Susobhanan, Abhimanyu and Swiggum, Joseph K. and Taylor, Jacob and Taylor, Stephen R. and Turner, Jacob E. and Unal, Caner and Vallisneri, Michele and Van Haasteren, Rutger and Vigeland, Sarah J. and Wahl, Haley M. and Wang, Qiaohong and Witt, Caitlin A. and Young, Olivia and {The NANOGrav Collaboration}},
  year = 2023,
  month = jul,
  journal = {ApJL},
  volume = {951},
  number = {1},
  pages = {L8},
  issn = {2041-8205, 2041-8213},
  doi = {10.3847/2041-8213/acdac6},
  urldate = {2026-04-16}
}

@article{Barnacka2018a,
  title = {Gravitational Lenses as High-Resolution Telescopes},
  author = {Barnacka, Anna},
  year = 2018,
  month = dec,
  journal = {Physics Reports},
  volume = {778--779},
  pages = {1--46},
  issn = {03701573},
  doi = {10.1016/j.physrep.2018.10.001},
  urldate = {2026-04-16},
  langid = {english}
}

@article{Barrows2016a,
  title = {{{SPATIALLY OFFSET ACTIVE GALACTIC NUCLEI}}. {{I}}. {{SELECTION AND SPECTROSCOPIC PROPERTIES}}},
  author = {Barrows, R. Scott and Comerford, Julia M. and Greene, Jenny E. and Pooley, David},
  year = 2016,
  month = sep,
  journal = {ApJ},
  volume = {829},
  number = {1},
  pages = {37},
  issn = {0004-637X, 1538-4357},
  doi = {10.3847/0004-637X/829/1/37},
  urldate = {2026-04-16}
}

@INPROCEEDINGS{Burke-Spolaor2018,
       author = {{Burke-Spolaor}, S. and {Blecha}, L. and {Bogdanovi{\'c}}, T. and {Comerford}, J.~M. and {Lazio}, J. and {Liu}, X. and {Maccarone}, T.~J. and {Pesce}, D. and {Shen}, Y. and {Taylor}, G.},
        title = "{Supermassive Black Hole Pairs and Binaries}",
    booktitle = {Science with a Next Generation Very Large Array},
         year = 2018,
       editor = {{Murphy}, Eric},
       series = {Astronomical Society of the Pacific Conference Series},
       volume = {517},
        month = dec,
        pages = {677},
       adsurl = {https://ui.adsabs.harvard.edu/abs/2018ASPC..517..677B}
}

@misc{Burke2025,
  title = {Sherpa/Sherpa: {{Sherpa}} 4.18.0},
  author = {Burke, Doug and Laurino, Omar and {wmclaugh} and G{\"u}nther, Hans Moritz and {Marie-Terrell} and {dtnguyen2} and Siemiginowska, Aneta and Cheer, Harlan and Budynkiewicz, Jamie and Aldcroft, Tom and {luzpaz} and Deil, Christoph and Sip{\H o}cz, Brigitta and {nplee} and Buchner, Johannes and Donath, Axel and Laginja, Iva and Leinweber, Katrin and {Todd}},
  year = 2025,
  month = oct,
  publisher = {Zenodo},
  doi = {10.5281/zenodo.17288983}
}

@article{Cash1979a,
  title = {Parameter Estimation in Astronomy through Application of the Likelihood Ratio},
  author = {Cash, W.},
  year = 1979,
  month = mar,
  journal = {ApJ},
  volume = {228},
  pages = {939},
  issn = {0004-637X, 1538-4357},
  doi = {10.1086/156922},
  urldate = {2026-04-16},
  langid = {english}
}

@inproceedings{Chambers2018,
  title = {The {{Pan-STARRS1 Surveys}}},
  booktitle = {American {{Astronomical Society Meeting Abstracts}} \#231},
  author = {Chambers, Kenneth and {Pan-STARRS Team}},
  year = 2018,
  month = jan,
  volume = {231},
  pages = {102.01},
  urldate = {2026-04-16}
}

@article{Chen2025c,
  title = {Varstrometry for {{Off-nucleus}} and {{Dual Sub-Kpc AGN}} ({{VODKA}}): {{Long-slit Optical Spectroscopic Follow-up}} with {{Gemini}}/{{GMOS}} and {{Hubble Space Telescope}}/{{STIS}}},
  shorttitle = {Varstrometry for {{Off-nucleus}} and {{Dual Sub-Kpc AGN}} ({{VODKA}})},
  author = {Chen, Yu-Ching and Gross, Arran C. and Liu, Xin and Shen, Yue and Zakamska, Nadia L. and Hwang, Hsiang-Chih and Zhuang, Ming-Yang},
  year = 2025,
  month = jul,
  journal = {ApJ},
  volume = {988},
  number = {1},
  pages = {126},
  issn = {0004-637X, 1538-4357},
  doi = {10.3847/1538-4357/ade23b},
  urldate = {2026-04-16}
}

@article{Collett2015a,
  title = {{{THE POPULATION OF GALAXY}}--{{GALAXY STRONG LENSES IN FORTHCOMING OPTICAL IMAGING SURVEYS}}},
  author = {Collett, Thomas E.},
  year = 2015,
  month = sep,
  journal = {ApJ},
  volume = {811},
  number = {1},
  pages = {20},
  issn = {1538-4357},
  doi = {10.1088/0004-637X/811/1/20},
  urldate = {2026-04-16},
  copyright = {http://iopscience.iop.org/info/page/text-and-data-mining}
}

@article{Collett2016a,
  title = {Compound Lensing: {{Einstein}} Zig-Zags and High-Multiplicity Lensed Images},
  shorttitle = {Compound Lensing},
  author = {Collett, Thomas E. and Bacon, David J.},
  year = 2016,
  month = feb,
  journal = {MNRAS},
  volume = {456},
  number = {2},
  pages = {2210--2220},
  issn = {0035-8711, 1365-2966},
  doi = {10.1093/mnras/stv2791},
  urldate = {2026-04-16},
  langid = {english}
}

@inproceedings{Davis2012,
  title = {Raytracing with {{MARX}}: X-Ray Observatory Design, Calibration, and Support},
  shorttitle = {Raytracing with {{MARX}}},
  booktitle = {{{SPIE Astronomical Telescopes}} + {{Instrumentation}}},
  author = {Davis, John E. and Bautz, Marshall W. and Dewey, Daniel and Heilmann, Ralf K. and Houck, John C. and Huenemoerder, David P. and Marshall, Herman L. and Nowak, Michael A. and Schattenburg, Mark L. and Schulz, Norbert S. and Smith, Randall K.},
  editor = {Takahashi, Tadayuki and Murray, Stephen S. and Den Herder, Jan-Willem A.},
  year = 2012,
  month = sep,
  pages = {84431A},
  address = {Amsterdam, Netherlands},
  doi = {10.1117/12.926937},
  urldate = {2026-05-08}
}

@article{Dux2025a,
  title = {J1721+8842: {{The}} First {{Einstein}} Zigzag Lens},
  shorttitle = {J1721+8842},
  author = {Dux, F. and Millon, M. and Lemon, C. and Schmidt, T. and Courbin, F. and Shajib, A. J. and Treu, T. and Birrer, S. and Wong, K. C. and Agnello, A. and Andrade, A. and Galan, A. and Hjorth, J. and Paic, E. and Schuldt, S. and Schweinfurth, A. and Sluse, D. and Smette, A. and Suyu, S. H.},
  year = 2025,
  month = feb,
  journal = {A\&A},
  volume = {694},
  pages = {A300},
  issn = {0004-6361, 1432-0746},
  doi = {10.1051/0004-6361/202452970},
  urldate = {2026-04-16},
  copyright = {https://creativecommons.org/licenses/by/4.0}
}

@article{ForsterSchreiber2020,
  title = {Star-{{Forming Galaxies}} at {{Cosmic Noon}}},
  author = {F{\"o}rster Schreiber, Natascha M. and Wuyts, Stijn},
  year = 2020,
  month = aug,
  journal = {Annu. Rev. Astron. Astrophys.},
  volume = {58},
  number = {1},
  pages = {661--725},
  issn = {0066-4146, 1545-4282},
  doi = {10.1146/annurev-astro-032620-021910},
  urldate = {2026-04-16},
  langid = {english}
}

@inproceedings{Fruscione2006a,
  title = {{{CIAO}}: {{Chandra}}'s Data Analysis System},
  shorttitle = {{{CIAO}}},
  booktitle = {{{SPIE Astronomical Telescopes}} + {{Instrumentation}}},
  author = {Fruscione, Antonella and McDowell, Jonathan C. and Allen, Glenn E. and Brickhouse, Nancy S. and Burke, Douglas J. and Davis, John E. and Durham, Nick and Elvis, Martin and Galle, Elizabeth C. and Harris, Daniel E. and Huenemoerder, David P. and Houck, John C. and Ishibashi, Bish and Karovska, Margarita and Nicastro, Fabrizio and Noble, Michael S. and Nowak, Michael A. and Primini, Frank A. and Siemiginowska, Aneta and Smith, Randall K. and Wise, Michael},
  editor = {Silva, David R. and Doxsey, Rodger E.},
  year = 2006,
  month = jun,
  pages = {62701V},
  address = {Orlando, Florida , USA},
  doi = {10.1117/12.671760},
  urldate = {2026-04-16}
}

@article{GaiaCollaboration2016,
  title = {The {{{\emph{Gaia}}}} Mission},
  author = {{Gaia Collaboration} and Prusti, T. and De Bruijne, J. H. J. and Brown, A. G. A. and Vallenari, A. and Babusiaux, C. and {Bailer-Jones}, C. A. L. and Bastian, U. and Biermann, M. and Evans, D. W. and Eyer, L. and Jansen, F. and Jordi, C. and Klioner, S. A. and Lammers, U. and Lindegren, L. and Luri, X. and Mignard, F. and Milligan, D. J. and Panem, C. and Poinsignon, V. and Pourbaix, D. and Randich, S. and Sarri, G. and Sartoretti, P. and Siddiqui, H. I. and Soubiran, C. and Valette, V. and Van Leeuwen, F. and Walton, N. A. and Aerts, C. and Arenou, F. and Cropper, M. and Drimmel, R. and H{\o}g, E. and Katz, D. and Lattanzi, M. G. and O'Mullane, W. and Grebel, E. K. and Holland, A. D. and Huc, C. and Passot, X. and Bramante, L. and Cacciari, C. and Casta{\~n}eda, J. and Chaoul, L. and Cheek, N. and De Angeli, F. and Fabricius, C. and Guerra, R. and Hern{\'a}ndez, J. and {Jean-Antoine-Piccolo}, A. and Masana, E. and Messineo, R. and Mowlavi, N. and Nienartowicz, K. and {Ord{\'o}{\~n}ez-Blanco}, D. and Panuzzo, P. and Portell, J. and Richards, P. J. and Riello, M. and Seabroke, G. M. and Tanga, P. and Th{\'e}venin, F. and Torra, J. and Els, S. G. and {Gracia-Abril}, G. and Comoretto, G. and {Garcia-Reinaldos}, M. and Lock, T. and Mercier, E. and Altmann, M. and Andrae, R. and Astraatmadja, T. L. and {Bellas-Velidis}, I. and Benson, K. and Berthier, J. and Blomme, R. and Busso, G. and Carry, B. and Cellino, A. and Clementini, G. and Cowell, S. and Creevey, O. and Cuypers, J. and Davidson, M. and De Ridder, J. and De Torres, A. and Delchambre, L. and Dell'Oro, A. and Ducourant, C. and Fr{\'e}mat, Y. and {Garc{\'i}a-Torres}, M. and Gosset, E. and Halbwachs, J.-L. and Hambly, N. C. and Harrison, D. L. and Hauser, M. and Hestroffer, D. and Hodgkin, S. T. and Huckle, H. E. and Hutton, A. and Jasniewicz, G. and Jordan, S. and Kontizas, M. and Korn, A. J. and Lanzafame, A. C. and Manteiga, M. and Moitinho, A. and Muinonen, K. and Osinde, J. and Pancino, E. and Pauwels, T. and Petit, J.-M. and {Recio-Blanco}, A. and Robin, A. C. and Sarro, L. M. and Siopis, C. and Smith, M. and Smith, K. W. and Sozzetti, A. and Thuillot, W. and Van Reeven, W. and Viala, Y. and Abbas, U. and Abreu Aramburu, A. and Accart, S. and Aguado, J. J. and Allan, P. M. and Allasia, W. and Altavilla, G. and {\'A}lvarez, M. A. and Alves, J. and Anderson, R. I. and Andrei, A. H. and Anglada Varela, E. and Antiche, E. and Antoja, T. and Ant{\'o}n, S. and Arcay, B. and Atzei, A. and Ayache, L. and Bach, N. and Baker, S. G. and {Balaguer-N{\'u}{\~n}ez}, L. and Barache, C. and Barata, C. and Barbier, A. and Barblan, F. and Baroni, M. and Barrado Y Navascu{\'e}s, D. and Barros, M. and Barstow, M. A. and Becciani, U. and Bellazzini, M. and Bellei, G. and Bello Garc{\'i}a, A. and Belokurov, V. and Bendjoya, P. and Berihuete, A. and Bianchi, L. and Bienaym{\'e}, O. and Billebaud, F. and Blagorodnova, N. and {Blanco-Cuaresma}, S. and Boch, T. and Bombrun, A. and Borrachero, R. and Bouquillon, S. and Bourda, G. and Bouy, H. and Bragaglia, A. and Breddels, M. A. and Brouillet, N. and Br{\"u}semeister, T. and Bucciarelli, B. and Budnik, F. and Burgess, P. and Burgon, R. and Burlacu, A. and Busonero, D. and Buzzi, R. and Caffau, E. and Cambras, J. and Campbell, H. and Cancelliere, R. and {Cantat-Gaudin}, T. and Carlucci, T. and Carrasco, J. M. and Castellani, M. and Charlot, P. and Charnas, J. and Charvet, P. and Chassat, F. and Chiavassa, A. and Clotet, M. and Cocozza, G. and Collins, R. S. and Collins, P. and Costigan, G. and Crifo, F. and Cross, N. J. G. and Crosta, M. and Crowley, C. and Dafonte, C. and Damerdji, Y. and Dapergolas, A. and David, P. and David, M. and De Cat, P. and De Felice, F. and De Laverny, P. and De Luise, F. and De March, R. and De Martino, D. and De Souza, R. and Debosscher, J. and Del Pozo, E. and Delbo, M. and Delgado, A. and Delgado, H. E. and Di Marco, F. and Di Matteo, P. and Diakite, S. and Distefano, E. and Dolding, C. and Dos Anjos, S. and Drazinos, P. and Dur{\'a}n, J. and Dzigan, Y. and Ecale, E. and Edvardsson, B. and Enke, H. and Erdmann, M. and Escolar, D. and Espina, M. and Evans, N. W. and Eynard Bontemps, G. and Fabre, C. and Fabrizio, M. and Faigler, S. and Falc{\~a}o, A. J. and Farr{\`a}s Casas, M. and Faye, F. and Federici, L. and Fedorets, G. and {Fern{\'a}ndez-Hern{\'a}ndez}, J. and Fernique, P. and Fienga, A. and Figueras, F. and Filippi, F. and Findeisen, K. and Fonti, A. and Fouesneau, M. and Fraile, E. and Fraser, M. and Fuchs, J. and Furnell, R. and Gai, M. and Galleti, S. and Galluccio, L. and Garabato, D. and {Garc{\'i}a-Sedano}, F. and Gar{\'e}, P. and Garofalo, A. and Garralda, N. and Gavras, P. and Gerssen, J. and Geyer, R. and Gilmore, G. and Girona, S. and Giuffrida, G. and Gomes, M. and {Gonz{\'a}lez-Marcos}, A. and {Gonz{\'a}lez-N{\'u}{\~n}ez}, J. and {Gonz{\'a}lez-Vidal}, J. J. and Granvik, M. and Guerrier, A. and Guillout, P. and Guiraud, J. and G{\'u}rpide, A. and {Guti{\'e}rrez-S{\'a}nchez}, R. and Guy, L. P. and Haigron, R. and Hatzidimitriou, D. and Haywood, M. and Heiter, U. and Helmi, A. and Hobbs, D. and Hofmann, W. and Holl, B. and Holland, G. and Hunt, J. A. S. and Hypki, A. and Icardi, V. and Irwin, M. and Jevardat De Fombelle, G. and Jofr{\'e}, P. and Jonker, P. G. and Jorissen, A. and Julbe, F. and Karampelas, A. and Kochoska, A. and Kohley, R. and Kolenberg, K. and Kontizas, E. and Koposov, S. E. and Kordopatis, G. and Koubsky, P. and Kowalczyk, A. and {Krone-Martins}, A. and Kudryashova, M. and Kull, I. and Bachchan, R. K. and {Lacoste-Seris}, F. and Lanza, A. F. and Lavigne, J.-B. and {Le Poncin-Lafitte}, C. and Lebreton, Y. and Lebzelter, T. and Leccia, S. and Leclerc, N. and {Lecoeur-Taibi}, I. and Lemaitre, V. and Lenhardt, H. and Leroux, F. and Liao, S. and Licata, E. and Lindstr{\o}m, H. E. P. and Lister, T. A. and Livanou, E. and Lobel, A. and L{\"o}ffler, W. and L{\'o}pez, M. and {Lopez-Lozano}, A. and Lorenz, D. and Loureiro, T. and MacDonald, I. and Magalh{\~a}es Fernandes, T. and Managau, S. and Mann, R. G. and Mantelet, G. and Marchal, O. and Marchant, J. M. and Marconi, M. and Marie, J. and Marinoni, S. and Marrese, P. M. and Marschalk{\'o}, G. and Marshall, D. J. and {Mart{\'i}n-Fleitas}, J. M. and Martino, M. and Mary, N. and Matijevi{\v c}, G. and Mazeh, T. and McMillan, P. J. and Messina, S. and Mestre, A. and Michalik, D. and Millar, N. R. and Miranda, B. M. H. and Molina, D. and Molinaro, R. and Molinaro, M. and Moln{\'a}r, L. and Moniez, M. and Montegriffo, P. and Monteiro, D. and Mor, R. and Mora, A. and Morbidelli, R. and Morel, T. and Morgenthaler, S. and Morley, T. and Morris, D. and Mulone, A. F. and Muraveva, T. and Musella, I. and Narbonne, J. and Nelemans, G. and Nicastro, L. and Noval, L. and Ord{\'e}novic, C. and {Ordieres-Mer{\'e}}, J. and Osborne, P. and Pagani, C. and Pagano, I. and Pailler, F. and Palacin, H. and Palaversa, L. and Parsons, P. and Paulsen, T. and Pecoraro, M. and Pedrosa, R. and Pentik{\"a}inen, H. and Pereira, J. and Pichon, B. and Piersimoni, A. M. and Pineau, F.-X. and Plachy, E. and Plum, G. and Poujoulet, E. and Pr{\v s}a, A. and Pulone, L. and Ragaini, S. and Rago, S. and Rambaux, N. and {Ramos-Lerate}, M. and Ranalli, P. and Rauw, G. and Read, A. and Regibo, S. and Renk, F. and Reyl{\'e}, C. and Ribeiro, R. A. and Rimoldini, L. and Ripepi, V. and Riva, A. and Rixon, G. and Roelens, M. and {Romero-G{\'o}mez}, M. and Rowell, N. and Royer, F. and Rudolph, A. and {Ruiz-Dern}, L. and Sadowski, G. and Sagrist{\`a} Sell{\'e}s, T. and Sahlmann, J. and Salgado, J. and Salguero, E. and Sarasso, M. and Savietto, H. and Schnorhk, A. and Schultheis, M. and Sciacca, E. and Segol, M. and Segovia, J. C. and Segransan, D. and Serpell, E. and Shih, I-C. and Smareglia, R. and Smart, R. L. and Smith, C. and Solano, E. and Solitro, F. and Sordo, R. and Soria Nieto, S. and Souchay, J. and Spagna, A. and Spoto, F. and Stampa, U. and Steele, I. A. and Steidelm{\"u}ller, H. and Stephenson, C. A. and Stoev, H. and Suess, F. F. and S{\"u}veges, M. and Surdej, J. and Szabados, L. and {Szegedi-Elek}, E. and Tapiador, D. and Taris, F. and Tauran, G. and Taylor, M. B. and Teixeira, R. and Terrett, D. and Tingley, B. and Trager, S. C. and Turon, C. and Ulla, A. and Utrilla, E. and Valentini, G. and Van Elteren, A. and Van Hemelryck, E. and Van Leeuwen, M. and Varadi, M. and Vecchiato, A. and Veljanoski, J. and Via, T. and Vicente, D. and Vogt, S. and Voss, H. and Votruba, V. and Voutsinas, S. and Walmsley, G. and Weiler, M. and Weingrill, K. and Werner, D. and Wevers, T. and Whitehead, G. and Wyrzykowski, {\L}. and Yoldas, A. and {\v Z}erjal, M. and Zucker, S. and Zurbach, C. and Zwitter, T. and Alecu, A. and Allen, M. and Allende Prieto, C. and Amorim, A. and {Anglada-Escud{\'e}}, G. and Arsenijevic, V. and Azaz, S. and Balm, P. and Beck, M. and Bernstein, H.-H. and Bigot, L. and Bijaoui, A. and Blasco, C. and Bonfigli, M. and Bono, G. and Boudreault, S. and Bressan, A. and Brown, S. and Brunet, P.-M. and Bunclark, P. and Buonanno, R. and Butkevich, A. G. and Carret, C. and Carrion, C. and Chemin, L. and Ch{\'e}reau, F. and Corcione, L. and Darmigny, E. and De Boer, K. S. and De Teodoro, P. and De Zeeuw, P. T. and Delle Luche, C. and Domingues, C. D. and Dubath, P. and Fodor, F. and Fr{\'e}zouls, B. and Fries, A. and Fustes, D. and Fyfe, D. and Gallardo, E. and Gallegos, J. and Gardiol, D. and Gebran, M. and Gomboc, A. and G{\'o}mez, A. and Grux, E. and Gueguen, A. and Heyrovsky, A. and Hoar, J. and Iannicola, G. and Isasi Parache, Y. and Janotto, A.-M. and Joliet, E. and Jonckheere, A. and Keil, R. and Kim, D.-W. and Klagyivik, P. and Klar, J. and Knude, J. and Kochukhov, O. and Kolka, I. and Kos, J. and Kutka, A. and Lainey, V. and LeBouquin, D. and Liu, C. and Loreggia, D. and Makarov, V. V. and Marseille, M. G. and Martayan, C. and {Martinez-Rubi}, O. and Massart, B. and Meynadier, F. and Mignot, S. and Munari, U. and Nguyen, A.-T. and Nordlander, T. and Ocvirk, P. and O'Flaherty, K. S. and Olias Sanz, A. and Ortiz, P. and Osorio, J. and Oszkiewicz, D. and Ouzounis, A. and Palmer, M. and Park, P. and Pasquato, E. and Peltzer, C. and Peralta, J. and P{\'e}turaud, F. and Pieniluoma, T. and Pigozzi, E. and Poels, J. and Prat, G. and Prod'homme, T. and Raison, F. and Rebordao, J. M. and Risquez, D. and {Rocca-Volmerange}, B. and Rosen, S. and {Ruiz-Fuertes}, M. I. and Russo, F. and Sembay, S. and Serraller Vizcaino, I. and Short, A. and Siebert, A. and Silva, H. and Sinachopoulos, D. and Slezak, E. and Soffel, M. and Sosnowska, D. and Strai{\v z}ys, V. and Ter Linden, M. and Terrell, D. and Theil, S. and Tiede, C. and Troisi, L. and Tsalmantza, P. and Tur, D. and Vaccari, M. and Vachier, F. and Valles, P. and Van Hamme, W. and Veltz, L. and Virtanen, J. and Wallut, J.-M. and Wichmann, R. and Wilkinson, M. I. and Ziaeepour, H. and Zschocke, S.},
  year = 2016,
  month = nov,
  journal = {A\&A},
  volume = {595},
  pages = {A1},
  issn = {0004-6361, 1432-0746},
  doi = {10.1051/0004-6361/201629272},
  urldate = {2026-04-16}
}

@article{GaiaCollaboration2023,
  title = {{\emph{Gaia}} {{Data Release}} 3: {{Summary}} of the Content and Survey Properties},
  shorttitle = {{\emph{Gaia}} {{Data Release}} 3},
  author = {{Gaia Collaboration} and Vallenari, A. and Brown, A. G. A. and Prusti, T. and De Bruijne, J. H. J. and Arenou, F. and Babusiaux, C. and Biermann, M. and Creevey, O. L. and Ducourant, C. and Evans, D. W. and Eyer, L. and Guerra, R. and Hutton, A. and Jordi, C. and Klioner, S. A. and Lammers, U. L. and Lindegren, L. and Luri, X. and Mignard, F. and Panem, C. and Pourbaix, D. and Randich, S. and Sartoretti, P. and Soubiran, C. and Tanga, P. and Walton, N. A. and {Bailer-Jones}, C. A. L. and Bastian, U. and Drimmel, R. and Jansen, F. and Katz, D. and Lattanzi, M. G. and Van Leeuwen, F. and Bakker, J. and Cacciari, C. and Casta{\~n}eda, J. and De Angeli, F. and Fabricius, C. and Fouesneau, M. and Fr{\'e}mat, Y. and Galluccio, L. and Guerrier, A. and Heiter, U. and Masana, E. and Messineo, R. and Mowlavi, N. and Nicolas, C. and Nienartowicz, K. and Pailler, F. and Panuzzo, P. and Riclet, F. and Roux, W. and Seabroke, G. M. and Sordo, R. and Th{\'e}venin, F. and {Gracia-Abril}, G. and Portell, J. and Teyssier, D. and Altmann, M. and Andrae, R. and Audard, M. and {Bellas-Velidis}, I. and Benson, K. and Berthier, J. and Blomme, R. and Burgess, P. W. and Busonero, D. and Busso, G. and C{\'a}novas, H. and Carry, B. and Cellino, A. and Cheek, N. and Clementini, G. and Damerdji, Y. and Davidson, M. and De Teodoro, P. and Nu{\~n}ez Campos, M. and Delchambre, L. and Dell'Oro, A. and Esquej, P. and {Fern{\'a}ndez-Hern{\'a}ndez}, J. and Fraile, E. and Garabato, D. and {Garc{\'i}a-Lario}, P. and Gosset, E. and Haigron, R. and Halbwachs, J.-L. and Hambly, N. C. and Harrison, D. L. and Hern{\'a}ndez, J. and Hestroffer, D. and Hodgkin, S. T. and Holl, B. and Jan{\ss}en, K. and Jevardat De Fombelle, G. and Jordan, S. and {Krone-Martins}, A. and Lanzafame, A. C. and L{\"o}ffler, W. and Marchal, O. and Marrese, P. M. and Moitinho, A. and Muinonen, K. and Osborne, P. and Pancino, E. and Pauwels, T. and {Recio-Blanco}, A. and Reyl{\'e}, C. and Riello, M. and Rimoldini, L. and Roegiers, T. and Rybizki, J. and Sarro, L. M. and Siopis, C. and Smith, M. and Sozzetti, A. and Utrilla, E. and Van Leeuwen, M. and Abbas, U. and {\'A}brah{\'a}m, P. and Abreu Aramburu, A. and Aerts, C. and Aguado, J. J. and Ajaj, M. and {Aldea-Montero}, F. and Altavilla, G. and {\'A}lvarez, M. A. and Alves, J. and Anders, F. and Anderson, R. I. and Anglada Varela, E. and Antoja, T. and Baines, D. and Baker, S. G. and {Balaguer-N{\'u}{\~n}ez}, L. and Balbinot, E. and Balog, Z. and Barache, C. and Barbato, D. and Barros, M. and Barstow, M. A. and Bartolom{\'e}, S. and Bassilana, J.-L. and Bauchet, N. and Becciani, U. and Bellazzini, M. and Berihuete, A. and Bernet, M. and Bertone, S. and Bianchi, L. and Binnenfeld, A. and {Blanco-Cuaresma}, S. and Blazere, A. and Boch, T. and Bombrun, A. and Bossini, D. and Bouquillon, S. and Bragaglia, A. and Bramante, L. and Breedt, E. and Bressan, A. and Brouillet, N. and Brugaletta, E. and Bucciarelli, B. and Burlacu, A. and Butkevich, A. G. and Buzzi, R. and Caffau, E. and Cancelliere, R. and {Cantat-Gaudin}, T. and Carballo, R. and Carlucci, T. and Carnerero, M. I. and Carrasco, J. M. and Casamiquela, L. and Castellani, M. and {Castro-Ginard}, A. and Chaoul, L. and Charlot, P. and Chemin, L. and Chiaramida, V. and Chiavassa, A. and Chornay, N. and Comoretto, G. and Contursi, G. and Cooper, W. J. and Cornez, T. and Cowell, S. and Crifo, F. and Cropper, M. and Crosta, M. and Crowley, C. and Dafonte, C. and Dapergolas, A. and David, M. and David, P. and De Laverny, P. and De Luise, F. and De March, R. and De Ridder, J. and De Souza, R. and De Torres, A. and Del Peloso, E. F. and Del Pozo, E. and Delbo, M. and Delgado, A. and Delisle, J.-B. and Demouchy, C. and Dharmawardena, T. E. and Di Matteo, P. and Diakite, S. and Diener, C. and Distefano, E. and Dolding, C. and Edvardsson, B. and Enke, H. and Fabre, C. and Fabrizio, M. and Faigler, S. and Fedorets, G. and Fernique, P. and Fienga, A. and Figueras, F. and Fournier, Y. and Fouron, C. and Fragkoudi, F. and Gai, M. and {Garcia-Gutierrez}, A. and {Garcia-Reinaldos}, M. and {Garc{\'i}a-Torres}, M. and Garofalo, A. and Gavel, A. and Gavras, P. and Gerlach, E. and Geyer, R. and Giacobbe, P. and Gilmore, G. and Girona, S. and Giuffrida, G. and Gomel, R. and Gomez, A. and {Gonz{\'a}lez-N{\'u}{\~n}ez}, J. and {Gonz{\'a}lez-Santamar{\'i}a}, I. and {Gonz{\'a}lez-Vidal}, J. J. and Granvik, M. and Guillout, P. and Guiraud, J. and {Guti{\'e}rrez-S{\'a}nchez}, R. and Guy, L. P. and Hatzidimitriou, D. and Hauser, M. and Haywood, M. and Helmer, A. and Helmi, A. and Sarmiento, M. H. and Hidalgo, S. L. and Hilger, T. and H{\l}adczuk, N. and Hobbs, D. and Holland, G. and Huckle, H. E. and Jardine, K. and Jasniewicz, G. and {Jean-Antoine Piccolo}, A. and {Jim{\'e}nez-Arranz}, {\'O}. and Jorissen, A. and Juaristi Campillo, J. and Julbe, F. and Karbevska, L. and Kervella, P. and Khanna, S. and Kontizas, M. and Kordopatis, G. and Korn, A. J. and K{\'o}sp{\'a}l, {\'A} and {Kostrzewa-Rutkowska}, Z. and Kruszy{\'n}ska, K. and Kun, M. and Laizeau, P. and Lambert, S. and Lanza, A. F. and Lasne, Y. and Le Campion, J.-F. and Lebreton, Y. and Lebzelter, T. and Leccia, S. and Leclerc, N. and {Lecoeur-Taibi}, I. and Liao, S. and Licata, E. L. and Lindstr{\o}m, H. E. P. and Lister, T. A. and Livanou, E. and Lobel, A. and Lorca, A. and Loup, C. and Madrero Pardo, P. and Magdaleno Romeo, A. and Managau, S. and Mann, R. G. and Manteiga, M. and Marchant, J. M. and Marconi, M. and Marcos, J. and Marcos Santos, M. M. S. and Mar{\'i}n Pina, D. and Marinoni, S. and Marocco, F. and Marshall, D. J. and Martin Polo, L. and {Mart{\'i}n-Fleitas}, J. M. and Marton, G. and Mary, N. and Masip, A. and Massari, D. and {Mastrobuono-Battisti}, A. and Mazeh, T. and McMillan, P. J. and Messina, S. and Michalik, D. and Millar, N. R. and Mints, A. and Molina, D. and Molinaro, R. and Moln{\'a}r, L. and Monari, G. and Mongui{\'o}, M. and Montegriffo, P. and Montero, A. and Mor, R. and Mora, A. and Morbidelli, R. and Morel, T. and Morris, D. and Muraveva, T. and Murphy, C. P. and Musella, I. and Nagy, Z. and Noval, L. and Oca{\~n}a, F. and Ogden, A. and Ordenovic, C. and Osinde, J. O. and Pagani, C. and Pagano, I. and Palaversa, L. and Palicio, P. A. and {Pallas-Quintela}, L. and Panahi, A. and {Payne-Wardenaar}, S. and Pe{\~n}alosa Esteller, X. and Penttil{\"a}, A. and Pichon, B. and Piersimoni, A. M. and Pineau, F.-X. and Plachy, E. and Plum, G. and Poggio, E. and Pr{\v s}a, A. and Pulone, L. and Racero, E. and Ragaini, S. and Rainer, M. and Raiteri, C. M. and Rambaux, N. and Ramos, P. and {Ramos-Lerate}, M. and Re Fiorentin, P. and Regibo, S. and Richards, P. J. and Rios Diaz, C. and Ripepi, V. and Riva, A. and Rix, H.-W. and Rixon, G. and Robichon, N. and Robin, A. C. and Robin, C. and Roelens, M. and Rogues, H. R. O. and Rohrbasser, L. and {Romero-G{\'o}mez}, M. and Rowell, N. and Royer, F. and Ruz Mieres, D. and Rybicki, K. A. and Sadowski, G. and S{\'a}ez N{\'u}{\~n}ez, A. and Sagrist{\`a} Sell{\'e}s, A. and Sahlmann, J. and Salguero, E. and Samaras, N. and Sanchez Gimenez, V. and Sanna, N. and Santove{\~n}a, R. and Sarasso, M. and Schultheis, M. and Sciacca, E. and Segol, M. and Segovia, J. C. and S{\'e}gransan, D. and Semeux, D. and Shahaf, S. and Siddiqui, H. I. and Siebert, A. and Siltala, L. and Silvelo, A. and Slezak, E. and Slezak, I. and Smart, R. L. and Snaith, O. N. and Solano, E. and Solitro, F. and Souami, D. and Souchay, J. and Spagna, A. and Spina, L. and Spoto, F. and Steele, I. A. and Steidelm{\"u}ller, H. and Stephenson, C. A. and S{\"u}veges, M. and Surdej, J. and Szabados, L. and {Szegedi-Elek}, E. and Taris, F. and Taylor, M. B. and Teixeira, R. and Tolomei, L. and Tonello, N. and Torra, F. and Torra, J. and Torralba Elipe, G. and Trabucchi, M. and Tsounis, A. T. and Turon, C. and Ulla, A. and Unger, N. and Vaillant, M. V. and Van Dillen, E. and Van Reeven, W. and Vanel, O. and Vecchiato, A. and Viala, Y. and Vicente, D. and Voutsinas, S. and Weiler, M. and Wevers, T. and Wyrzykowski, {\L}. and Yoldas, A. and Yvard, P. and Zhao, H. and Zorec, J. and Zucker, S. and Zwitter, T.},
  year = 2023,
  month = jun,
  journal = {A\&A},
  volume = {674},
  pages = {A1},
  issn = {0004-6361, 1432-0746},
  doi = {10.1051/0004-6361/202243940},
  urldate = {2026-04-16},
  copyright = {https://creativecommons.org/licenses/by/4.0}
}

@article{Harris2003a,
  title = {Flaring {{X-Ray Emission}} from {{HST-1}}, a {{Knot}} in the {{M87 Jet}}},
  author = {Harris, D. E. and Biretta, J. A. and Junor, W. and Perlman, E. S. and Sparks, W. B. and Wilson, A. S.},
  year = 2003,
  month = mar,
  journal = {ApJ},
  volume = {586},
  number = {1},
  pages = {L41-L44},
  issn = {0004637X, 15384357},
  doi = {10.1086/374773},
  urldate = {2026-04-16}
}

@article{HI4PICollaboration:2016,
  title = {{{HI4PI}}: A Full-Sky {{H}} i Survey Based on {{EBHIS}} and {{GASS}}},
  shorttitle = {{{HI4PI}}},
  author = {{HI4PI Collaboration} and Ben Bekhti, N. and Fl{\"o}er, L. and Keller, R. and Kerp, J. and Lenz, D. and Winkel, B. and Bailin, J. and Calabretta, M. R. and Dedes, L. and Ford, H. A. and Gibson, B. K. and Haud, U. and Janowiecki, S. and Kalberla, P. M. W. and Lockman, F. J. and {McClure-Griffiths}, N. M. and Murphy, T. and Nakanishi, H. and Pisano, D. J. and {Staveley-Smith}, L.},
  year = 2016,
  month = oct,
  journal = {A\&A},
  volume = {594},
  pages = {A116},
  issn = {0004-6361, 1432-0746},
  doi = {10.1051/0004-6361/201629178},
  urldate = {2026-04-16}
}

@article{Hou2020a,
  title = {A {{Chandra X-Ray Survey}} of {{Optically Selected AGN Pairs}}},
  author = {Hou, Meicun and Li, Zhiyuan and Liu, Xin},
  year = 2020,
  month = sep,
  journal = {ApJ},
  volume = {900},
  number = {1},
  pages = {79},
  issn = {0004-637X, 1538-4357},
  doi = {10.3847/1538-4357/aba4a7},
  urldate = {2026-04-16}
}

@inproceedings{Jerius2004a,
  title = {The Role of Modeling in the Calibration of the {{Chandra}}'s Optics},
  booktitle = {Optical {{Science}} and {{Technology}}, {{SPIE}}'s 48th {{Annual Meeting}}},
  author = {Jerius, Diab H. and Cohen, Lester and Edgar, Richard J. and Freeman, Mark and Gaetz, Terrance J. and Hughes, John P. and Nguyen, Dan and Podgorski, William A. and Tibbetts, Michael and Van Speybroeck, Leon P. and Zhao, Ping},
  year = 2004,
  month = feb,
  pages = {402},
  address = {San Diego, California, USA},
  doi = {10.1117/12.509378},
  urldate = {2026-04-16}
}

@inproceedings{Joye2003a,
  title = {New {{Features}} of {{SAOImage DS9}}},
  booktitle = {Astronomical {{Data Analysis Software}} and {{Systems XII}}},
  author = {Joye, W. A. and Mandel, E.},
  year = 2003,
  month = jan,
  volume = {295},
  pages = {489},
  urldate = {2026-04-16}
}

@misc{Keeton2001a,
  title = {Computational {{Methods}} for {{Gravitational Lensing}}},
  author = {Keeton, Charles R.},
  year = 2001,
  publisher = {arXiv},
  doi = {10.48550/ARXIV.ASTRO-PH/0102340},
  urldate = {2026-04-16},
  copyright = {Assumed arXiv.org perpetual, non-exclusive license to distribute this article for submissions made before January 2004}
}

@article{Lemon2018a,
  title = {Gravitationally Lensed Quasars in {{Gaia}} -- {{II}}. {{Discovery}} of 24 Lensed Quasars},
  author = {Lemon, Cameron A and Auger, Matthew W and McMahon, Richard G and Ostrovski, Fernanda},
  year = 2018,
  month = oct,
  journal = {MNRAS},
  volume = {479},
  number = {4},
  pages = {5060--5074},
  issn = {0035-8711, 1365-2966},
  doi = {10.1093/mnras/sty911},
  urldate = {2026-04-16},
  copyright = {http://academic.oup.com/journals/pages/about\_us/legal/notices},
  langid = {english}
}

@article{Lemon2022a,
  title = {J1721+8842: A Gravitationally Lensed Binary Quasar with a Proximate Damped {{Lyman-}} {\emph{{$\alpha$}}} Absorber},
  shorttitle = {J1721+8842},
  author = {Lemon, C. and Millon, M. and Sluse, D. and Courbin, F. and Auger, M. and Chan, J. H. H. and Paic, E. and Agnello, A.},
  year = 2022,
  month = jan,
  journal = {A\&A},
  volume = {657},
  pages = {A113},
  issn = {0004-6361, 1432-0746},
  doi = {10.1051/0004-6361/202142138},
  urldate = {2026-04-16},
  copyright = {https://www.edpsciences.org/en/authors/copyright-and-licensing}
}

@article{Mangat2021a,
  title = {{{PS J1721}}+8842: A Gravitationally Lensed Dual {{AGN}} System at Redshift 2.37 with Two Radio Components},
  shorttitle = {{{PS J1721}}+8842},
  author = {Mangat, C S and McKean, J P and Brilenkov, R and Hartley, P and Stacey, H R and Vegetti, S and Wen, D},
  year = 2021,
  month = oct,
  journal = {MNRAS},
  volume = {508},
  number = {1},
  pages = {L64-L68},
  issn = {1745-3925, 1745-3933},
  doi = {10.1093/mnrasl/slab106},
  urldate = {2026-04-16},
  copyright = {https://academic.oup.com/journals/pages/open\_access/funder\_policies/chorus/standard\_publication\_model},
  langid = {english}
}

@article{McCarty2025,
  title = {Strong Gravitational Lensing with Upcoming Wide-Field Radio Surveys},
  author = {McCarty, Samuel and Connor, Liam},
  year = 2025,
  month = aug,
  journal = {MNRAS},
  volume = {542},
  number = {3},
  pages = {2494--2509},
  issn = {0035-8711, 1365-2966},
  doi = {10.1093/mnras/staf1381},
  urldate = {2026-05-06},
  copyright = {https://creativecommons.org/licenses/by/4.0/},
  langid = {english}
}

@article{PlanckCollaboration2016,
  title = {{\emph{Planck}} 2015 Results: {{XIII}}. {{Cosmological}} Parameters},
  shorttitle = {{\emph{Planck}} 2015 Results},
  author = {{Planck Collaboration} and Ade, P. A. R. and Aghanim, N. and Arnaud, M. and Ashdown, M. and Aumont, J. and Baccigalupi, C. and Banday, A. J. and Barreiro, R. B. and Bartlett, J. G. and Bartolo, N. and Battaner, E. and Battye, R. and Benabed, K. and Beno{\^i}t, A. and {Benoit-L{\'e}vy}, A. and Bernard, J.-P. and Bersanelli, M. and Bielewicz, P. and Bock, J. J. and Bonaldi, A. and Bonavera, L. and Bond, J. R. and Borrill, J. and Bouchet, F. R. and Boulanger, F. and Bucher, M. and Burigana, C. and Butler, R. C. and Calabrese, E. and Cardoso, J.-F. and Catalano, A. and Challinor, A. and Chamballu, A. and Chary, R.-R. and Chiang, H. C. and Chluba, J. and Christensen, P. R. and Church, S. and Clements, D. L. and Colombi, S. and Colombo, L. P. L. and Combet, C. and Coulais, A. and Crill, B. P. and Curto, A. and Cuttaia, F. and Danese, L. and Davies, R. D. and Davis, R. J. and De Bernardis, P. and De Rosa, A. and De Zotti, G. and Delabrouille, J. and D{\'e}sert, F.-X. and Di Valentino, E. and Dickinson, C. and Diego, J. M. and Dolag, K. and Dole, H. and Donzelli, S. and Dor{\'e}, O. and Douspis, M. and Ducout, A. and Dunkley, J. and Dupac, X. and Efstathiou, G. and Elsner, F. and En{\ss}lin, T. A. and Eriksen, H. K. and Farhang, M. and Fergusson, J. and Finelli, F. and Forni, O. and Frailis, M. and Fraisse, A. A. and Franceschi, E. and Frejsel, A. and Galeotta, S. and Galli, S. and Ganga, K. and Gauthier, C. and Gerbino, M. and Ghosh, T. and Giard, M. and {Giraud-H{\'e}raud}, Y. and Giusarma, E. and Gjerl{\o}w, E. and {Gonz{\'a}lez-Nuevo}, J. and G{\'o}rski, K. M. and Gratton, S. and Gregorio, A. and Gruppuso, A. and Gudmundsson, J. E. and Hamann, J. and Hansen, F. K. and Hanson, D. and Harrison, D. L. and Helou, G. and {Henrot-Versill{\'e}}, S. and {Hern{\'a}ndez-Monteagudo}, C. and Herranz, D. and Hildebrandt, S. R. and Hivon, E. and Hobson, M. and Holmes, W. A. and Hornstrup, A. and Hovest, W. and Huang, Z. and Huffenberger, K. M. and Hurier, G. and Jaffe, A. H. and Jaffe, T. R. and Jones, W. C. and Juvela, M. and Keih{\"a}nen, E. and Keskitalo, R. and Kisner, T. S. and Kneissl, R. and Knoche, J. and Knox, L. and Kunz, M. and {Kurki-Suonio}, H. and Lagache, G. and L{\"a}hteenm{\"a}ki, A. and Lamarre, J.-M. and Lasenby, A. and Lattanzi, M. and Lawrence, C. R. and Leahy, J. P. and Leonardi, R. and Lesgourgues, J. and Levrier, F. and Lewis, A. and Liguori, M. and Lilje, P. B. and {Linden-V{\o}rnle}, M. and {L{\'o}pez-Caniego}, M. and Lubin, P. M. and {Mac{\'i}as-P{\'e}rez}, J. F. and Maggio, G. and Maino, D. and Mandolesi, N. and Mangilli, A. and Marchini, A. and Maris, M. and Martin, P. G. and Martinelli, M. and {Mart{\'i}nez-Gonz{\'a}lez}, E. and Masi, S. and Matarrese, S. and McGehee, P. and Meinhold, P. R. and Melchiorri, A. and Melin, J.-B. and Mendes, L. and Mennella, A. and Migliaccio, M. and Millea, M. and Mitra, S. and {Miville-Desch{\^e}nes}, M.-A. and Moneti, A. and Montier, L. and Morgante, G. and Mortlock, D. and Moss, A. and Munshi, D. and Murphy, J. A. and Naselsky, P. and Nati, F. and Natoli, P. and Netterfield, C. B. and {N{\o}rgaard-Nielsen}, H. U. and Noviello, F. and Novikov, D. and Novikov, I. and Oxborrow, C. A. and Paci, F. and Pagano, L. and Pajot, F. and Paladini, R. and Paoletti, D. and Partridge, B. and Pasian, F. and Patanchon, G. and Pearson, T. J. and Perdereau, O. and Perotto, L. and Perrotta, F. and Pettorino, V. and Piacentini, F. and Piat, M. and Pierpaoli, E. and Pietrobon, D. and Plaszczynski, S. and Pointecouteau, E. and Polenta, G. and Popa, L. and Pratt, G. W. and Pr{\'e}zeau, G. and Prunet, S. and Puget, J.-L. and Rachen, J. P. and Reach, W. T. and Rebolo, R. and Reinecke, M. and Remazeilles, M. and Renault, C. and Renzi, A. and Ristorcelli, I. and Rocha, G. and Rosset, C. and Rossetti, M. and Roudier, G. and {Rouill{\'e} d'Orfeuil}, B. and {Rowan-Robinson}, M. and {Rubi{\~n}o-Mart{\'i}n}, J. A. and Rusholme, B. and Said, N. and Salvatelli, V. and Salvati, L. and Sandri, M. and Santos, D. and Savelainen, M. and Savini, G. and Scott, D. and Seiffert, M. D. and Serra, P. and Shellard, E. P. S. and Spencer, L. D. and Spinelli, M. and Stolyarov, V. and Stompor, R. and Sudiwala, R. and Sunyaev, R. and Sutton, D. and {Suur-Uski}, A.-S. and Sygnet, J.-F. and Tauber, J. A. and Terenzi, L. and Toffolatti, L. and Tomasi, M. and Tristram, M. and Trombetti, T. and Tucci, M. and Tuovinen, J. and T{\"u}rler, M. and Umana, G. and Valenziano, L. and Valiviita, J. and Van Tent, F. and Vielva, P. and Villa, F. and Wade, L. A. and Wandelt, B. D. and Wehus, I. K. and White, M. and White, S. D. M. and Wilkinson, A. and Yvon, D. and Zacchei, A. and Zonca, A.},
  year = 2016,
  month = oct,
  journal = {A\&A},
  volume = {594},
  pages = {A13},
  issn = {0004-6361, 1432-0746},
  doi = {10.1051/0004-6361/201525830},
  urldate = {2026-04-20}
}

@article{Rogers2025a,
  title = {Milliarcsecond {{X-Ray Positions}} and {{X-Ray Varstrometry}} for the {{Strongly Lensed Active Galactic Nucleus HE}} 0435-1223},
  author = {Rogers, Alysa and Schwartz, Daniel and Spingola, Cristiana and Barnacka, Anna},
  year = 2025,
  month = jul,
  journal = {ApJ},
  volume = {987},
  number = {1},
  pages = {75},
  issn = {0004-637X, 1538-4357},
  doi = {10.3847/1538-4357/add71a},
  urldate = {2026-04-16}
}

@article{Schmidt2025a,
  title = {{{TDCOSMO}}: {{XVIII}}. {{Strong}} Lens Model and Time-Delay Predictions for {{J1721}}+8842, the First {{Einstein}} Zigzag Lens},
  shorttitle = {{{TDCOSMO}}},
  author = {Schmidt, T. and Treu, T. and Birrer, S. and Millon, M. and Sluse, D. and Galan, A. and Shajib, A. J. and Lemon, C. and Dux, F. and Courbin, F.},
  year = 2025,
  month = aug,
  journal = {A\&A},
  volume = {700},
  pages = {A92},
  issn = {0004-6361, 1432-0746},
  doi = {10.1051/0004-6361/202449984},
  urldate = {2026-04-16},
  copyright = {https://creativecommons.org/licenses/by/4.0}
}

@article{Schwartz2021a,
  title = {Resolving {{Complex Inner X-Ray Structure}} of the {{Gravitationally Lensed AGN MG B2016}}+112},
  author = {Schwartz, Daniel and Spingola, Cristiana and Barnacka, Anna},
  year = 2021,
  month = aug,
  journal = {ApJ},
  volume = {917},
  number = {1},
  pages = {26},
  issn = {0004-637X, 1538-4357},
  doi = {10.3847/1538-4357/ac0909},
  urldate = {2026-04-16}
}

@article{Sisk-Reynes2026a,
  title = {Determining the {{Spatial Origin}} of {{X-Ray}} and {{Optical Emission}} in the z = 3.1 {{Strongly Lensed Radio-quiet Quasar GraL J065904}}.1+162909 to {{Hundreds}} of {{Parsecs}}},
  author = {{Sisk-Reyn{\'e}s}, J{\'u}lia M. and Schwartz, Daniel A. and Barnacka, Anna and Spingola, Cristiana and Migliori, Giulia},
  year = 2026,
  month = mar,
  journal = {ApJ},
  volume = {999},
  number = {2},
  pages = {215},
  issn = {0004-637X, 1538-4357},
  doi = {10.3847/1538-4357/ae3f2c},
  urldate = {2026-04-16}
}

@article{Spingola2019a,
  title = {Proper Motion in Lensed Radio Jets at Redshift 3: {{A}} Possible Dual Super-Massive Black Hole System in the Early {{Universe}}},
  shorttitle = {Proper Motion in Lensed Radio Jets at Redshift 3},
  author = {Spingola, C. and McKean, J. P. and Massari, D. and Koopmans, L. V. E.},
  year = 2019,
  month = oct,
  journal = {A\&A},
  volume = {630},
  pages = {A108},
  issn = {0004-6361, 1432-0746},
  doi = {10.1051/0004-6361/201935427},
  urldate = {2026-04-16},
  copyright = {https://www.edpsciences.org/en/authors/copyright-and-licensing}
}

@article{Spingola2020a,
  title = {Constraining {{VLBI}}-optical Offsets in High Redshift Galaxies Using Strong Gravitational Lensing},
  author = {Spingola, Cristiana and Barnacka, Anna},
  year = 2020,
  month = may,
  journal = {MNRAS},
  volume = {494},
  number = {2},
  pages = {2312--2326},
  issn = {0035-8711, 1365-2966},
  doi = {10.1093/mnras/staa870},
  urldate = {2026-04-16},
  copyright = {https://academic.oup.com/journals/pages/open\_access/funder\_policies/chorus/standard\_publication\_model},
  langid = {english}
}

@article{Spingola2022a,
  title = {Milliarcsecond {{X-Ray Astrometry}} to {{Resolve Inner Regions}} of {{AGN}} at z {$>$} 1 {{Using Gravitational Lensing}}},
  author = {Spingola, Cristiana and Schwartz, Daniel and Barnacka, Anna},
  year = 2022,
  month = may,
  journal = {ApJ},
  volume = {931},
  number = {1},
  pages = {68},
  issn = {0004-637X, 1538-4357},
  doi = {10.3847/1538-4357/ac68eb},
  urldate = {2026-04-16}
}

@article{Stern2012a,
  title = {{{MID-INFRARED SELECTION OF ACTIVE GALACTIC NUCLEI WITH THE}} {{{\emph{WIDE-FIELD INFRARED SURVEY EXPLORER}}}} . {{I}}. {{CHARACTERIZING}} {{{\emph{WISE}}}} -{{SELECTED ACTIVE GALACTIC NUCLEI IN COSMOS}}},
  author = {Stern, Daniel and Assef, Roberto J. and Benford, Dominic J. and Blain, Andrew and Cutri, Roc and Dey, Arjun and Eisenhardt, Peter and Griffith, Roger L. and Jarrett, T. H. and Lake, Sean and Masci, Frank and Petty, Sara and Stanford, S. A. and Tsai, Chao-Wei and Wright, E. L. and Yan, Lin and Harrison, Fiona and Madsen, Kristin},
  year = 2012,
  month = jul,
  journal = {ApJ},
  volume = {753},
  number = {1},
  pages = {30},
  issn = {0004-637X, 1538-4357},
  doi = {10.1088/0004-637X/753/1/30},
  urldate = {2026-04-16}
}

@article{Harris2006,
    title = {The {Outburst} of {HST}‐1 in the {M87} {Jet}},
    volume = {640},
    issn = {0004-637X, 1538-4357},
    url = {https://iopscience.iop.org/article/10.1086/500081},
    doi = {10.1086/500081},
    language = {en},
    number = {1},
    urldate = {2026-04-20},
    journal = {The Astrophysical Journal},
    author = {Harris, D. E. and Cheung, C. C. and Biretta, J. A. and Sparks, W. B. and Junor, W. and Perlman, E. S. and Wilson, A. S.},
    month = mar,
    year = {2006},
    pages = {211--218},
}

@article{Blecha2011,
    title = {Recoiling black holes in merging galaxies: relationship to active galactic nucleus lifetimes, starbursts and the {MBH}-σ* relation: {Recoiling} black holes in merging galaxies},
    volume = {412},
    issn = {00358711},
    shorttitle = {Recoiling black holes in merging galaxies},
    url = {https://academic.oup.com/mnras/article-lookup/doi/10.1111/j.1365-2966.2010.18042.x},
    doi = {10.1111/j.1365-2966.2010.18042.x},
    language = {en},
    number = {4},
    urldate = {2026-05-14},
    journal = {Monthly Notices of the Royal Astronomical Society},
    author = {Blecha, Laura and Cox, Thomas J. and Loeb, Abraham and Hernquist, Lars},
    month = apr,
    year = {2011},
    pages = {2154--2182},
}

@article{Foord2021a,
    title = {{AGN} {Triality} of {Triple} {Mergers}: {Detection} of {Faint} {X}-{Ray} {Point} {Sources}},
    volume = {907},
    issn = {0004-637X, 1538-4357},
    shorttitle = {{AGN} {Triality} of {Triple} {Mergers}},
    url = {https://iopscience.iop.org/article/10.3847/1538-4357/abce5d},
    doi = {10.3847/1538-4357/abce5d},
    number = {2},
    urldate = {2026-04-16},
    journal = {The Astrophysical Journal},
    author = {Foord, Adi and Gültekin, Kayhan and Runnoe, Jessie C. and Koss, Michael J.},
    month = feb,
    year = {2021},
    pages = {71},
}

@article{Foord2020a,
    title = {A {Second} {Look} at 12 {Candidate} {Dual} {AGNs} {Using} {BAYMAX}},
    volume = {892},
    issn = {0004-637X, 1538-4357},
    url = {https://iopscience.iop.org/article/10.3847/1538-4357/ab72fa},
    doi = {10.3847/1538-4357/ab72fa},
    number = {1},
    urldate = {2026-04-16},
    journal = {The Astrophysical Journal},
    author = {Foord, Adi and Gültekin, Kayhan and Nevin, Rebecca and Comerford, Julia M. and Hodges-Kluck, Edmund and Barrows, R. Scott and Goulding, Andrew D. and Greene, Jenny E.},
    month = mar,
    year = {2020},
    pages = {29},
}

@article{Koss2012,
    title = {{UNDERSTANDING} {DUAL} {ACTIVE} {GALACTIC} {NUCLEUS} {ACTIVATION} {IN} {THE} {NEARBY} {UNIVERSE}},
    volume = {746},
    issn = {2041-8205, 2041-8213},
    url = {https://iopscience.iop.org/article/10.1088/2041-8205/746/2/L22},
    doi = {10.1088/2041-8205/746/2/L22},
    number = {2},
    urldate = {2026-05-24},
    journal = {The Astrophysical Journal Letters},
    author = {Koss, Michael and Mushotzky, Richard and Treister, Ezequiel and Veilleux, Sylvain and Vasudevan, Ranjan and Trippe, Margaret},
    month = feb,
    year = {2012},
    pages = {L22},
}

@article{Blecha2016,
    title = {Recoiling black holes: prospects for detection and implications of spin alignment},
    volume = {456},
    issn = {0035-8711, 1365-2966},
    shorttitle = {Recoiling black holes},
    url = {https://academic.oup.com/mnras/article-lookup/doi/10.1093/mnras/stv2646},
    doi = {10.1093/mnras/stv2646},
    language = {en},
    number = {1},
    urldate = {2026-05-24},
    journal = {Monthly Notices of the Royal Astronomical Society},
    author = {Blecha, Laura and Sijacki, Debora and Kelley, Luke Zoltan and Torrey, Paul and Vogelsberger, Mark and Nelson, Dylan and Springel, Volker and Snyder, Gregory and Hernquist, Lars},
    month = feb,
    year = {2016},
    pages = {961--989},
}

@article{Kataoka2008,
    title = {\textit{{Chandra}} {Reveals} {Twin} {X}‐{Ray} {Jets} in the {Powerful} {FR} {II} {Radio} {Galaxy} {3C} 353},
    volume = {685},
    issn = {0004-637X, 1538-4357},
    url = {https://iopscience.iop.org/article/10.1086/591024},
    doi = {10.1086/591024},
    language = {en},
    number = {2},
    urldate = {2026-05-24},
    journal = {The Astrophysical Journal},
    author = {Kataoka, J. and Stawarz, L. and Harris, D. E. and Siemiginowska, A. and Ostrowski, M. and Swain, M. R. and Hardcastle, M. J. and Goodger, J. L. and Iwasawa, K. and Edwards, P. G.},
    month = oct,
    year = {2008},
    pages = {839--857},
}

@article{Wyithe2003,
    title = {Self‐regulated {Growth} of {Supermassive} {Black} {Holes} in {Galaxies} as the {Origin} of the {Optical} and {X}‐{Ray} {Luminosity} {Functions} of {Quasars}},
    volume = {595},
    issn = {0004-637X, 1538-4357},
    url = {https://iopscience.iop.org/article/10.1086/377475},
    doi = {10.1086/377475},
    language = {en},
    number = {2},
    urldate = {2026-05-25},
    journal = {The Astrophysical Journal},
    author = {Wyithe, J. Stuart B. and Loeb, Abraham},
    month = oct,
    year = {2003},
    pages = {614--623},
}

@article{Lasker2008,
    title = {{THE} {SECOND}-{GENERATION} {GUIDE} {STAR} {CATALOG}: {DESCRIPTION} {AND} {PROPERTIES}},
    volume = {136},
    issn = {0004-6256, 1538-3881},
    shorttitle = {{THE} {SECOND}-{GENERATION} {GUIDE} {STAR} {CATALOG}},
    url = {https://iopscience.iop.org/article/10.1088/0004-6256/136/2/735},
    doi = {10.1088/0004-6256/136/2/735},
    number = {2},
    urldate = {2026-05-29},
    journal = {The Astronomical Journal},
    author = {Lasker, Barry M. and Lattanzi, Mario G. and McLean, Brian J. and Bucciarelli, Beatrice and Drimmel, Ronald and Garcia, Jorge and Greene, Gretchen and Guglielmetti, Fabrizia and Hanley, Christopher and Hawkins, George and Laidler, Victoria G. and Loomis, Charles and Meakes, Michael and Mignani, Roberto and Morbidelli, Roberto and Morrison, Jane and Pannunzio, Renato and Rosenberg, Amy and Sarasso, Maria and Smart, Richard L. and Spagna, Alessandro and Sturch, Conrad R. and Volpicelli, Antonio and White, Richard L. and Wolfe, David and Zacchei, Andrea},
    month = aug,
    year = {2008},
    pages = {735--766},
}

@article{Auchettl2015,
    title = {X-{RAY} {ANALYSIS} {OF} {THE} {PROPER} {MOTION} {AND} {PULSAR} {WIND} {NEBULA} {FOR} {PSR} {J1741}-2054},
    volume = {802},
    copyright = {http://iopscience.iop.org/info/page/text-and-data-mining},
    issn = {1538-4357},
    url = {https://iopscience.iop.org/article/10.1088/0004-637X/802/1/68},
    doi = {10.1088/0004-637X/802/1/68},
    number = {1},
    urldate = {2026-06-08},
    journal = {The Astrophysical Journal},
    author = {Auchettl, Katie and Slane, Patrick and Romani, Roger W. and Posselt, Bettina and Pavlov, George G. and Kargaltsev, Oleg and Ng, C-Y. and Temim, Tea and Weisskopf, Martin. C. and Bykov, Andrei and Swartz, Douglas A.},
    month = mar,
    year = {2015},
    pages = {68},
}

@ARTICLE{Fruscione2026,
       author = {{Fruscione}, Antonella and {McDowell}, Jonathan and {Burke}, Douglas and {Cresitello Dittmar}, Mark and {Evans}, Ian N. and {Evans}, Janet D. and {Glotfelty}, Kenny and {Guenther}, Hans Moritz and {Huenemoerder}, David and {Joye}, William and {Lee}, Nicholas P. and {McLaughlin}, Warren and {Miller}, Joseph B. and {Nynka}, Melania and {Principe}, David A. and {Siemiginowska}, Aneta},
        title = "{CIAO: Chandra's Data Analysis System for X-Ray Astronomy and Beyond}",
      journal = {arXiv e-prints},
         year = 2026,
        month = may,
          eid = {arXiv:2605.14144},
        pages = {arXiv:2605.14144},
          doi = {10.48550/arXiv.2605.14144},
archivePrefix = {arXiv},
       eprint = {2605.14144},
 primaryClass = {astro-ph.IM},
       adsurl = {https://ui.adsabs.harvard.edu/abs/2026arXiv260514144F}
}

@article{Gu2025,
  title = {Delving into the Depths of {{NGC}} 3783 with {{XRISM}}: {{III}}. {{Birth}} of an Ultrafast Outflow during a Soft Flare},
  shorttitle = {Delving into the Depths of {{NGC}} 3783 with {{XRISM}}},
  author = {Gu, Liyi and Fukumura, Keigo and Kaastra, Jelle and Eckart, Megan and Ballhausen, Ralf and Behar, Ehud and Diez, Camille and Guainazzi, Matteo and Kallman, Timothy and Kara, Erin and Li, Chen and Mehdipour, Missagh and Mizumoto, Misaki and Ogawa, Shoji and Panagiotou, Christos and Signorini, Matilde and Tanimoto, Atsushi and Zhao, Keqin and Noda, Hirofumi and Miller, Jon and Yamada, Satoshi},
  year = 2025,
  month = dec,
  journal = {A\&A},
  volume = {704},
  pages = {A146},
  issn = {0004-6361, 1432-0746},
  doi = {10.1051/0004-6361/202557189},
  urldate = {2026-06-17},
  copyright = {https://creativecommons.org/licenses/by/4.0}
}

@article{Schmidt2022,
    title = {{STRIDES}: automated uniform models for 30 quadruply imaged quasars},
    volume = {518},
    copyright = {https://academic.oup.com/journals/pages/open\_access/funder\_policies/chorus/standard\_publication\_model},
    issn = {0035-8711, 1365-2966},
    shorttitle = {{STRIDES}},
    url = {https://academic.oup.com/mnras/article/518/1/1260/6786285},
    doi = {10.1093/mnras/stac2235},
    language = {en},
    number = {1},
    urldate = {2026-06-18},
    journal = {Monthly Notices of the Royal Astronomical Society},
    author = {Schmidt, T and Treu, T and Birrer, S and Shajib, A J and Lemon, C and Millon, M and Sluse, D and Agnello, A and Anguita, T and Auger-Williams, M W and McMahon, R G and Motta, V and Schechter, P and Spiniello, C and Kayo, I and Courbin, F and Ertl, S and Fassnacht, C D and Frieman, J A and More, A and Schuldt, S and Suyu, S H and Aguena, M and Andrade-Oliveira, F and Annis, J and Bacon, D and Bertin, E and Brooks, D and Burke, D L and Carnero Rosell, A and Carrasco Kind, M and Carretero, J and Conselice, C and Costanzi, M and da Costa, L N and Pereira, M E S and De Vicente, J and Desai, S and Doel, P and Everett, S and Ferrero, I and Friedel, D and García-Bellido, J and Gaztanaga, E and Gruen, D and Gruendl, R A and Gschwend, J and Gutierrez, G and Hinton, S R and Hollowood, D L and Honscheid, K and James, D J and Kuehn, K and Lahav, O and Menanteau, F and Miquel, R and Palmese, A and Paz-Chinchón, F and Pieres, A and Plazas Malagón, A A and Prat, J and Rodriguez-Monroy, M and Romer, A K and Sanchez, E and Scarpine, V and Sevilla-Noarbe, I and Smith, M and Suchyta, E and Tarle, G and To, C and Varga, T N and {(DES Collaboration)}},
    month = nov,
    year = {2022},
    pages = {1260--1300},
}

@article{Ertl2023,
    title = {{TDCOSMO}: {X}. {Automated} modeling of nine strongly lensed quasars and comparison between lens-modeling software},
    volume = {672},
    copyright = {https://creativecommons.org/licenses/by/4.0},
    issn = {0004-6361, 1432-0746},
    shorttitle = {{TDCOSMO}},
    url = {https://www.aanda.org/10.1051/0004-6361/202244909},
    doi = {10.1051/0004-6361/202244909},
    urldate = {2026-06-18},
    journal = {Astronomy \& Astrophysics},
    author = {Ertl, S. and Schuldt, S. and Suyu, S. H. and Schmidt, T. and Treu, T. and Birrer, S. and Shajib, A. J. and Sluse, D.},
    month = apr,
    year = {2023},
    pages = {A2},
}

@article{Reeves2000,
    title = {X-ray spectra of a large sample of quasars with {ASCA}},
    volume = {316},
    issn = {0035-8711, 1365-2966},
    url = {https://academic.oup.com/mnras/article/316/2/234/1019716},
    doi = {10.1046/j.1365-8711.2000.03510.x},
    language = {en},
    number = {2},
    urldate = {2026-06-20},
    journal = {Monthly Notices of the Royal Astronomical Society},
    author = {Reeves, J. N. and Turner, M. J. L.},
    month = aug,
    year = {2000},
    pages = {234--248},
}

@article{Abbott2016,
    title = {Observation of {Gravitational} {Waves} from a {Binary} {Black} {Hole} {Merger}},
    volume = {116},
    copyright = {http://creativecommons.org/licenses/by/3.0/},
    issn = {0031-9007, 1079-7114},
    url = {https://link.aps.org/doi/10.1103/PhysRevLett.116.061102},
    doi = {10.1103/PhysRevLett.116.061102},
    language = {en},
    number = {6},
    urldate = {2026-06-22},
    journal = {Physical Review Letters},
    author = {Abbott, B. P. and Abbott, R. and Abbott, T. D. and Abernathy, M. R. and Acernese, F. and Ackley, K. and Adams, C. and Adams, T. and Addesso, P. and Adhikari, R. X. and Adya, V. B. and Affeldt, C. and Agathos, M. and Agatsuma, K. and Aggarwal, N. and Aguiar, O. D. and Aiello, L. and Ain, A. and Ajith, P. and Allen, B. and Allocca, A. and Altin, P. A. and Anderson, S. B. and Anderson, W. G. and Arai, K. and Arain, M. A. and Araya, M. C. and Arceneaux, C. C. and Areeda, J. S. and Arnaud, N. and Arun, K. G. and Ascenzi, S. and Ashton, G. and Ast, M. and Aston, S. M. and Astone, P. and Aufmuth, P. and Aulbert, C. and Babak, S. and Bacon, P. and Bader, M. K. M. and Baker, P. T. and Baldaccini, F. and Ballardin, G. and Ballmer, S. W. and Barayoga, J. C. and Barclay, S. E. and Barish, B. C. and Barker, D. and Barone, F. and Barr, B. and Barsotti, L. and Barsuglia, M. and Barta, D. and Bartlett, J. and Barton, M. A. and Bartos, I. and Bassiri, R. and Basti, A. and Batch, J. C. and Baune, C. and Bavigadda, V. and Bazzan, M. and Behnke, B. and Bejger, M. and Belczynski, C. and Bell, A. S. and Bell, C. J. and Berger, B. K. and Bergman, J. and Bergmann, G. and Berry, C. P. L. and Bersanetti, D. and Bertolini, A. and Betzwieser, J. and Bhagwat, S. and Bhandare, R. and Bilenko, I. A. and Billingsley, G. and Birch, J. and Birney, I. A. and Birnholtz, O. and Biscans, S. and Bisht, A. and Bitossi, M. and Biwer, C. and Bizouard, M. A. and Blackburn, J. K. and Blair, C. D. and Blair, D. G. and Blair, R. M. and Bloemen, S. and Bock, O. and Bodiya, T. P. and Boer, M. and Bogaert, G. and Bogan, C. and Bohe, A. and Bojtos, P. and Bond, C. and Bondu, F. and Bonnand, R. and Boom, B. A. and Bork, R. and Boschi, V. and Bose, S. and Bouffanais, Y. and Bozzi, A. and Bradaschia, C. and Brady, P. R. and Braginsky, V. B. and Branchesi, M. and Brau, J. E. and Briant, T. and Brillet, A. and Brinkmann, M. and Brisson, V. and Brockill, P. and Brooks, A. F. and Brown, D. A. and Brown, D. D. and Brown, N. M. and Buchanan, C. C. and Buikema, A. and Bulik, T. and Bulten, H. J. and Buonanno, A. and Buskulic, D. and Buy, C. and Byer, R. L. and Cabero, M. and Cadonati, L. and Cagnoli, G. and Cahillane, C. and Bustillo, J. Calderón and Callister, T. and Calloni, E. and Camp, J. B. and Cannon, K. C. and Cao, J. and Capano, C. D. and Capocasa, E. and Carbognani, F. and Caride, S. and Diaz, J. Casanueva and Casentini, C. and Caudill, S. and Cavaglià, M. and Cavalier, F. and Cavalieri, R. and Cella, G. and Cepeda, C. B. and Baiardi, L. Cerboni and Cerretani, G. and Cesarini, E. and Chakraborty, R. and Chalermsongsak, T. and Chamberlin, S. J. and Chan, M. and Chao, S. and Charlton, P. and Chassande-Mottin, E. and Chen, H. Y. and Chen, Y. and Cheng, C. and Chincarini, A. and Chiummo, A. and Cho, H. S. and Cho, M. and Chow, J. H. and Christensen, N. and Chu, Q. and Chua, S. and Chung, S. and Ciani, G. and Clara, F. and Clark, J. A. and Cleva, F. and Coccia, E. and Cohadon, P.-F. and Colla, A. and Collette, C. G. and Cominsky, L. and Constancio, M. and Conte, A. and Conti, L. and Cook, D. and Corbitt, T. R. and Cornish, N. and Corsi, A. and Cortese, S. and Costa, C. A. and Coughlin, M. W. and Coughlin, S. B. and Coulon, J.-P. and Countryman, S. T. and Couvares, P. and Cowan, E. E. and Coward, D. M. and Cowart, M. J. and Coyne, D. C. and Coyne, R. and Craig, K. and Creighton, J. D. E. and Creighton, T. D. and Cripe, J. and Crowder, S. G. and Cruise, A. M. and Cumming, A. and Cunningham, L. and Cuoco, E. and Canton, T. Dal and Danilishin, S. L. and D’Antonio, S. and Danzmann, K. and Darman, N. S. and Da Silva Costa, C. F. and Dattilo, V. and Dave, I. and Daveloza, H. P. and Davier, M. and Davies, G. S. and Daw, E. J. and Day, R. and De, S. and DeBra, D. and Debreczeni, G. and Degallaix, J. and De Laurentis, M. and Deléglise, S. and Del Pozzo, W. and Denker, T. and Dent, T. and Dereli, H. and Dergachev, V. and DeRosa, R. T. and De Rosa, R. and DeSalvo, R. and Dhurandhar, S. and Díaz, M. C. and Di Fiore, L. and Di Giovanni, M. and Di Lieto, A. and Di Pace, S. and Di Palma, I. and Di Virgilio, A. and Dojcinoski, G. and Dolique, V. and Donovan, F. and Dooley, K. L. and Doravari, S. and Douglas, R. and Downes, T. P. and Drago, M. and Drever, R. W. P. and Driggers, J. C. and Du, Z. and Ducrot, M. and Dwyer, S. E. and Edo, T. B. and Edwards, M. C. and Effler, A. and Eggenstein, H.-B. and Ehrens, P. and Eichholz, J. and Eikenberry, S. S. and Engels, W. and Essick, R. C. and Etzel, T. and Evans, M. and Evans, T. M. and Everett, R. and Factourovich, M. and Fafone, V. and Fair, H. and Fairhurst, S. and Fan, X. and Fang, Q. and Farinon, S. and Farr, B. and Farr, W. M. and Favata, M. and Fays, M. and Fehrmann, H. and Fejer, M. M. and Feldbaum, D. and Ferrante, I. and Ferreira, E. C. and Ferrini, F. and Fidecaro, F. and Finn, L. S. and Fiori, I. and Fiorucci, D. and Fisher, R. P. and Flaminio, R. and Fletcher, M. and Fong, H. and Fournier, J.-D. and Franco, S. and Frasca, S. and Frasconi, F. and Frede, M. and Frei, Z. and Freise, A. and Frey, R. and Frey, V. and Fricke, T. T. and Fritschel, P. and Frolov, V. V. and Fulda, P. and Fyffe, M. and Gabbard, H. A. G. and Gair, J. R. and Gammaitoni, L. and Gaonkar, S. G. and Garufi, F. and Gatto, A. and Gaur, G. and Gehrels, N. and Gemme, G. and Gendre, B. and Genin, E. and Gennai, A. and George, J. and Gergely, L. and Germain, V. and Ghosh, Abhirup and Ghosh, Archisman and Ghosh, S. and Giaime, J. A. and Giardina, K. D. and Giazotto, A. and Gill, K. and Glaefke, A. and Gleason, J. R. and Goetz, E. and Goetz, R. and Gondan, L. and González, G. and Castro, J. M. Gonzalez and Gopakumar, A. and Gordon, N. A. and Gorodetsky, M. L. and Gossan, S. E. and Gosselin, M. and Gouaty, R. and Graef, C. and Graff, P. B. and Granata, M. and Grant, A. and Gras, S. and Gray, C. and Greco, G. and Green, A. C. and Greenhalgh, R. J. S. and Groot, P. and Grote, H. and Grunewald, S. and Guidi, G. M. and Guo, X. and Gupta, A. and Gupta, M. K. and Gushwa, K. E. and Gustafson, E. K. and Gustafson, R. and Hacker, J. J. and Hall, B. R. and Hall, E. D. and Hammond, G. and Haney, M. and Hanke, M. M. and Hanks, J. and Hanna, C. and Hannam, M. D. and Hanson, J. and Hardwick, T. and Harms, J. and Harry, G. M. and Harry, I. W. and Hart, M. J. and Hartman, M. T. and Haster, C.-J. and Haughian, K. and Healy, J. and Heefner, J. and Heidmann, A. and Heintze, M. C. and Heinzel, G. and Heitmann, H. and Hello, P. and Hemming, G. and Hendry, M. and Heng, I. S. and Hennig, J. and Heptonstall, A. W. and Heurs, M. and Hild, S. and Hoak, D. and Hodge, K. A. and Hofman, D. and Hollitt, S. E. and Holt, K. and Holz, D. E. and Hopkins, P. and Hosken, D. J. and Hough, J. and Houston, E. A. and Howell, E. J. and Hu, Y. M. and Huang, S. and Huerta, E. A. and Huet, D. and Hughey, B. and Husa, S. and Huttner, S. H. and Huynh-Dinh, T. and Idrisy, A. and Indik, N. and Ingram, D. R. and Inta, R. and Isa, H. N. and Isac, J.-M. and Isi, M. and Islas, G. and Isogai, T. and Iyer, B. R. and Izumi, K. and Jacobson, M. B. and Jacqmin, T. and Jang, H. and Jani, K. and Jaranowski, P. and Jawahar, S. and Jiménez-Forteza, F. and Johnson, W. W. and Johnson-McDaniel, N. K. and Jones, D. I. and Jones, R. and Jonker, R. J. G. and Ju, L. and Haris, K. and Kalaghatgi, C. V. and Kalogera, V. and Kandhasamy, S. and Kang, G. and Kanner, J. B. and Karki, S. and Kasprzack, M. and Katsavounidis, E. and Katzman, W. and Kaufer, S. and Kaur, T. and Kawabe, K. and Kawazoe, F. and Kéfélian, F. and Kehl, M. S. and Keitel, D. and Kelley, D. B. and Kells, W. and Kennedy, R. and Keppel, D. G. and Key, J. S. and Khalaidovski, A. and Khalili, F. Y. and Khan, I. and Khan, S. and Khan, Z. and Khazanov, E. A. and Kijbunchoo, N. and Kim, C. and Kim, J. and Kim, K. and Kim, Nam-Gyu and Kim, Namjun and Kim, Y.-M. and King, E. J. and King, P. J. and Kinzel, D. L. and Kissel, J. S. and Kleybolte, L. and Klimenko, S. and Koehlenbeck, S. M. and Kokeyama, K. and Koley, S. and Kondrashov, V. and Kontos, A. and Koranda, S. and Korobko, M. and Korth, W. Z. and Kowalska, I. and Kozak, D. B. and Kringel, V. and Krishnan, B. and Królak, A. and Krueger, C. and Kuehn, G. and Kumar, P. and Kumar, R. and Kuo, L. and Kutynia, A. and Kwee, P. and Lackey, B. D. and Landry, M. and Lange, J. and Lantz, B. and Lasky, P. D. and Lazzarini, A. and Lazzaro, C. and Leaci, P. and Leavey, S. and Lebigot, E. O. and Lee, C. H. and Lee, H. K. and Lee, H. M. and Lee, K. and Lenon, A. and Leonardi, M. and Leong, J. R. and Leroy, N. and Letendre, N. and Levin, Y. and Levine, B. M. and Li, T. G. F. and Libson, A. and Littenberg, T. B. and Lockerbie, N. A. and Logue, J. and Lombardi, A. L. and London, L. T. and Lord, J. E. and Lorenzini, M. and Loriette, V. and Lormand, M. and Losurdo, G. and Lough, J. D. and Lousto, C. O. and Lovelace, G. and Lück, H. and Lundgren, A. P. and Luo, J. and Lynch, R. and Ma, Y. and MacDonald, T. and Machenschalk, B. and MacInnis, M. and Macleod, D. M. and Magaña-Sandoval, F. and Magee, R. M. and Mageswaran, M. and Majorana, E. and Maksimovic, I. and Malvezzi, V. and Man, N. and Mandel, I. and Mandic, V. and Mangano, V. and Mansell, G. L. and Manske, M. and Mantovani, M. and Marchesoni, F. and Marion, F. and Márka, S. and Márka, Z. and Markosyan, A. S. and Maros, E. and Martelli, F. and Martellini, L. and Martin, I. W. and Martin, R. M. and Martynov, D. V. and Marx, J. N. and Mason, K. and Masserot, A. and Massinger, T. J. and Masso-Reid, M. and Matichard, F. and Matone, L. and Mavalvala, N. and Mazumder, N. and Mazzolo, G. and McCarthy, R. and McClelland, D. E. and McCormick, S. and McGuire, S. C. and McIntyre, G. and McIver, J. and McManus, D. J. and McWilliams, S. T. and Meacher, D. and Meadors, G. D. and Meidam, J. and Melatos, A. and Mendell, G. and Mendoza-Gandara, D. and Mercer, R. A. and Merilh, E. and Merzougui, M. and Meshkov, S. and Messenger, C. and Messick, C. and Meyers, P. M. and Mezzani, F. and Miao, H. and Michel, C. and Middleton, H. and Mikhailov, E. E. and Milano, L. and Miller, J. and Millhouse, M. and Minenkov, Y. and Ming, J. and Mirshekari, S. and Mishra, C. and Mitra, S. and Mitrofanov, V. P. and Mitselmakher, G. and Mittleman, R. and Moggi, A. and Mohan, M. and Mohapatra, S. R. P. and Montani, M. and Moore, B. C. and Moore, C. J. and Moraru, D. and Moreno, G. and Morriss, S. R. and Mossavi, K. and Mours, B. and Mow-Lowry, C. M. and Mueller, C. L. and Mueller, G. and Muir, A. W. and Mukherjee, Arunava and Mukherjee, D. and Mukherjee, S. and Mukund, N. and Mullavey, A. and Munch, J. and Murphy, D. J. and Murray, P. G. and Mytidis, A. and Nardecchia, I. and Naticchioni, L. and Nayak, R. K. and Necula, V. and Nedkova, K. and Nelemans, G. and Neri, M. and Neunzert, A. and Newton, G. and Nguyen, T. T. and Nielsen, A. B. and Nissanke, S. and Nitz, A. and Nocera, F. and Nolting, D. and Normandin, M. E. N. and Nuttall, L. K. and Oberling, J. and Ochsner, E. and O’Dell, J. and Oelker, E. and Ogin, G. H. and Oh, J. J. and Oh, S. H. and Ohme, F. and Oliver, M. and Oppermann, P. and Oram, Richard J. and O’Reilly, B. and O’Shaughnessy, R. and Ott, C. D. and Ottaway, D. J. and Ottens, R. S. and Overmier, H. and Owen, B. J. and Pai, A. and Pai, S. A. and Palamos, J. R. and Palashov, O. and Palomba, C. and Pal-Singh, A. and Pan, H. and Pan, Y. and Pankow, C. and Pannarale, F. and Pant, B. C. and Paoletti, F. and Paoli, A. and Papa, M. A. and Paris, H. R. and Parker, W. and Pascucci, D. and Pasqualetti, A. and Passaquieti, R. and Passuello, D. and Patricelli, B. and Patrick, Z. and Pearlstone, B. L. and Pedraza, M. and Pedurand, R. and Pekowsky, L. and Pele, A. and Penn, S. and Perreca, A. and Pfeiffer, H. P. and Phelps, M. and Piccinni, O. and Pichot, M. and Pickenpack, M. and Piergiovanni, F. and Pierro, V. and Pillant, G. and Pinard, L. and Pinto, I. M. and Pitkin, M. and Poeld, J. H. and Poggiani, R. and Popolizio, P. and Post, A. and Powell, J. and Prasad, J. and Predoi, V. and Premachandra, S. S. and Prestegard, T. and Price, L. R. and Prijatelj, M. and Principe, M. and Privitera, S. and Prix, R. and Prodi, G. A. and Prokhorov, L. and Puncken, O. and Punturo, M. and Puppo, P. and Pürrer, M. and Qi, H. and Qin, J. and Quetschke, V. and Quintero, E. A. and Quitzow-James, R. and Raab, F. J. and Rabeling, D. S. and Radkins, H. and Raffai, P. and Raja, S. and Rakhmanov, M. and Ramet, C. R. and Rapagnani, P. and Raymond, V. and Razzano, M. and Re, V. and Read, J. and Reed, C. M. and Regimbau, T. and Rei, L. and Reid, S. and Reitze, D. H. and Rew, H. and Reyes, S. D. and Ricci, F. and Riles, K. and Robertson, N. A. and Robie, R. and Robinet, F. and Rocchi, A. and Rolland, L. and Rollins, J. G. and Roma, V. J. and Romano, J. D. and Romano, R. and Romanov, G. and Romie, J. H. and Rosińska, D. and Rowan, S. and Rüdiger, A. and Ruggi, P. and Ryan, K. and Sachdev, S. and Sadecki, T. and Sadeghian, L. and Salconi, L. and Saleem, M. and Salemi, F. and Samajdar, A. and Sammut, L. and Sampson, L. M. and Sanchez, E. J. and Sandberg, V. and Sandeen, B. and Sanders, G. H. and Sanders, J. R. and Sassolas, B. and Sathyaprakash, B. S. and Saulson, P. R. and Sauter, O. and Savage, R. L. and Sawadsky, A. and Schale, P. and Schilling, R. and Schmidt, J. and Schmidt, P. and Schnabel, R. and Schofield, R. M. S. and Schönbeck, A. and Schreiber, E. and Schuette, D. and Schutz, B. F. and Scott, J. and Scott, S. M. and Sellers, D. and Sengupta, A. S. and Sentenac, D. and Sequino, V. and Sergeev, A. and Serna, G. and Setyawati, Y. and Sevigny, A. and Shaddock, D. A. and Shaffer, T. and Shah, S. and Shahriar, M. S. and Shaltev, M. and Shao, Z. and Shapiro, B. and Shawhan, P. and Sheperd, A. and Shoemaker, D. H. and Shoemaker, D. M. and Siellez, K. and Siemens, X. and Sigg, D. and Silva, A. D. and Simakov, D. and Singer, A. and Singer, L. P. and Singh, A. and Singh, R. and Singhal, A. and Sintes, A. M. and Slagmolen, B. J. J. and Smith, J. R. and Smith, M. R. and Smith, N. D. and Smith, R. J. E. and Son, E. J. and Sorazu, B. and Sorrentino, F. and Souradeep, T. and Srivastava, A. K. and Staley, A. and Steinke, M. and Steinlechner, J. and Steinlechner, S. and Steinmeyer, D. and Stephens, B. C. and Stevenson, S. P. and Stone, R. and Strain, K. A. and Straniero, N. and Stratta, G. and Strauss, N. A. and Strigin, S. and Sturani, R. and Stuver, A. L. and Summerscales, T. Z. and Sun, L. and Sutton, P. J. and Swinkels, B. L. and Szczepańczyk, M. J. and Tacca, M. and Talukder, D. and Tanner, D. B. and Tápai, M. and Tarabrin, S. P. and Taracchini, A. and Taylor, R. and Theeg, T. and Thirugnanasambandam, M. P. and Thomas, E. G. and Thomas, M. and Thomas, P. and Thorne, K. A. and Thorne, K. S. and Thrane, E. and Tiwari, S. and Tiwari, V. and Tokmakov, K. V. and Tomlinson, C. and Tonelli, M. and Torres, C. V. and Torrie, C. I. and Töyrä, D. and Travasso, F. and Traylor, G. and Trifirò, D. and Tringali, M. C. and Trozzo, L. and Tse, M. and Turconi, M. and Tuyenbayev, D. and Ugolini, D. and Unnikrishnan, C. S. and Urban, A. L. and Usman, S. A. and Vahlbruch, H. and Vajente, G. and Valdes, G. and Vallisneri, M. and Van Bakel, N. and Van Beuzekom, M. and Van Den Brand, J. F. J. and Van Den Broeck, C. and Vander-Hyde, D. C. and Van Der Schaaf, L. and Van Heijningen, J. V. and Van Veggel, A. A. and Vardaro, M. and Vass, S. and Vasúth, M. and Vaulin, R. and Vecchio, A. and Vedovato, G. and Veitch, J. and Veitch, P. J. and Venkateswara, K. and Verkindt, D. and Vetrano, F. and Viceré, A. and Vinciguerra, S. and Vine, D. J. and Vinet, J.-Y. and Vitale, S. and Vo, T. and Vocca, H. and Vorvick, C. and Voss, D. and Vousden, W. D. and Vyatchanin, S. P. and Wade, A. R. and Wade, L. E. and Wade, M. and Waldman, S. J. and Walker, M. and Wallace, L. and Walsh, S. and Wang, G. and Wang, H. and Wang, M. and Wang, X. and Wang, Y. and Ward, H. and Ward, R. L. and Warner, J. and Was, M. and Weaver, B. and Wei, L.-W. and Weinert, M. and Weinstein, A. J. and Weiss, R. and Welborn, T. and Wen, L. and Weßels, P. and Westphal, T. and Wette, K. and Whelan, J. T. and Whitcomb, S. E. and White, D. J. and Whiting, B. F. and Wiesner, K. and Wilkinson, C. and Willems, P. A. and Williams, L. and Williams, R. D. and Williamson, A. R. and Willis, J. L. and Willke, B. and Wimmer, M. H. and Winkelmann, L. and Winkler, W. and Wipf, C. C. and Wiseman, A. G. and Wittel, H. and Woan, G. and Worden, J. and Wright, J. L. and Wu, G. and Yablon, J. and Yakushin, I. and Yam, W. and Yamamoto, H. and Yancey, C. C. and Yap, M. J. and Yu, H. and Yvert, M. and Zadrożny, A. and Zangrando, L. and Zanolin, M. and Zendri, J.-P. and Zevin, M. and Zhang, F. and Zhang, L. and Zhang, M. and Zhang, Y. and Zhao, C. and Zhou, M. and Zhou, Z. and Zhu, X. J. and Zucker, M. E. and Zuraw, S. E. and Zweizig, J. and {LIGO Scientific Collaboration and Virgo Collaboration}},
    month = feb,
    year = {2016},
    pages = {061102},
}

@article{Burke-Spolaor2019,
    title = {The astrophysics of nanohertz gravitational waves},
    volume = {27},
    issn = {0935-4956, 1432-0754},
    url = {http://link.springer.com/10.1007/s00159-019-0115-7},
    doi = {10.1007/s00159-019-0115-7},
    language = {en},
    number = {1},
    urldate = {2026-06-22},
    journal = {The Astronomy and Astrophysics Review},
    author = {Burke-Spolaor, Sarah and Taylor, Stephen R. and Charisi, Maria and Dolch, Timothy and Hazboun, Jeffrey S. and Holgado, A. Miguel and Kelley, Luke Zoltan and Lazio, T. Joseph W. and Madison, Dustin R. and McMann, Natasha and Mingarelli, Chiara M. F. and Rasskazov, Alexander and Siemens, Xavier and Simon, Joseph J. and Smith, Tristan L.},
    month = dec,
    year = {2019},
    pages = {5},
}

@article{Gross2025,
    title = {Varstrometry for {Off}-nucleus and {Dual} {Sub}-{Kpc} {AGN} ({VODKA}): {A} {Mix} of {Singles}, {Lenses}, and {True} {Duals} at {Cosmic} {Noon}},
    volume = {989},
    issn = {0004-637X, 1538-4357},
    shorttitle = {Varstrometry for {Off}-nucleus and {Dual} {Sub}-{Kpc} {AGN} ({VODKA})},
    url = {https://iopscience.iop.org/article/10.3847/1538-4357/ade671},
    doi = {10.3847/1538-4357/ade671},
    number = {1},
    urldate = {2026-06-24},
    journal = {The Astrophysical Journal},
    author = {Gross, Arran C. and Chen, Yu-Ching and Oguri, Masamune and Nolan, Liam and Liu, Xin and Shen, Yue and Zhuang, Ming-Yang and Li, Junyao and Zakamska, Nadia L. and Hwang, Hsiang-Chih and Ishikawa, Yuzo},
    month = aug,
    year = {2025},
    pages = {112},
}

@article{Katz2019,
    title = {Probing {Massive} {Black} {Hole} {Binary} {Populations} with {LISA}},
    copyright = {https://academic.oup.com/journals/pages/open\_access/funder\_policies/chorus/standard\_publication\_model},
    issn = {0035-8711, 1365-2966},
    url = {https://academic.oup.com/mnras/advance-article/doi/10.1093/mnras/stz3102/5614512},
    doi = {10.1093/mnras/stz3102},
    language = {en},
    urldate = {2026-06-24},
    journal = {Monthly Notices of the Royal Astronomical Society},
    author = {Katz, Michael L and Kelley, Luke Zoltan and Dosopoulou, Fani and Berry, Samantha and Blecha, Laura and Larson, Shane L},
    month = nov,
    year = {2019},
    pages = {stz3102},
}

@article{Enoki2004,
    title = {Gravitational {Waves} from {Supermassive} {Black} {Hole} {Coalescence} in a {Hierarchical} {Galaxy} {Formation} {Model}},
    volume = {615},
    issn = {0004-637X, 1538-4357},
    url = {https://iopscience.iop.org/article/10.1086/424475},
    doi = {10.1086/424475},
    language = {en},
    number = {1},
    urldate = {2026-06-24},
    journal = {The Astrophysical Journal},
    author = {Enoki, Motohiro and Inoue, Kaiki T. and Nagashima, Masahiro and Sugiyama, Naoshi},
    month = nov,
    year = {2004},
    pages = {19--28},
}

@article{Reeves2026,
    title = {Winds of {Change}: {XRISM} {Resolve} {X}-{Ray} {Spectroscopy} of {NGC} 4051},
    volume = {1001},
    issn = {0004-637X, 1538-4357},
    shorttitle = {Winds of {Change}},
    url = {https://iopscience.iop.org/article/10.3847/1538-4357/ae5527},
    doi = {10.3847/1538-4357/ae5527},
    number = {2},
    urldate = {2026-06-26},
    journal = {The Astrophysical Journal},
    author = {Reeves, James N. and Ogawa, Shoji and Turner, Tracey J. and Braito, Valentina and Yamada, Satoshi and Kraemer, Steven B. and Noda, Hirofumi and Falcão, Anna Trindade and Elvis, Martin and Fabbiano, Giuseppina},
    month = apr,
    year = {2026},
    pages = {137},
}

@article{Xiang2025,
    title = {{XRISM} {Spectroscopy} of {Accretion}-driven {Wind} {Feedback} in {NGC} 4151},
    volume = {988},
    issn = {2041-8205, 2041-8213},
    url = {https://iopscience.iop.org/article/10.3847/2041-8213/adee9b},
    doi = {10.3847/2041-8213/adee9b},
    number = {2},
    urldate = {2026-06-26},
    journal = {The Astrophysical Journal Letters},
    author = {Xiang, Xin and Miller, Jon M. and Behar, Ehud and Boissay-Malaquin, Rozenn and Brenneman, Laura and Buhariwalla, Margaret and Byun, Doyee and Done, Chris and Gallo, Luigi and Gerolymatou, Dimitra and Hagen, Scott and Kaastra, Jelle and Paltani, Stephane and Porter, Frederick S. and Mushotzky, Richard and Noda, Hirofumi and Mehdipour, Missagh and Minezaki, Takeo and Tashiro, Makoto and Zoghbi, Abderahmen},
    month = aug,
    year = {2025},
    pages = {L54},
}

@article{Virtanen2020,
    title = {{SciPy} 1.0: fundamental algorithms for scientific computing in {Python}},
    volume = {17},
    issn = {1548-7091, 1548-7105},
    shorttitle = {{SciPy} 1.0},
    url = {https://www.nature.com/articles/s41592-019-0686-2},
    doi = {10.1038/s41592-019-0686-2},
    language = {en},
    number = {3},
    urldate = {2026-06-29},
    journal = {Nature Methods},
    author = {Virtanen, Pauli and Gommers, Ralf and Oliphant, Travis E. and Haberland, Matt and Reddy, Tyler and Cournapeau, David and Burovski, Evgeni and Peterson, Pearu and Weckesser, Warren and Bright, Jonathan and Van Der Walt, Stéfan J. and Brett, Matthew and Wilson, Joshua and Millman, K. Jarrod and Mayorov, Nikolay and Nelson, Andrew R. J. and Jones, Eric and Kern, Robert and Larson, Eric and Carey, C J and Polat, İlhan and Feng, Yu and Moore, Eric W. and VanderPlas, Jake and Laxalde, Denis and Perktold, Josef and Cimrman, Robert and Henriksen, Ian and Quintero, E. A. and Harris, Charles R. and Archibald, Anne M. and Ribeiro, Antônio H. and Pedregosa, Fabian and Van Mulbregt, Paul and {SciPy 1.0 Contributors} and Vijaykumar, Aditya and Bardelli, Alessandro Pietro and Rothberg, Alex and Hilboll, Andreas and Kloeckner, Andreas and Scopatz, Anthony and Lee, Antony and Rokem, Ariel and Woods, C. Nathan and Fulton, Chad and Masson, Charles and Häggström, Christian and Fitzgerald, Clark and Nicholson, David A. and Hagen, David R. and Pasechnik, Dmitrii V. and Olivetti, Emanuele and Martin, Eric and Wieser, Eric and Silva, Fabrice and Lenders, Felix and Wilhelm, Florian and Young, G. and Price, Gavin A. and Ingold, Gert-Ludwig and Allen, Gregory E. and Lee, Gregory R. and Audren, Hervé and Probst, Irvin and Dietrich, Jörg P. and Silterra, Jacob and Webber, James T and Slavič, Janko and Nothman, Joel and Buchner, Johannes and Kulick, Johannes and Schönberger, Johannes L. and De Miranda Cardoso, José Vinícius and Reimer, Joscha and Harrington, Joseph and Rodríguez, Juan Luis Cano and Nunez-Iglesias, Juan and Kuczynski, Justin and Tritz, Kevin and Thoma, Martin and Newville, Matthew and Kümmerer, Matthias and Bolingbroke, Maximilian and Tartre, Michael and Pak, Mikhail and Smith, Nathaniel J. and Nowaczyk, Nikolai and Shebanov, Nikolay and Pavlyk, Oleksandr and Brodtkorb, Per A. and Lee, Perry and McGibbon, Robert T. and Feldbauer, Roman and Lewis, Sam and Tygier, Sam and Sievert, Scott and Vigna, Sebastiano and Peterson, Stefan and More, Surhud and Pudlik, Tadeusz and Oshima, Takuya and Pingel, Thomas J. and Robitaille, Thomas P. and Spura, Thomas and Jones, Thouis R. and Cera, Tim and Leslie, Tim and Zito, Tiziano and Krauss, Tom and Upadhyay, Utkarsh and Halchenko, Yaroslav O. and Vázquez-Baeza, Yoshiki},
    month = mar,
    year = {2020},
    pages = {261--272},
}

@article{Luhtaru2021,
    title = {What {Makes} {Quadruply} {Lensed} {Quasars} {Quadruple}?},
    volume = {915},
    issn = {0004-637X, 1538-4357},
    url = {https://iopscience.iop.org/article/10.3847/1538-4357/abfda1},
    doi = {10.3847/1538-4357/abfda1},
    number = {1},
    urldate = {2026-06-29},
    journal = {The Astrophysical Journal},
    author = {Luhtaru, Richard and Schechter, Paul L. and De Soto, Kaylee M.},
    month = jul,
    year = {2021},
    pages = {4},
}

@article{Schechter2024,
    title = {Witt’s {Hyperbola} {Is} {Both} {Predicted} and {Observed} to {Pass} {Close} to the {Lensing} {Galaxies} in {Quadruple} {Quasars}},
    volume = {975},
    issn = {0004-637X, 1538-4357},
    url = {https://iopscience.iop.org/article/10.3847/1538-4357/ad7f4d},
    doi = {10.3847/1538-4357/ad7f4d},
    number = {1},
    urldate = {2026-06-29},
    journal = {The Astrophysical Journal},
    author = {Schechter, Paul L. and Luhtaru, Richard},
    month = nov,
    year = {2024},
    pages = {62},
}

@article{Vernardos2024,
    title = {Microlensing of {Strongly} {Lensed} {Quasars}},
    volume = {220},
    issn = {0038-6308, 1572-9672},
    url = {https://link.springer.com/10.1007/s11214-024-01043-8},
    doi = {10.1007/s11214-024-01043-8},
    language = {en},
    number = {1},
    urldate = {2026-06-29},
    journal = {Space Science Reviews},
    author = {Vernardos, G. and Sluse, D. and Pooley, D. and Schmidt, R. W. and Millon, M. and Weisenbach, L. and Motta, V. and Anguita, T. and Saha, P. and O’Dowd, M. and Peel, A. and Schechter, P. L.},
    month = feb,
    year = {2024},
    pages = {14},
}
\bibliographystyle{aasjournalv7}
\clearpage
\appendix

\section{Model Checks}
\subsection{Results with Image F Removed}\label{app:F_removed}
\begin{figure}[!h]
    \centering
    \begin{subfigure}{0.48\linewidth}
        \centering
        \includegraphics[width=\linewidth]{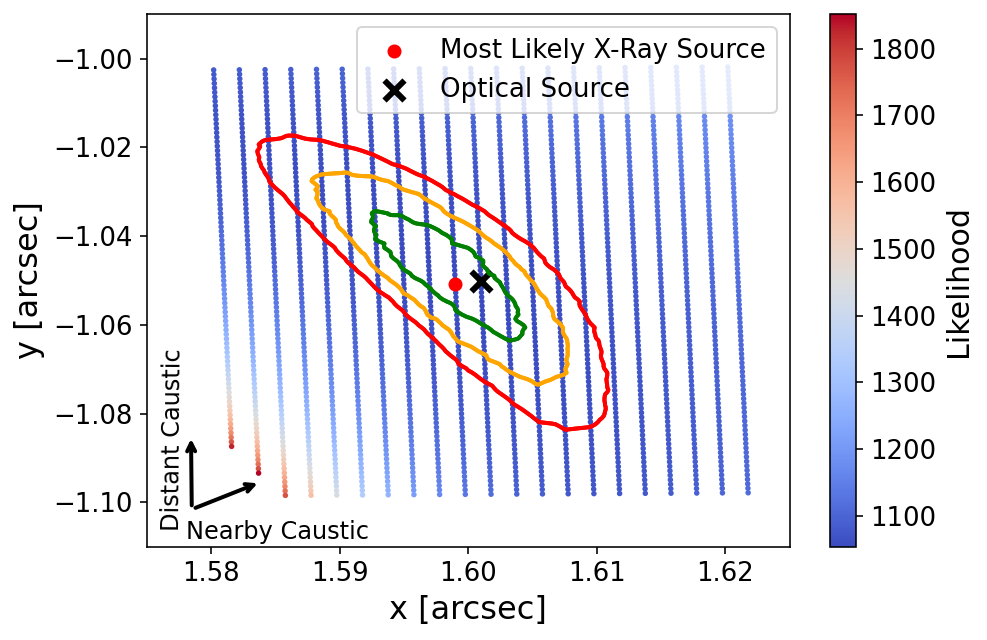}
        \caption{Broadband $0.5-7\ \mathrm{keV}$}
        \label{fig:contour_plot_noF}
    \end{subfigure}
    \hfill
    \begin{subfigure}{0.48\linewidth}
        \centering
        \includegraphics[width=\linewidth]{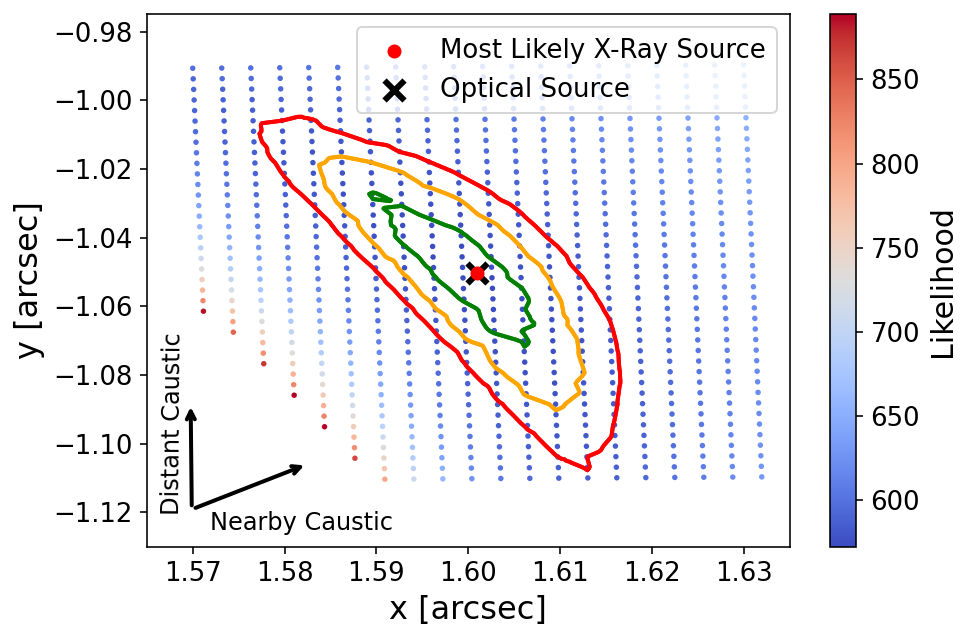}
        \caption{Hard Band $2-7\ \mathrm{keV}$}
        \label{fig:hard_contour_plot_noF}
    \end{subfigure}

    \vspace{0.5cm}

    \begin{subfigure}{0.48\linewidth}
        \centering
        \includegraphics[width=\linewidth]{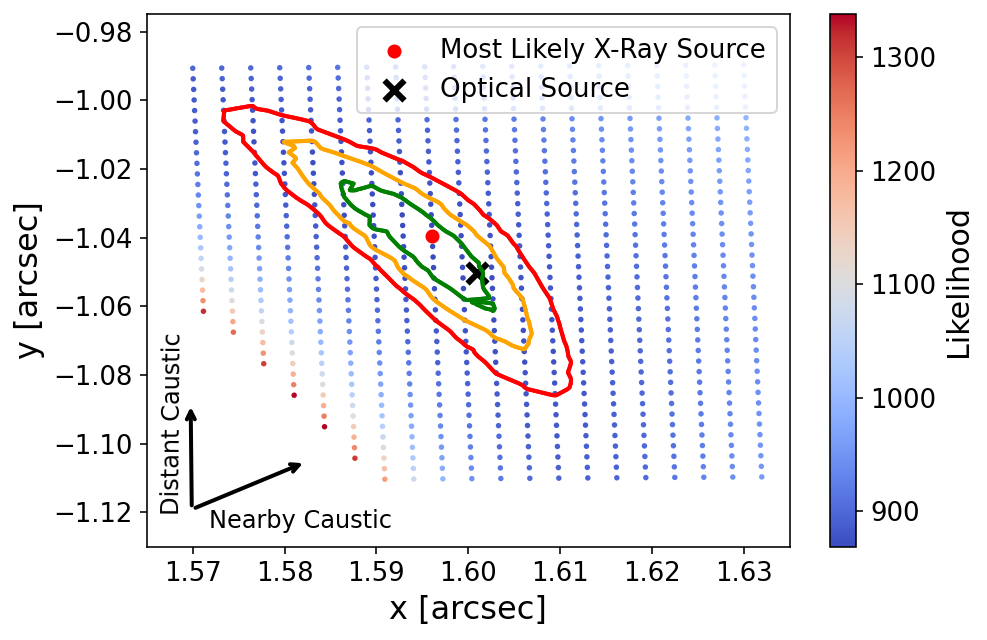}
        \caption{Soft Band $0.5-2\ \mathrm{keV}$}
        \label{fig:soft_contour_plot_noF}
    \end{subfigure}

    \caption{Source-plane reconstructions for the X-ray emission relative to the optical source with image F removed from the ML calculation. The contours are at 68\% (green), 95.5\% (orange), and 99.7\% (red) confidence levels. Each trial source in the grid is color-coded by the ML resulting from the analysis of both ObsIDs (Equation \ref{eq:Cash_stat}), with the most likely position for the X-ray source corresponding to the point with the lowest likelihood value. In the hard band, the most likely X-ray source is the optical source position found from the lens model.}
    \label{fig:contour_noF}
\end{figure}

The most likely X-ray source position in each band remains cospatial with the optical source within the $1\sigma$ contour when image F is not used in the double ML calculation. Therefore, as expected from Equation \ref{eq:Cash_stat}, the choice to either include or not include image F does not significantly alter the most likely X-ray position due to its low number of X-ray counts.

\clearpage
\subsection{Results with Image F Moved to Lens Model Position}\label{app:F_moved}
\begin{figure}[!h]
    \centering
    \begin{subfigure}{0.48\linewidth}
        \centering
        \includegraphics[width=\linewidth]{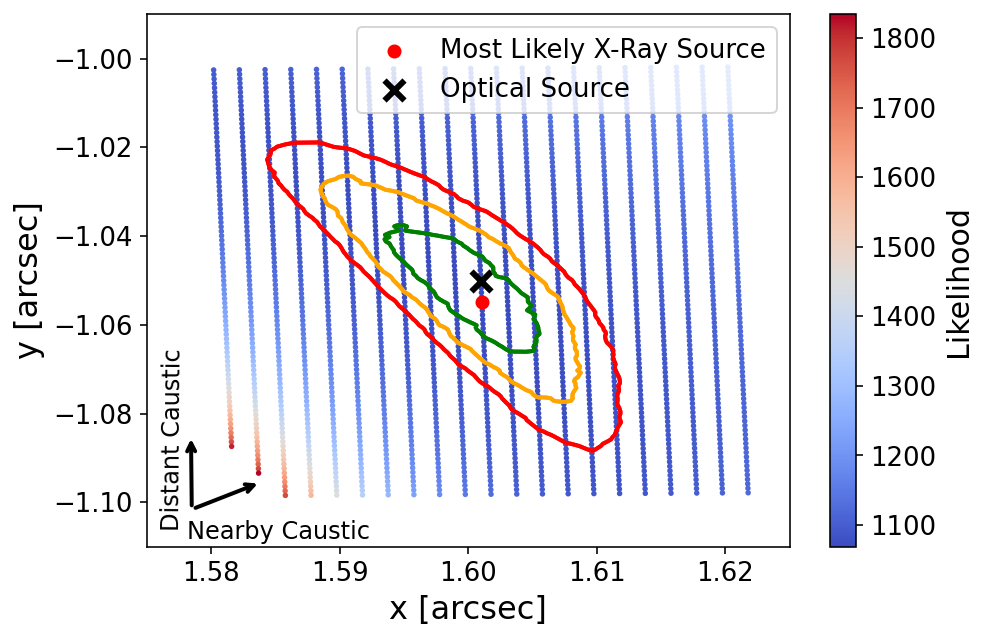}
        \caption{Broadband $0.5-7\ \mathrm{keV}$}
        \label{fig:contour_plot_modelF}
    \end{subfigure}
    \hfill
    \begin{subfigure}{0.48\linewidth}
        \centering
        \includegraphics[width=\linewidth]{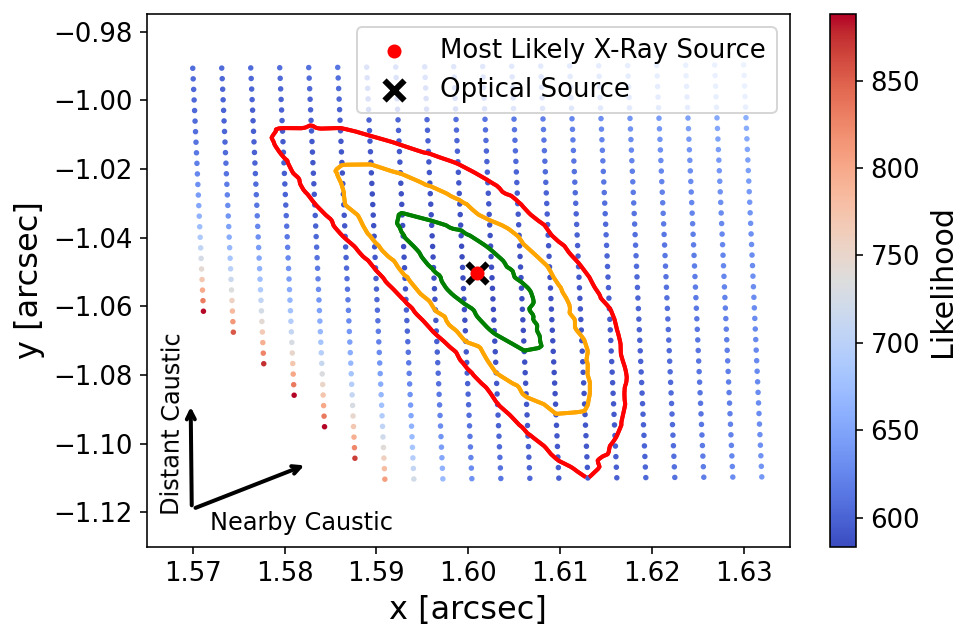}
        \caption{Hard Band $2-7\ \mathrm{keV}$}
        \label{fig:hard_contour_plot_modelF}
    \end{subfigure}

    \vspace{0.5cm}

    \begin{subfigure}{0.48\linewidth}
        \centering
        \includegraphics[width=\linewidth]{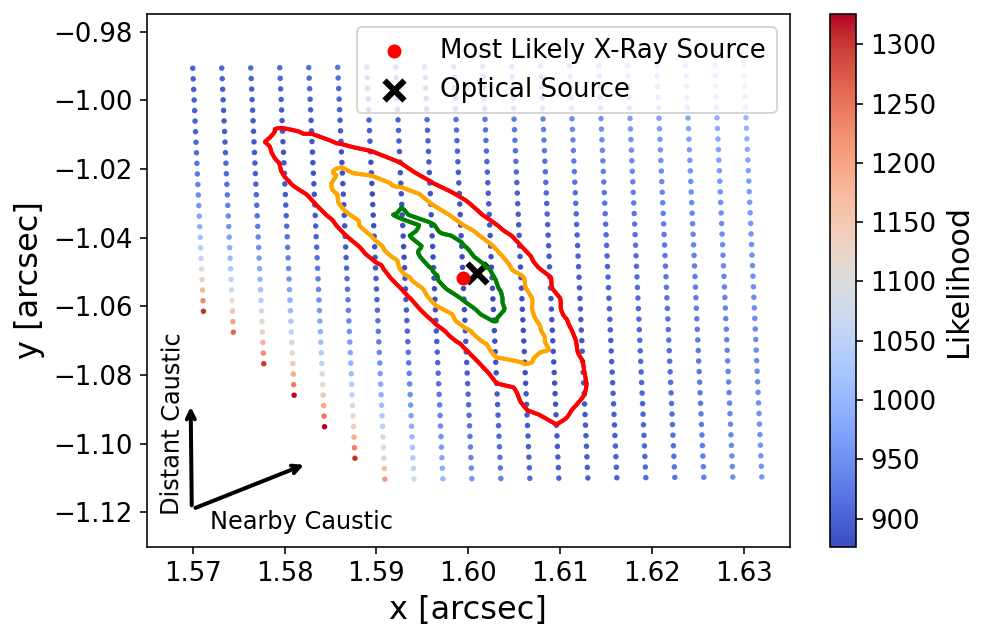}
        \caption{Soft Band $0.5-2\ \mathrm{keV}$}
        \label{fig:soft_contour_plot_modelF}
    \end{subfigure}

    \caption{Source-plane reconstructions for the X-ray emission relative to the optical source with the `observed' position of image F moved to the lens model-predicted position. The contours are at 68\% (green), 95.5\% (orange), and 99.7\% (red) confidence levels. Each trial source in the grid is color-coded by the ML resulting from the analysis of both ObsIDs (Equation \ref{eq:Cash_stat}), with the most likely position for the X-ray source corresponding to the point with the lowest likelihood value. In the hard band, the most likely X-ray source is the optical source position found from the lens model.}
    \label{fig:contour_modelF}
\end{figure}

The most likely X-ray source position in each band remains cospatial with the optical source within the $1\sigma$ contour when the `observed' position of image F is changed to the lens model-predicted position in the double ML calculation. Therefore, as expected from Equation \ref{eq:Cash_stat}, this shift in position for image F does not significantly alter the most likely X-ray position due to its low number of X-ray counts.

\clearpage
\section{Lens Model Parameter Distributions}\label{app:parameter_distributions}

\vspace{0.5cm}

\begin{center}
\begin{minipage}{0.69\linewidth}
    \includegraphics[width=0.85\linewidth]{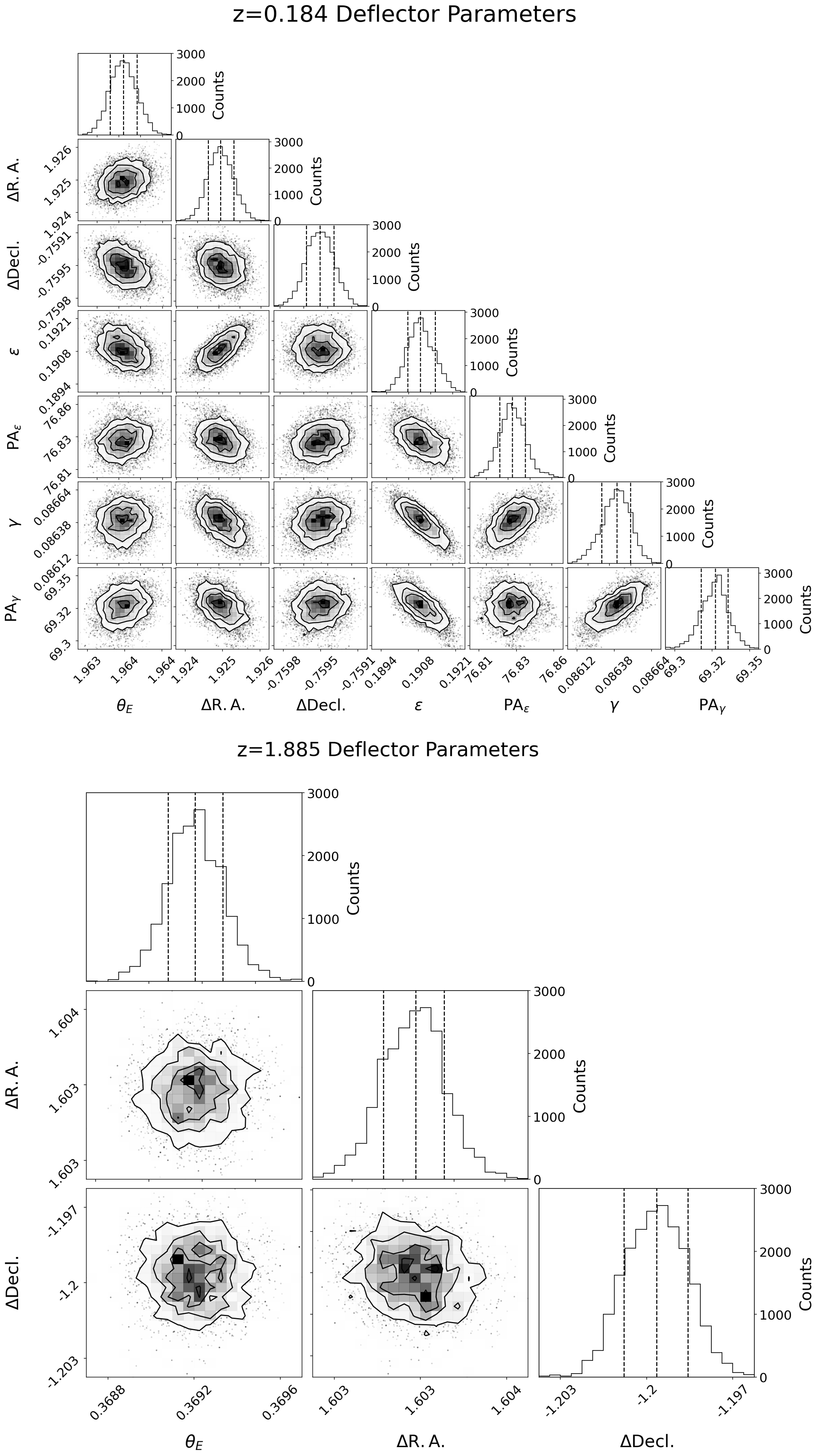}
    \captionof{figure}{The lens model parameters and distributions for the deflectors. The dashed lines on the histograms represent the median and $\pm1\sigma$ regions. The contours represent the $0.5\sigma$, $1\sigma$, $1.5\sigma$, and $2\sigma$ levels. The $1\sigma$ uncertainties are used in Table \ref{tab:params_table}.}
    \label{fig:corner_6image}
\end{minipage}
\end{center}

The lens model parameters and distributions for both deflectors are shown in Figure \ref{fig:corner_6image}.  The distributions were generated from an MCMC sampling in \gravlens as described in Section \ref{sec:lens_model_method_opt} and illustrate correlations between the model parameters.

\clearpage
\section{$z=1.885$ Deflector Model}\label{app:distant_deflector}

\vspace{0.5cm}

\begin{center}
\begin{minipage}{\linewidth}
    \centering
    \includegraphics[width=1\linewidth]{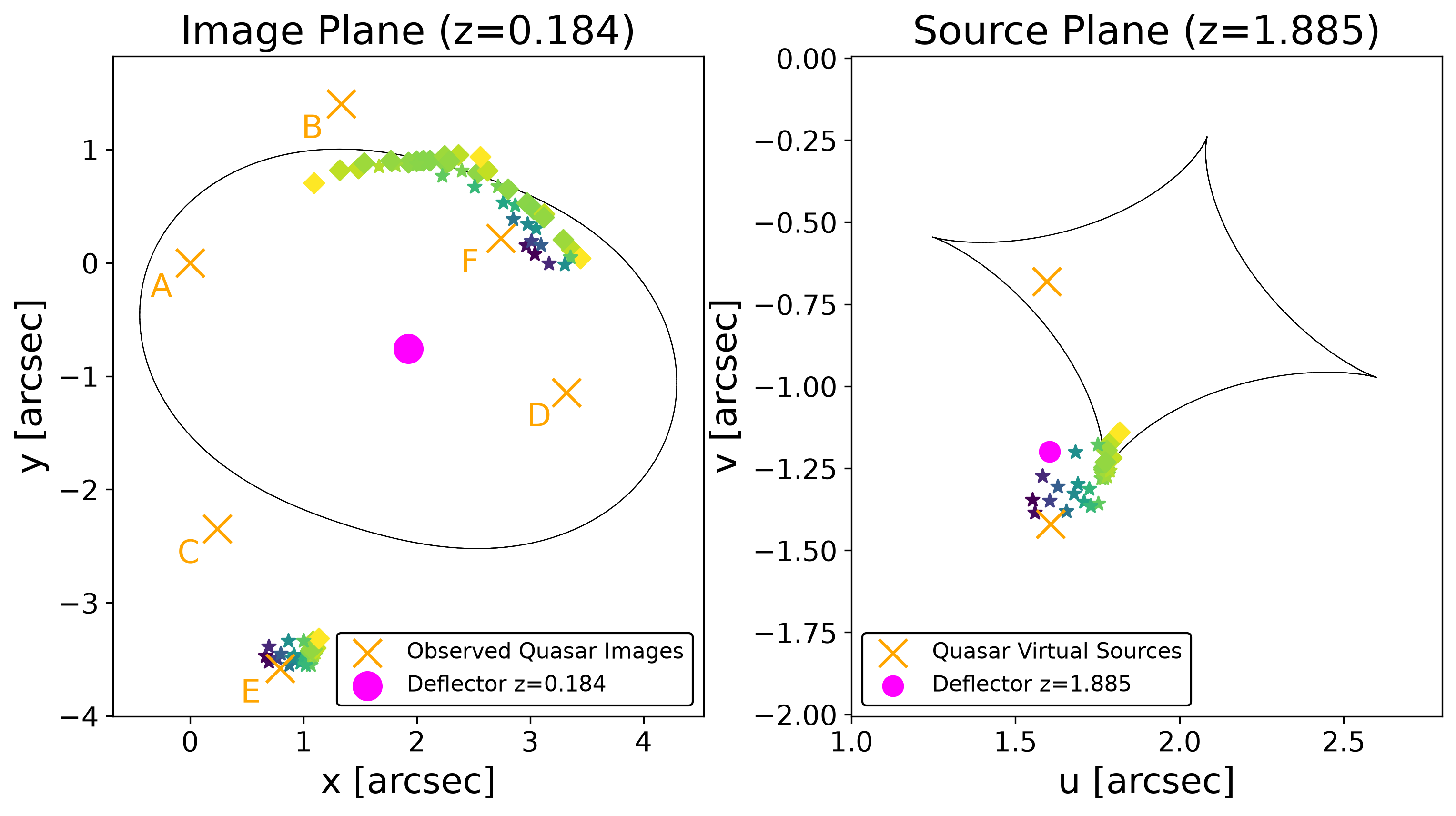}
    \captionof{figure}{The F814W \HST image was used to determine the positions of the points along the arcs as it provides better spatial resolution than the F160W filter, despite the arcs being brighter in the latter. The critical curve ($z=0.184$ plane) and inner caustic ($z=1.885$ plane) of the nearby deflector are shown by the black lines in the left and right panels, respectively. The points selected along the arcs, together with their corresponding source positions and additional images, are marked by stars and diamonds, with each set of images and its associated source shown in the same color. The diamonds represent the quadruply lensed sources along with their predicted image-plane positions, while the stars represent the doubly lensed sources and their respective image-plane positions.}
    \label{fig:arc_sources}
\end{minipage}
\end{center}

As a self-consistency check for our mass model, we place several ``test'' image-plane positions along the red arcs to infer their corresponding source-plane position (i.e.~that of the $z=1.885$ deflector) and assess its consistency with the source-plane position of the $z=1.885$ deflector predicted by our mass model. We selected 23 points along the larger arc near image F and 4 points along the smaller arc near image E and used the \texttt{findsrc} command in \gravlens to determine their corresponding source positions and identify any additional images, keeping the mass model fixed (Figure \ref{fig:arc_sources}). Of the 27 points selected along the arcs, 8 correspond to source positions that produced four images (diamonds in Figure \ref{fig:arc_sources}), while the remaining 19 correspond to source positions that produced two images (stars in Figure \ref{fig:arc_sources}), as the extended source crosses the inner caustic of the nearby deflector.~As indicated in Section \ref{sec:lens_model}, the source-plane position of the distant deflector inferred from the arcs is slightly offset from the position predicted by the lens model, with offsets ranging from $\sim0\farcs05-0\farcs2$ across all arc sources. This could be due to the visible light from the galaxy not being perfectly aligned with the center of its mass distribution. This could also be due to regions of the $z = 1.885$ deflector galaxy being less highly magnified, resulting in arcs that are too faint to be detected.~In the latter scenario,~we would not be reconstructing the full extent of the galaxy. Deeper observations of J1721 may reveal additional images or arcs associated with the distant deflector, providing an interesting opportunity for future investigation.

\end{document}